\documentclass[showpacs,amsmath,amssymb,twocolumn,superscriptaddress,notitlepage,preprintnumbers,pra,floatfix]{revtex4-1}

\usepackage{qcircuit}
\usepackage[dvips]{graphicx}
\usepackage{amsmath,amssymb,amsthm,mathrsfs,amsfonts,dsfont}
\usepackage{subfigure, epsfig}
\usepackage{braket}
\usepackage{bm}
\usepackage{enumerate}
\usepackage{color}
\usepackage{graphicx}
\usepackage{algorithm}
\usepackage{algpseudocode}

\newcommand{\tr}{\mathrm{Tr}}

\newcommand{\mc}[1]{\mathcal{#1}}

\newcommand{\sun}[1]{\textcolor{black}{#1}}

\begin{document}

\title{Mitigating realistic noise in practical noisy intermediate-scale quantum devices}

\date{\today}
\author{Jinzhao Sun}
\email{jinzhao.sun@physics.ox.ac.uk}
\affiliation{Clarendon Laboratory, University of Oxford, Parks Road, Oxford OX1 3PU, United Kingdom}

\author{Xiao Yuan}
\email{xiaoyuan@pku.edu.cn}
\affiliation{Center on Frontiers of Computing Studies, Department of Computer Science, Peking University, Beijing 100871, China}
\affiliation{Stanford Institute for Theoretical Physics, Stanford University, Stanford California 94305, USA}
\affiliation{Department of Materials, University of Oxford, Parks Road, Oxford OX1 3PH, United Kingdom}

\author{{Takahiro Tsunoda}}
\affiliation{Clarendon Laboratory, University of Oxford, Parks Road, Oxford OX1 3PU, United Kingdom}

\author{Vlatko Vedral}
\affiliation{Clarendon Laboratory, University of Oxford, Parks Road, Oxford OX1 3PU, United Kingdom}
\affiliation{Centre for Quantum Technologies, National University of Singapore, Singapore 117543,Singapore}

\author{Simon C. Benjamin}
\affiliation{Department of Materials, University of Oxford, Parks Road, Oxford OX1 3PH, United Kingdom}

\author{Suguru Endo }
\email{suguru.endou.uc@hco.ntt.co.jp}
\affiliation{Department of Materials, University of Oxford, Parks Road, Oxford OX1 3PH, United Kingdom}
\affiliation{NTT Secure Platform Laboratories, NTT Corporation, Musashino 180-8585, Japan}


\begin{abstract}
Quantum error mitigation (QEM) is vital for  noisy intermediate-scale quantum (NISQ) devices. While most conventional QEM schemes assume discrete gate-based circuits with noise appearing either before or after each gate, the assumptions are inappropriate for describing realistic noise that may have strong gate-dependence and complicated nonlocal effects, and general computing models such as analog quantum simulators. To address these challenges, we first extend the scenario, where each computation process, being either digital or analog, is described by a continuous time evolution. For noise from imperfections of the engineered Hamiltonian or additional noise operators, we show it can be effectively suppressed by a novel stochastic QEM method. Since our method only assumes accurate single qubit controls, it is applicable to all digital quantum computers and various analog simulators.
Meanwhile, errors in the mitigation procedure can be suppressed by leveraging the Richardson extrapolation method. As we numerically test our method with various Hamiltonians under energy relaxation and dephasing noise and digital quantum circuits with additional two-qubit crosstalk, we show an improvement of simulation accuracy by two orders. We assess the resource cost of our scheme and conclude the feasibility of accurate quantum computing with NISQ devices.

\end{abstract}

\maketitle

\section{Introduction}
With the experimental demonstration of quantum supremacy~\cite{arute2019quantum}, whether current or near-future noisy intermediate-scale quantum (NISQ) devices are sufficient for realising quantum advantages in practical problems becomes one of the most exciting challenges in quantum computing~\cite{preskill2018quantum}. Since NISQ devices have insufficient qubits to implement fault-tolerance, effective quantum error mitigation (QEM) schemes are crucial for suppressing errors to guarantee the calculation accuracy to surpass the classical limit. Among different QEM schemes via different post-processing mechanisms~\cite{li2017efficient,temme2017error,endo2018practical,PhysRevA.95.042308,huo2018temporally,bonet2018low,dumitrescu2018cloud,otten2019recovering,mcardle2019error,sagastizabal2019experimental,huggins2019efficient,PhysRevX.8.011021,otten2019accounting,mcclean2020decoding,giurgica2020digital,keen2020quantum,cai2020multi,he2020resource,maciejewski2020mitigation,chen2019detector,kwon2020hybrid,strikis2020learning,bravyi2020mitigating,czarnik2020error,zlokapa2020deep,endo2020hybrid,cerezo2020variationalreview}, the probabilistic QEM method is one of the most effective techniques~\cite{temme2017error,endo2018practical}, which fully inverts noise effect by requiring a full tomography of the noise process and assuming noise independently appears either before or after each gate in a digital gate-based quantum computer. While these assumptions are adopted for many QEM schemes, realistic noise is more complicated. Specifically, since every gate is experimentally realised via the time evolution of quantum controls~\cite{kandala2019error,kjaergaard2019superconducting,krantz2019quantum,sheldon2016procedure,arute2019quantum,plantenberg2007demonstration,neill2018blueprint,corcoles2013process,chow2012universal,rigetti2010fully}, noise happens along with the evolution, whose effect inevitably mixes with the gate or process and even scramble nonlocally~(See Appendix). For example, as one of the major noise in superconducting qubits, crosstalk of multi-qubit gates originates from the imperfect time evolution with unwanted interactions~\cite{kjaergaard2019superconducting,krantz2019quantum,takita2017experimental,reagor2018demonstration,sheldon2016procedure,rigetti2010fully}. Therefore, such inherent dynamics-based and nonlocal noise effects make conventional QEM schemes less effective for practical NISQ devices.
Meanwhile, a more natural and noise-robust computation model is via analog quantum simulators~\cite{eisert2015quantum,bohnet2016quantum,an2015experimental,schindler2013quantum,landsman2019verified,li2017measuring,garttner2017measuring,houck2012chip,harris2018phase,garttner2017measuring,li2014experimental,friedenauer2008simulating,zhang2017observation,hart2015observation,gong2013prethermalization,struck2011quantum,wu2016understanding,bernien2017probing,jurcevic2017direct}, which directly emulate the target system without even implementing gates. It also remains an important open challenge to suppress errors for reliable medium- or large-scale analog quantum simulators~\cite{altman2019quantum,georgescu2014quantum,poggi2020quantifying,hauke2012can}.

In this work, we present QEM schemes without assumptions of gate-based circuits or simplified local noise models of each gate. Specifically, we introduce \emph{stochastic error mitigation} for a continuous evolution with noise described by imperfections of the engineered Hamiltonian or super-operators induced from the interaction with the environment~\cite{georgescu2014quantum,houck2012chip,hauke2012can,bylander2011noise}.
Compared to existing methods, such as dynamical decoupling, which are generally limited to low frequency noise and small simulations~\cite{bertet2005dephasing,martinis2003decoherence,walther2012controlling,pokharel2018demonstration}, our work introduces a universal way to mitigate realistic noise under experiment-friendly assumptions.
Our work considers continuous evolution of the system and assumes accurate single-qubit operations, which is applicable to all digital quantum simulators and various analog simulators.
Our method is compatible with existing QEMs, and its combination with Richardson extrapolation can be further leveraged to suppress errors in inaccurate model estimations and recovery operations. We numerically test our scheme for various Hamiltonians with energy relaxation and dephasing noise and a quantum circuit with two-qubit crosstalk noise. We conduct a resource estimation for near-term devices involving up to 100 qubits and show the feasibility of our QEM scheme in the NISQ regime.

\vspace{0.2cm}

\section{Background and framework}
We first introduce the background of analog quantum simulation (AQS) and digital quantum simulation (DQS) with noisy operations.
In a digital gate-based quantum computer, the effect of noise is usually simplified as a quantum channel appearing either before or after each gate, whereas realistic noise occurring in the experimental apparatus is more complicated. Specifically, every gate in digital circuits or every process in analog simulation is physically realised via a continuous real time evolution of a Hamiltonian and therefore errors can either inherently mix with the evolution~---~making it strongly gate or process dependent, or act on multiple number of qubits~---~leading to highly nonlocal correlated effects (crosstalks). For instance, dominant errors in superconducting qubits  are inherent system dephasing or relaxation,  and coherent errors (or crosstalk) when applying entangling gates.
While AQS are believed to be less prone to noise, this holds true mostly comparing to DQS, and when considering an intermediate simulation scale~\cite{altman2019quantum}, outcomes of AQS could be sensitive to noise (for example, see theoretical studies on the sensitivity to errors~\cite{poggi2020quantifying,hauke2012can} and noisy simulation result~\cite{zhang2017observation} of AQS).


Since conventional quantum error mitigation methods are restricted to gate-based digital quantum computers and over-simplified noise models, they fail to work for realistic errors and general continuous quantum processes.
For instance, owing to the restricted set of allowed operations in analog quantum simulators, it is challenging to suppress or correct errors {of a continuous process} in this context~\cite{poggi2020quantifying} and it remains an important open challenge to suppress errors for reliable medium- or large-scale quantum simulators~\cite{altman2019quantum}.
Our work addresses this problem by considering a more general scenario of a continuous process with realistic noise models occurring in quantum simulators.

In particular, we introduce the model that describes either gate syntheses or continuous processes in digital or analog simulation. We consider the ideal evolution of state $\rho_I(t)$ with a target Hamiltonian $H_{\textrm{sys}}$ as
\begin{align}\label{eqn:goodeqn}
\frac{d \rho_I(t)}{dt}=-i[H_{\rm sys}(t), \rho_I(t)].
\end{align}
In practice, we map $H_{\textrm{sys}}$ to a noisy controllable quantum hardware $H_{\rm sim}$, whose time evolution is described by the Lindblad master equation of the noisy state  $\rho_N(t)$  as
\begin{align}\label{eqn:noisyeqn}
\frac{d \rho_N(t)}{dt}=-i[H_{\rm sim}(t), \rho_N(t)]+\lambda \mathcal{L}_{\mathrm{exp}}\big[\rho _N(t)\big].
\end{align}
Here,  $H_{\rm sim}=H_{\rm sys}+\delta H$ corresponds to coherent errors (such as crosstalk or imperfections of Hamiltonian) and $\mathcal{L}_{\mathrm{exp}}[\rho]= \sum_k (2L_k \rho L_k^\dag -L_k^\dag L_k \rho -\rho L_k^\dag L_k )$ is the  noise superoperator with error strength $\lambda$  that discribes inherent coupling with the environment  (such as dephasing and damping)~\cite{houck2012chip,georgescu2014quantum}. Instead of assuming local single qubit noise channel of each gate in conventional QEM, we consider local noise model by assuming local Lindblad terms. We note that local noise operators at instant time $t$ can easily propagate to become global noise after integrating time, which may cause nonlocal noise effects in reality.

Suppose we are interested in measuring the state at time $T$ with an observable ${O}$. The task of QEM is to recover the noiseless measurement outcome $\braket{O}_I=\tr[{O}\rho_I(t)]$ via noisy process.
In general, it would be difficult to efficiently mitigate arbitrary noise with any noise strength. Here, we assume that the noise operators act weakly, locally {and time-independently} on small subsystems.
We note that even though the coherent error $\delta H$ and the Lindblad operators $L_k$ act locally on the quantum system, the effect of errors propagates to the entire system after the evolution. Therefore, such global effects of noise cannot be effectively mitigated using the conventional quasi-probability method, which assumes simple gate-independent error model described by single- or two-qubit error channels before or after each gate.
\sun{
We also assume that accurate individual single-qubit controls are allowed, which holds for digital NISQ devices where single-qubit operations can achieve averaged fidelity of 99.9999\%~\cite{PhysRevLett.113.220501} whereas the record for two-qubit fidelity is three orders lower~\cite{PhysRevLett.117.060504,PhysRevLett.117.060505}. While not all analog quantum simulators support individual single qubit controls, they can indeed be achieved in various platforms with superconducting qubits~\cite{kjaergaard2019superconducting,mezzacapo2014digital,garcia2015fermion,asaad2016independent,weber2017coherent}, ion trap systems~\cite{zhang2017observation,haffner2005scalable,blatt2012quantum}, and Rydberg atoms~\cite{saffman2016quantum}.
Therefore, our framework is compatible with various practical NISQ devices. In the following, we focus on qubit systems and assume time-independent noise. We note that the discussion can be naturally generalised to multi-level systems, as well as general time-dependent noise (See Appendix~\ref{section:continuouserror} and~\ref{appendix: timedepsimulation}).
}



\vspace{0.2cm}

\section{Continuous QEM}

\begin{figure}[t]
\centering
\includegraphics[width=0.45\textwidth]{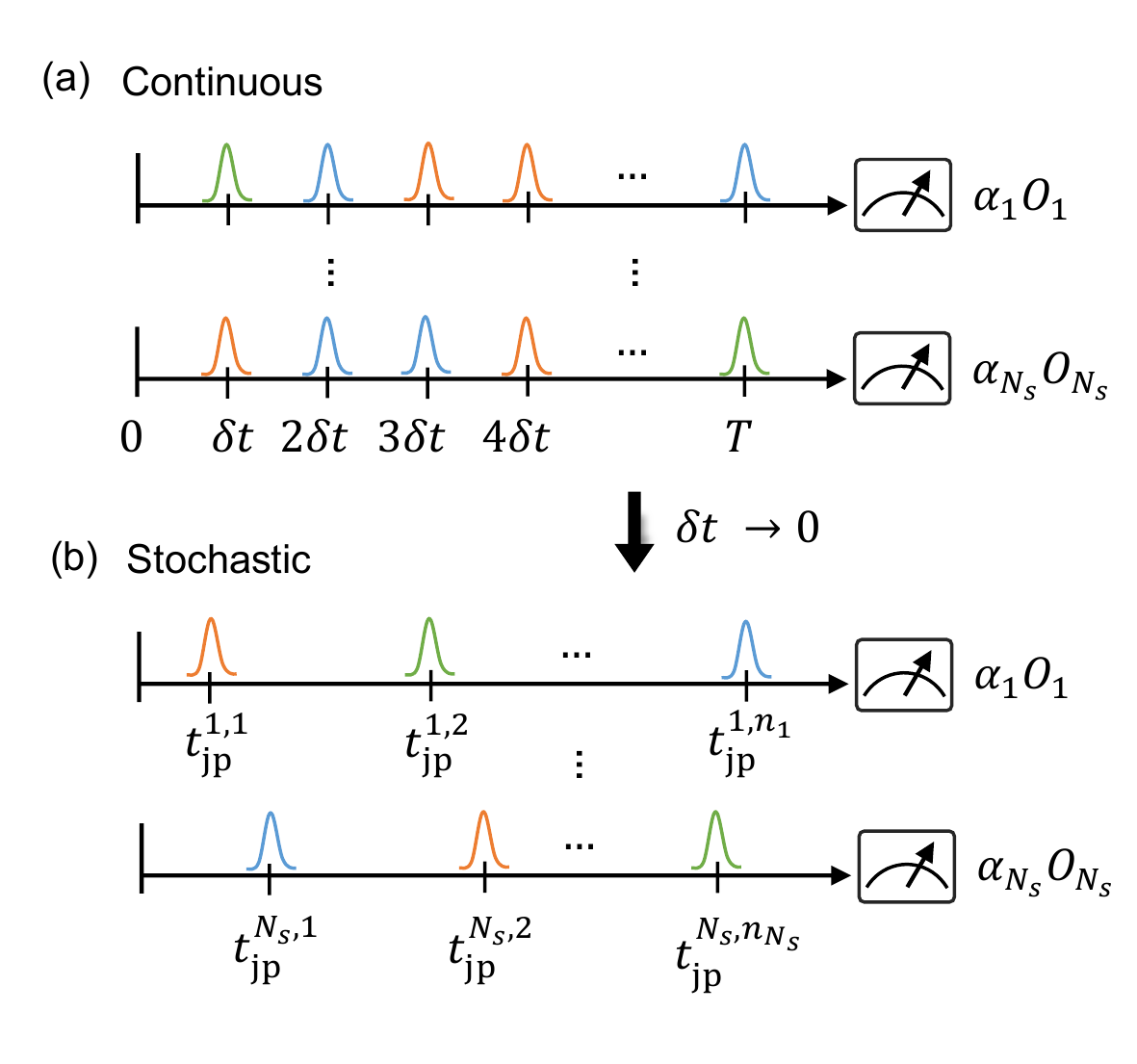}
\caption{
(a) Continuous QEM. With discretised time step $\delta t$, each recovery operation is weakly and `continuously' acted after each noisy evolution of time $\delta t$. Here different color represents different recovery operations. The output state is measured and repeated to obtain $N_s$ outcomes $\{O_m\}$, and their average corresponds to the error mitigated outcome.
(b) Stochastic QEM. We can equivalently realise (a) by $\delta t\rightarrow 0^+$ and randomly applying a small number $n_m$ of strong recovery operations as in Algorithm~\ref{alg1}. The time $\{t^{m,k}_{\rm jp}\}_m$ to apply recovery operations of the $m$th run are predetermined, which can be further pre-engineered  into the original evolution via a noisy time evolution of a modified Hamiltonian.
 }
\label{fig:scheme}
\end{figure}

Quantum gate in digital circuits or joint evolution process in analog simulation is physically realised via a continuous real time evolution of a Hamiltonian.
In this section, we extend the QEM method to a more practical scenario and show how to mitigate errors for these inherent dynamics-based and nonlocal noise in practical noisy quantum devices.
 We first introduce `continuous' QEM as a preliminary scheme as shown in Fig.~\ref{fig:scheme}(a).
Consider a small time step $\delta t$, the discretised evolution of Eqs.~\eqref{eqn:goodeqn} and \eqref{eqn:noisyeqn} can be represented as
\begin{equation}
    {\rho_{\alpha}(t+\delta t)} = \mc E_{\alpha}(t)	{\rho_{\alpha}(t)}.
\end{equation}
Here $\alpha=I,N$ and $\mc E_{\alpha}(t)$ denotes the ideal ($\alpha=I$) or noisy ($\alpha=N$) channel that evolves the state from $t$ to $t+\delta t$ within small $\delta t$. We can find a recovery operation $\mc E_Q$ that approximately maps the noisy evolution back to the noiseless one as
\begin{equation}
\mc E_{I}(t)=\mc E_{Q}  \mc E_{N}(t)+\mc O(\delta t^2).
\end{equation}
The operation $\mc E_{Q}$ is in general not completely positive, hence cannot be physically realised by a quantum channel. Nevertheless, similar to probabilistic QEM for discrete gates~\cite{temme2017error,endo2018practical}, we can efficiently decompose $\mc E_{Q}$ as a linear sum of {a polynomial number} of physical operators $\{\mathcal{B}_j\}$ that are tensor products of qubit operators,
\begin{equation}
\label{subsystem}
\begin{aligned}
\mc E_Q = c \sum_{j} \alpha_j p_j \mathcal{B}_j,
\end{aligned}
\end{equation}
with coefficients $c=1+\mc O(\delta t)$, $\alpha_j=\pm1$, and a normalised probability distribution $p_j$.
We refer to Sec.~\ref{section:Decomposition_optimisation} and Appendix~\ref{appendix: basisoperation} for details of the decomposition and its optimisation via linear programming.  Under this decomposition, the whole ideal evolution from $0$ to $T$ can be mathematically decomposed as
~
~
\begin{equation}
\label{eq:expansiion1213}
\begin{aligned}
\prod_{k=0}^{n-1} \mc E_{I}(k\delta t) = C \sum_{\vec j}\alpha_{\vec j} p_{\vec j} \prod_{k=0}^{n-1} \mathcal{B}_{j_k}\mc E_{N}(k\delta t)+\mc O(T\delta t),
\end{aligned}
\end{equation}
where $n=T/\delta t$, $ C= c^n$, $\alpha_{\vec j}=\prod_{k=0}^{n-1}  \alpha_{j_k}$, $p_{\vec j}=\prod_{k=0}^{n-1}  p_{j_k}$, and $\vec j = (j_1, \dots j_{n-1})$. Denote the ideally evolved state as
$
    \rho_I(T)=\prod_{k=0}^{n-1} \mc E_{I}(k\delta t) \rho(0)
$
and the noisily evolved and corrected state as $\rho_{Q,\vec j}(T) = \prod_{k=0}^{n-1} \mathcal{B}_{j_k} \mc E_{N}(k\delta t)\rho(0)$, we can approximate the ideal state $\rho_I(T)$ as a linear sum of noisy states as
\begin{equation}
    \rho_I(T) = C \sum_{\vec j}\alpha_{\vec j} p_{\vec j}  \rho_{Q,\vec j}(T)+ \mc O(T\delta t).
\end{equation}
When measuring an observable $O$ of the ideal state, the ideal measurement outcome $\braket{O}_I=\tr[\rho_I(T) O]$ is also approximated as a linear sum of the noisy measurement outcomes $\braket{O}_{Q,\vec j}=\tr[\rho_{Q,\vec j}(T) O]$ as
\begin{equation}
\begin{aligned}
\braket{O}_I = C \sum_{\vec j}\alpha_{\vec j} p_{\vec j} \braket{O}_{Q,\vec j}+ \mc O(T\delta t).
\end{aligned}
\end{equation}
In practice, we can randomly prepare $\rho_{Q,\vec j}(T)$ with probability $p_{\vec j}$, measure the observable $O$, and multiply the outcome with the coefficient $C\alpha_{\vec j}$. Then the average measurement outcome $\braket{O}_{Q,\vec j}$ of the noisy and corrected states $\rho_{Q,\vec j}$ approximates the noiseless measurement outcome.

To measure the average outcome to an additive error $\varepsilon$ with failure probability $\delta$, we need $T \propto {C^2}\log(\delta^{-1})/{\varepsilon^2}$ samples according to the Hoeffding inequality. Since the number of samples needed given access to  $\rho_I(T)$ is $T_0 \propto \log(\delta^{-1})/{\varepsilon^2}$, the error mitigation scheme introduces a sampling overhead $C^2$, {which can be regarded as a resource cost for the stochastic QEM scheme. The overhead scales as $C^2(T) =\exp(\mc O(\lambda T ))$ given noisy strength $\lambda$ and evolution time $T$. Here we choose a normalisation $\lambda$ so that the contribution from $\mc L_{\textrm{exp}}$ is bounded by a constant. Therefore the condition that the scheme works efficiently with a constant resource cost is $\lambda T = \mc O(1)$. By regarding $\lambda$ as the error rate, the condition can be intuitively interpreted as that the total noise rate is a constant, aligning with the result for conventional QEM.}

We remark that this error mitigation scheme works for general errors and we can mitigate correlated stochastic noise and unwanted interactions between (a small number of) multiple qubits. In Appendix~\ref{section:continuouserror}, we provide more details of the continuous error mitigation, including the decomposition of recovery operations and the resource cost for this method.  {In addition, this scheme can be naturally applied to multi-level systems when we can prepare the basis operations $\{\mathcal{B}_j\}$ for them.}

\vspace{0.2cm}

\section{Stochastic QEM}
In practice, it could be challenging to `continuously' interchange the noisy evolution and the recovery operation within a sufficiently small time step $\delta t$.
Since $\mc E_I(t)\approx \mc E_N(t)$ and the recovery operation at each time is almost an identity operation
\begin{equation}
\begin{aligned}
\mc E_Q &= (1+q_0 \delta t) \mc I +\sum_{j \geq 1} q_{j}\delta t\mc B_j\\
&=c \bigg( p_0 \mc I + \sum_{j\ge 1} \alpha_j \tilde p_j \delta t \mathcal{B}_j \bigg),
\label{eq:decomp_operation}
\end{aligned}
\end{equation}
with  $\mc B_0$ being the identity channel $\mc I$.
The probability to generate the identity operation $\mathcal{I}$ and $\mathcal{B}_i ~(i \geq 1)$ is $p_0 = 1-\sum_{j\ge 1} \tilde p_j \delta t= 1-\mc O(\delta t)$ and $\tilde p_j=p_j/\delta t=\mc O(1)$,
$c=1+(q_0+\sum_{j\ge 1} |q_j|)\delta t$. In addition, the parity $\alpha_0$ for $\mathcal{B}_0=\mathcal{I}$ is always unity, and the parity $\alpha_i$ corresponding to $\mathcal{B}_j ~(i \geq 1)$ equals to $\mathrm{sign}(q_j)$.

{We can further apply the Monte Carlo  method to stochastically realise the continuous recovery  operations as shown in Fig.~\ref{fig:scheme}(b). }
Specifically, we initialise $\alpha=1$ and randomly generate $q \in [0,1]$ at time $t=0$. Then evolve the state according to the noisy evolution $\mc E_N$ until time $t_{\mathrm{jp}}$ by solving $p(t_{\mathrm{jp}})=q$ with $p(t)=\mathrm{exp}\left(-\Gamma(t) \right)$ and \sun{$\Gamma(t)=  t\sum_{j \geq 1} \tilde p_j$}.
At time $t_{\mathrm{jp}}$, we generate another uniformly distributed random number $q' \in [0,1]$, apply the recovery operation $\mathcal{B}_j$ if $q' \in [s_{j-1},s_j]$, and update the coefficient as $\alpha=\alpha_j \alpha$. Here $s_j(t)=(\sum_{i=1}^j \tilde p_i) / (\sum_{i=1}^{N_{\rm op}} \tilde p_i)$, $N_{\rm op}$ is the number of basis operations, and the sum omits the identity channel. Then, we randomly initialise $q$, and repeat this procedure until time reaches $T$. \sun{On average, we prove in section~\ref{section: Equivalence} that the \emph{stochastic} QEM scheme is equivalent to the `continuous' one. While differently, the stochastic QEM does not assume time discretisation and it only requires to randomly apply a few recovery operations, scaling linearly to the total noise strength as $O(\lambda T)$ (See Appendix~\ref{section:continuouserror}). We can insert the recovery operations by `pausing' the original noisy evolution.  Alternatively, since we can determine the time $t_{\rm jp}$ and the recovery operations before the experiment, they can be pre-engineered into the original evolution. Therefore, we can effectively implement stochastic QEM via the noisy time evolution of Eq.~\eqref{eqn:noisyeqn} with an adjusted Hamiltonian.}
We discuss its implementation for both analog quantum simulation and digital gate-based quantum simulation in the section~\ref{section:implementation}.

The stochastic error mitigation scheme is summarised as follows.

\begin{algorithm}[H]
\begin{algorithmic}[1]
\State{Get $C$, $\{\alpha_j\}$, and $\{p_j\}$ of Eq.~\eqref{subsystem}, set $\bigg\{s_j=\frac{\sum_{i=1}^j \tilde p_i}{ \sum_{i=1}^{N_{\rm op}} \tilde p_i}$\bigg\}.}
\For {$m=1$ to $N_s$}
\State {Randomly generate $q_0\in[0,1]$, set $t=0$, $n=0$, $\alpha=1$.}
\While{$t\le T$}
\State{Get $t_{\textrm{jp}}^{n}$ by solving $\mathrm{exp}\left(- \Gamma(t^n_{\mathrm{jp}}) \right)=q_n$.}
\State{Randomly generate $q'_n\in[0,1]$.}
\State{Set $j_n=j$ if $q'_n \in [s_{j-1},s_j]$ and update $\alpha=\alpha_{j_n}\cdot \alpha$.}
\State{Update $t=t+t_{\textrm{jp}}^{n}$ and $n=n+1$.}
\EndWhile
\State{Set $\rho_Q=\rho(0)$ and $\bar O=0$.}
\For{$k = 0:n-1$}
\State{Evolve $\rho_Q$ under $\mc E_N$ for time $t_{\textrm{jp}}^{k}$ and apply $\mc B_{j_k}$.}
\EndFor
\State{Evolve $\rho_Q$ under $\mc E_N$ for time $T-\sum_{k=0}^{n-1}t_{\textrm{jp}}^{k}$.
\State{Measure $O$ of $\rho_Q$ to get $O_m$.}
\State{Update $\bar{O}=\bar O+C\alpha O_m/N_s$}
}
\EndFor
\end{algorithmic}
\caption{Stochastic error mitigation. \\Input: initial state $\rho(0)$, number of samples $N_s$, noisy evolution  $\mc E_N$, basis
operations $\mc B_j$; Output: $\bar O$.  }
\label{alg1}
\end{algorithm}

\section{Equivalence between continuous error mitigation and stochastic error mitigation}
\label{section: Equivalence}
We now prove the equivalence of stochastic error mitigation to continuous error mitigation.
Provided the recovery operation in Eq. (\ref{eq:decomp_operation}), we can interpret it as with probability $1-\sum_i\tilde p_{i}\delta t$ we do nothing, and with probability $\tilde p_{i}\delta t$ we apply a corresponding correction operation. We also multiply $c\cdot \alpha_i$ to the output measurement.
We can regard the event that applies the correction operations as a jump similar to the stochastic Schr\"odinger equation approach.

Starting at time $t=0$, the probability that there is no jump until time $t$ is
\begin{equation}
	Q(t) = \lim_ {\delta t\rightarrow 0}\prod_{i=0}^{t/\delta t} \bigg(1-\sum_{i \geq 1} \tilde p_{i}\delta t\bigg) =e^{-\int_0^t \Gamma(t')dt'}
\end{equation}
where $\Gamma(t)=\sum_{i\geq 1} \tilde p_i(t)$.
The probability to have jump in the time interval $[t, t+dt]$ is
\begin{align}\label{Eq:appdixPdt}
P(t)dt = \Gamma(t) e^{-\int_0^t \Gamma(t')dt'} dt.
\end{align}
Now suppose we generate a uniformly distributed random variable $q \in [0,1]$ and solve
\begin{align}
q = e^{-\int_0^{t_{\mathrm{jp}}} \Gamma(t')dt'},
\label{eqn:jumptime}
\end{align}
to determine the jump time $t_{\mathrm{jp}}$. Then the probability that jump happens at time time $t_{\mathrm{jp}}$ or in particular between $[t_{\mathrm{jp}},t_{\mathrm{jp}}+dt]$ is
\begin{equation}
	dq = P(t) dt,
\end{equation}
which agrees with Eq.~\eqref{Eq:appdixPdt}.
We can thus use the uniformly distributed random variable $q$ to determine the jump time to equivalently simulate the original continuous error mitigation process.


Now, at the jump time $t_{\mathrm{jp}}$, we apply the basis operation other than the identity operation. We can determine the basis operation by generating another uniformly distributed random number $q' \in [0,1]$. If $q' \in [s_{k-1},s_k]$, we set the basis operation
to $\mathcal{B}_k$, where $s_k(t)=(\sum_{j=1}^k \tilde p_j) / (\sum_{j=1}^{N_{\rm op}} \tilde p_j)$ and $N_{\rm op}$ is the number of the basis operations.

\sun{We can pre-determine the jump time ${t_{\rm jp}^1}, t_{\rm jp}^2, ..., t_{\rm jp}^k$ from Eq. (\ref{eqn:jumptime}).
For time-independent noise, the jump time can be simply determined as $t_{\rm jp} = -\log(q)/\sum_{i \geq 1} {\tilde p_i}$ with $q$ randomly generated from $[0,1]$. Given evolution time $T$, the average number of recovery operations is proportional to $\mathcal{O}(\lambda T)$. In the numerics, the average number of recovery operations is about $0.3$ times per evolution on average given a realistic noise model and simulation task.
}

\section{Decomposition of the recovery operation and optimisation}
\label{section:Decomposition_optimisation}
We now discuss the decomposition of the recovery operation into local basis operations.
We denote the complete basis operations as $\{\mathcal{B}_i\}$. For multiple qubit systems, tensor products of single qubit operations, e.g., $\mathcal{B}_i \otimes \mathcal{B}_j$ also forms a complete basis set for composite systems. Therefore, if we can implement the complete basis operations for a single qubit, we can also emulate arbitrary operations for multiple qubits systems.
In Ref~\cite{endo2018practical}, it is shown that every single qubit operation can be emulated by using $16$ basis operations. This is because every single qubit operation (including projective measurements) can be expressed with square matrices with $4 \times 4 = 16$ elements by using the Pauli transfer representation~\cite{greenbaum2015introduction}. Therefore, $16$ linearly independent operations are sufficient to emulate arbitrary single qubit operations. In Table~\ref{tab:bases} of Appendix~\ref{appendix: basisoperation}, we show one efficient set of basis operations for a single qubit.

We show in the Appendix~\ref{section:continuouserror} that the recovery operations without Hamiltonian error can be analytically expressed as
\begin{equation}
    \mathcal{E}_Q=\mathcal{I}-\lambda\mathcal{L}\delta t.
    \label{eq:Lindblad_recovery}
\end{equation}
where $\mathcal{L}$ represents the noise superoperator and $\lambda$ is the noise strength. From Eq. (\ref{eq:Lindblad_recovery}), we can analytically decompose the general noise into local basis operations.
In the Appendix~\ref{appendix: basisoperation}, we provide the recovery operations for several typical Markovian processes, including depolarising, dephasing and amplitude damping, during the quantum simulation.
It is worth mentioning that by using only observables within spatial domain, we can recover the Lindbladian acting on this domain and reconstruct the local Markovian dynamics~\cite{bairey2020learning,da2011practical}.


Over-complete basis can be used to further reduce the resource cost for the stochastic error mitigation scheme.
In general, the target quasi-probability operation $\mathcal{E}_Q$ can be decomposed as a linear combination of unitary channels and projective measurements by using Pauli transfer matrix representation. The quasi-probability operation $\mathcal{E}_Q$ can be decomposed into a complete basis $\{ \mathcal{B}_i \}$ as
\begin{equation}
\mathcal{E}_Q=\sum_{i} q_{i} \mathcal{B}_{i},
\end{equation}
where we set $\mathcal{B}_0=\mathcal{I}$. Given the target quasi-probability operation $ \mathcal{E}_Q$, the overall resource cost for quasi-probability scheme is given by $C(T)=\exp \left( C_{1} T \right)$ with $C_1 =q_0 +\sum_{i\ge 1} |q_i|$.

In order to minimise the resource cost, we aim to reduce $C_1$.
Consider an over-complete basis $\{ \mathcal{B}_i^{\prime} \}$ which includes the complete basis $\{\mathcal{B}_i \}$ and also
other randomly generated unitary operators and projective measurements. Then the quasi-probability operation $ E_Q$ is decomposed into this over-complete basis $\{\mathcal{B}_i^{\prime} \}$ as
\begin{equation}
\mathcal{E}_Q=\sum_{i} q_{i}^{\prime} \mathcal{B}_{i}^{\prime}.
\label{eqn:overcompleteApp}
\end{equation}
Minimising $C_1 = q_0 +\sum_{i\ge 1} |q_i|$ can be further rewritten as a linear programming as follows,
\begin{equation}
\begin{aligned}
\min C_1 &= q_0+\sum_{i\ge 1} (q^+_i - q^-_i),\\
s.t.~&\mathcal{E}_Q=\sum_{i} (q^+_i - q^-_i) \mathcal{B}_{i}^{\prime},\\
&q^+_i, q^-_i\ge 0.
\end{aligned}
\end{equation}
The overall resource cost $C(T)$ for stochastic error mitigation
scheme can therefore be reduced by this linear programming optimisation method.

\section{Implementation of the scheme with analog and digital quantum simulators  }
\label{section:implementation}



\begin{figure}[htb]
\centering
\includegraphics[width=0.46\textwidth]{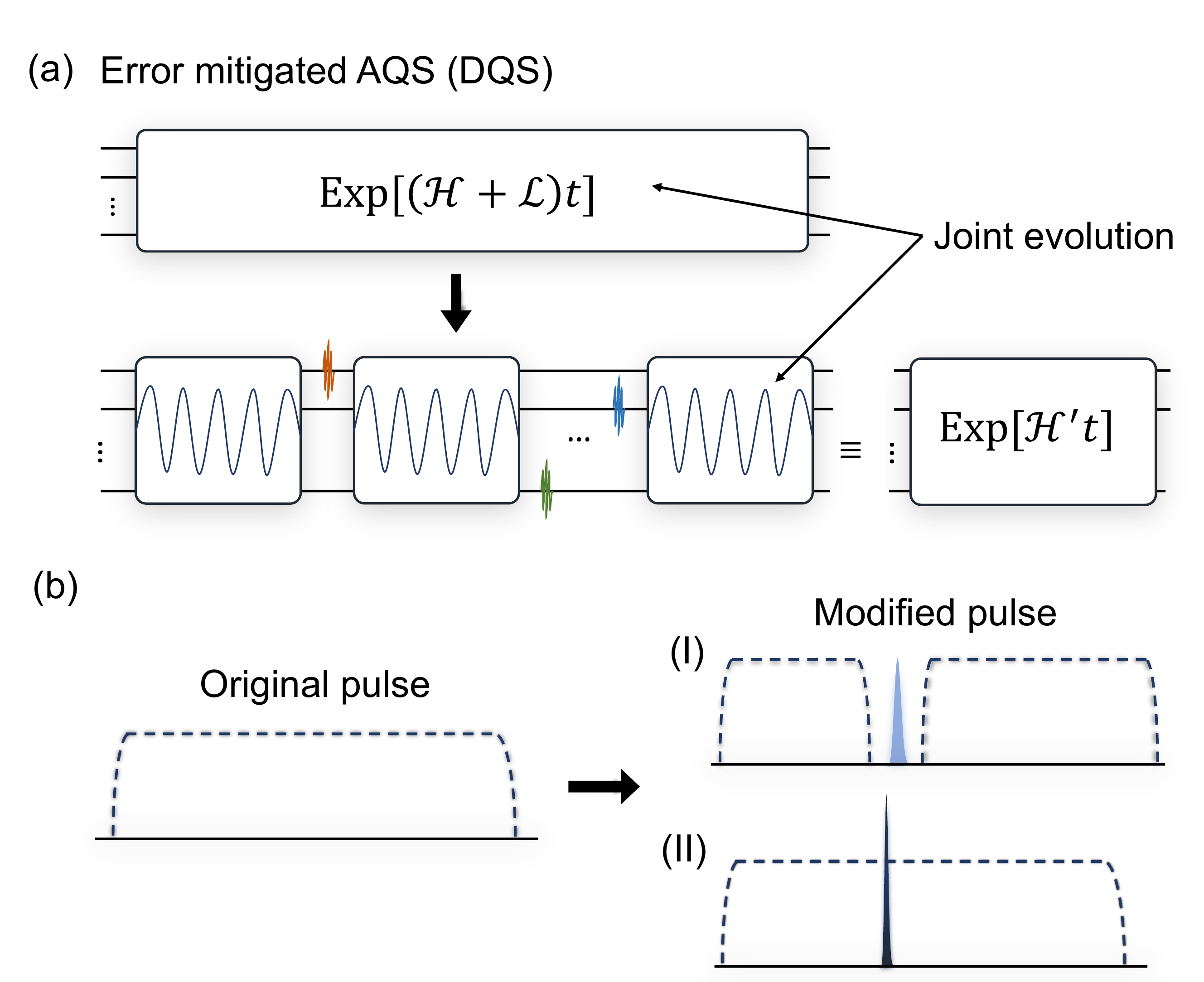}
\caption{(a) Schematic diagram of the error mitigated AQS or DQS with controllable single qubit operations. The AQS or DQS is realised by a continuous process under the ideal driving Hamiltonian.  We denote the noisy unitary and stochastic processes by  $\mathcal H$ and $\mathcal{L}$ with the superoperator formalism.
Our work considers joint dynamics of all qubits sandwiched with a small number ($\mathcal O(1)$) of pre-engineered single qubit dynamics to mitigate the errors accumulated in the evolution, which can also be regarded as a modified evolution $\mathcal{H}'$.
  (b) Two schemes of the error-mitigated process or gate with modified pulse sequences. Dashed lines represent the original pulse that constructs the target process or multi-qubit gates. Provided controllable drive that could be freely turned on/off, we can synthesise the error mitigated process/gate by modifying the original pulse sequence as shown in scheme I, which corresponds to (a). In the case of restricted driving operations, we can alternatively apply a strong and fast single qubit pulse to the original pulse to mitigate either process errors or gate errors as shown in scheme II. We note that scheme II could similarly be applied in (a), which introduces a negligible error of $\mathcal O(\delta t^3)$ when each single qubit gate is implemented in $\delta t\ll1$.
 }
\label{fig:AQS}
\end{figure}

In this section, we discuss the implementation of our scheme with analog and digital quantum simulators. To implement stochastic error mitigation with an analog quantum simulator, we insert the single-qubit recovery operations at each jump time, which is pre-determined by Algorithm $1$. As the evolution of most quantum simulators are based on external pulses, such as trapped ions and superconducting qubits, it would be practically feasible to interrupt the continuous evolution by simply turning off the external pulses and then turning on the single qubit recovery pulses. The joint evolution and the single qubit dynamics can be pre-engineered as a modified evolution of AQS, as shown in Fig.~\ref{fig:AQS}.
In practice, when turning on/off the joint evolution cannot be realised in a short time, we can alternatively apply single qubit recovery pulses with a short duration and a sufficiently strong intensity compared to the parameters of the AQS Hamiltonian, as shown in Fig.~\ref{fig:AQS} (b) II.
This is similar to the banged analog-digital quantum computing protocol introduced in Refs.~\cite{parra2020digital,martin2020digital}, which implements single qubit gates without turning off the background Hamiltonian. In this case, when single-qubit rotations are performed in a time $\delta t$ that is much smaller than the timescale of the joint evolution, the additional error per single-qubit rotation introduced by the background evolution of the Hamiltonian is on the order of $\mathcal{O}(\delta t^3)$.
Therefore, errors induced from the error mitigation procedure could be very small, and they could be further mitigated via the hybrid approach in the Section~\ref{section: modelError}.



On the other hand, the stochastic error mitigation scheme could be naturally implemented on a digital gate-based quantum computer.
We note that digital gates are experimentally realised via continuous pulse sequences~\cite{kandala2019error,kjaergaard2019superconducting,krantz2019quantum,sheldon2016procedure,arute2019quantum,plantenberg2007demonstration,neill2018blueprint,corcoles2013process,chow2012universal,rigetti2010fully}, thus we can construct the error mitigated gates by modifying the original pulse sequence with the pre-determined pulses (recovery operations). Similar process has been experimentally demonstrated in Ref.~\cite{corcoles2013process}, where the effect of the new pulse sequence is to effectively mitigate the unwanted terms in the driving Hamiltonian. Our QEM method can be used to eliminate the general coherent and incoherent errors of the gates to achieve high gate fidelity. Therefore, provided control of pulse sequence, we can engineer the pulse sequence as shown in Fig.~\ref{fig:AQS} (b), and prepare the error mitigated gates to perform quantum computing or quantum simulation tasks. We also note that if the driving operations are restricted, we can similarly apply the fast single qubit pulse to the original pulse to mitigate the errors, where the additional error induced in this process is on the order of $\mathcal{O}(\delta t^3)$, as shown in the bottom of Fig.~\ref{fig:AQS} (b).

As shown in Fig.~\ref{fig:AQS}, the exact quantum hardware that we need to implement our error mitigation scheme actually lies in between fully analog simulation (we can only control all qubits with a predetermined Hamiltonian) and fully digital quantum computer (we can apply any operation on a small number of qubits).
The scenario can be regarded as a modified analog-digital quantum simulator, where we only need to apply strong local gates along the background dynamics (no matter whether it is digital or analog).
Compared to a fully AQS, we only need to insert single qubit operations
and joint evolution and the single qubit dynamics can be pre-engineered by using Algorithm $1$.
The readers could regard it as analog-digital quantum simulation or a time-dependent Hamiltonian dynamics (although we only require controllable single qubit operations instead of arbitrary Hamiltonian simulation).
 Compared to a fully DQS, our scheme does not need to apply any two-qubit gate and hence it significantly avoids crosstalks.
It is worth noting that given evolution time $T$ in AQS or pulse sequence of the target gate in DQS, the average number of recovery operations is linear to $\lambda T$.  In practice, the average number of recovery operations could a small number (for instance around $1$ in our numerical simulation), and therefore easy to implement in reality.
To summarise, our scheme can be implemented on all digital and most analog quantum simulators, and analog-digital quantum simulators we describe above, as long as accurate and fast single qubit operations are allowed.


 \begin{figure*}[htb]
\centering
\includegraphics[width=1\textwidth]{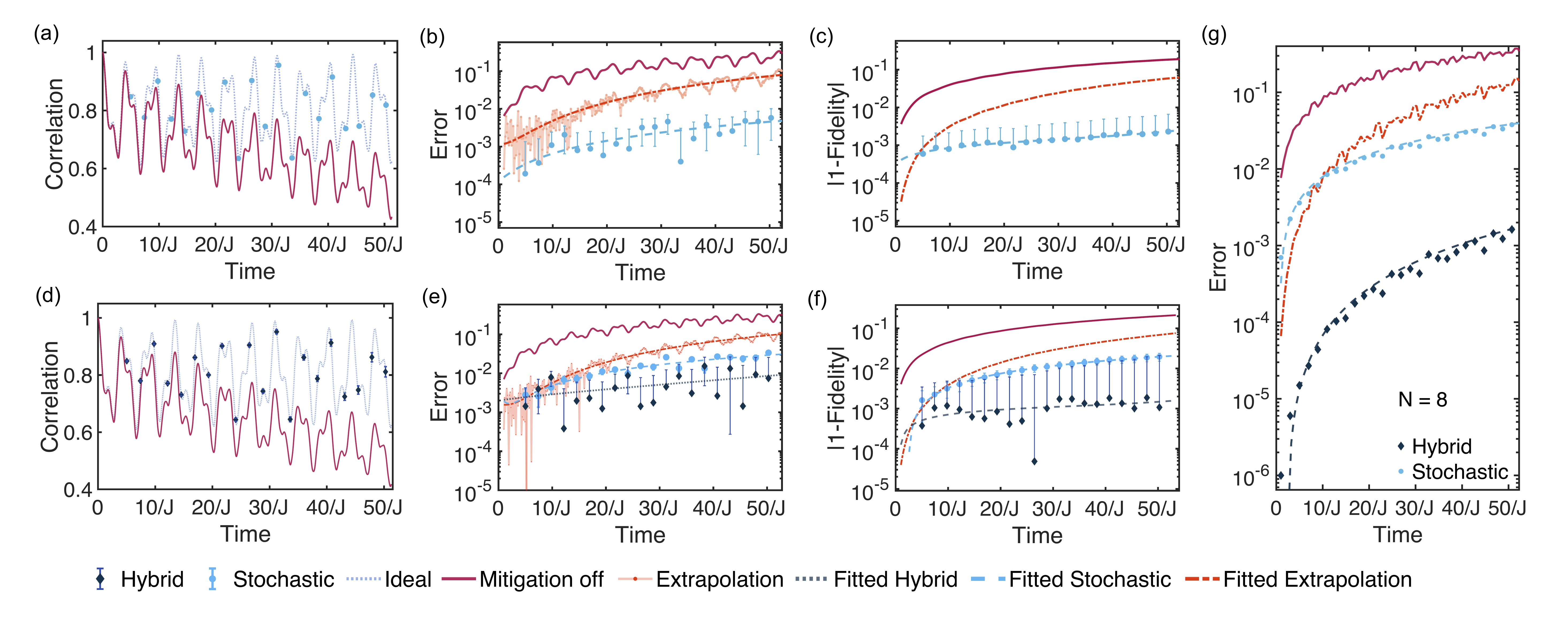}
\caption{Numerical test of the QEM schemes without  ((a)(b)(c)) and with $10\%$ model estimation error $\lambda_{\textrm{exp}}  = 1.1 \lambda_{\textrm{est}}$ ((d)(e)(f)(g)).
We consider the dynamics of 2D anisotropic Heisenberg Hamiltonian with energy relaxation and dephasing noise. (a)$-$(f) consider a four-qubit Hamiltonian with finite $(10^6)$ number of samples. (a) and (d) compares the time evolved nearest-neighbour correlation function
$\sum_{\langle ij\rangle}   \sigma_x^{(i)}  \sigma_x^{(j)}/4$.
(b) and (e) shows the error between the exact value and the error-mitigated value.
(c) and (f) shows the fidelity of the effective density matrix $\rho^{\alpha}_{\textrm{eff}}$ and the ideal one $\rho_{I}$ under different error mitigation scheme $\alpha$.
(g) consider an eight-qubit Hamiltonian with infinite number of samples. The hybrid error mitigation scheme suppresses the error up to about four orders of magnitude even with $10\%$ model estimation error.}
\label{fig:Allmitig}
\end{figure*}

\section{Reduction of model estimation error}
\label{section: modelError}
For systems with finite-range interactions, local Markovian dynamical process can be reconstructed by using only local measurements and we refer to Refs.~\cite{bairey2020learning,da2011practical} for details. Given a  prior knowledge of the noise model, the above stochastic QEM schemes can eliminate the physical errors by applying basis operations at jump time. Nonetheless, the realistic noise $\mathcal{L}_{\rm exp}$ and the estimated noise $\mathcal{L}_{\rm est}$ may differ due to imprecise estimation of the noise model.
Here we combine the extrapolation QEM method~\cite{li2017efficient,temme2017error} to mitigate model estimation error and the errors associated with imperfect recovery operations.

We first show how to boost model estimation error, which will be used for its mitigation as a preliminary.
Assume that the evolution of the quantum system is described by the open-system master equation
\begin{align}
\frac{d}{dt} \rho_{\lambda} = -i[H(t), \rho_{\lambda}]+\lambda \mathcal{L}_{\mathrm{exp}}\left[\rho_{\lambda}\right].
\end{align}
We show in the Appendix~\ref{appendix: extra} that by evolving the state $\rho_{\lambda}^{\prime}$ under the re-scaled Hamiltonian $\frac{1}{r} H\big(\frac{t}{r}\big)$ for time $rt$, we can effectively boost physical errors of quantum systems, which can be expressed as $\rho_{\lambda}^{\prime}(r t)=\rho_{r \lambda}(t)$. Here, we assume that the noise superoperator $\mathcal{L}$ is invariant under rescaling, and the initial conditions holds $\rho_{\lambda}^{\prime}(0)=\rho_{r \lambda}(0)$.

The effective evolution after applying the error mitigation method with $\mathcal{L}_{\rm est}$ is
\begin{equation}
\label{appendix:noisy}
\frac{d }{dt} \rho_{\lambda}^{(Q)}(t)  =-i[H(t), \rho_{\lambda}^{(Q)}(t)] + \lambda \Delta \mc L \big[ \rho_{\lambda}^{(Q)}(t) \big],
\end{equation}
where $\rho_{\lambda}^{(Q)}(t)$ is the effective density matrix and $\Delta \mc L=\mathcal{L}_{\rm exp}-\mathcal{L}_{\rm est}$. By re-scaling $H(t) \rightarrow \frac{1}{r} H(\frac{t}{r})$, the evolution for rescaled time $rt$ is
\begin{equation}\label{noisy}
\frac{d }{dt} \rho_{r \lambda}^{ (Q)}(t)=-i[H(t), \rho_{r \lambda}^{(Q)}(t)] +r\lambda  \Delta  \mc L  \big[ \rho_{r\lambda}^{(Q)}(t) \big],
\end{equation}
 which can be implemented by re-running the error-mitigated experiment for a re-scaled time $rt$ under the re-scaled Hamiltonian.

As the value of $r \ge 1$ can be tuned, we choose several different values of $r$ and suppress the model estimation error via Richardson extrapolation.
Specifically, with more than two values of $r$ denoted as $\{r_j\}$ and constants  $\beta_j=\prod_{l \neq j}r_l (r_l-r_j)^{-1}$, we have
\begin{equation}
\braket{O}_I = \sum_{j=0}^n \beta_j \braket{O}_{r_j\lambda}+\mc O\left( \frac{\gamma_n\left(r_{\max}\lambda T\left\|\Delta \mathcal{L}\right\|_{1}\right)^{n+1}}{(n+1)!}\right).
\label{eq:RichardsonIdeal}
\end{equation}
where $\braket{O}_{r \lambda}$ is the measurement outcome after stochastic error mitigation, corresponding to $\rho_{r \lambda}^{(Q)}(T)$, $\gamma_n=\sum_j|\beta_j|$, $r_{\max}=\max_jr_j$, and $\|\Delta \mathcal{L}\|_{1}=\max_{\rho}\tr|\mathcal{L}(\rho)|$. Therefore, in addition to $\lambda T = \mc O(1)$, the scheme is efficient provided
\begin{equation}
 r_{\max}\|\Delta \mathcal{L}\|_{1}=\mc O(1).
\end{equation}
We refer to Appendix~\ref{appendix: model} for the derivation of Eq. (\ref{eq:RichardsonIdeal}).

From Eq.(\ref{eq:RichardsonIdeal}), the deviation between the ideal measurement outcome and the error-mitigated one is bounded independently with the Hamiltonian,
i.e., the to-be-simulated problem. The bound only relies on the noise model, the evolution time, the number of samples, and the parameters used in extrapolation.
Moreover, since imperfections of the basis operations $\mathcal{B}_i$ lead to deviation of $\mathcal{L}_{\mathrm{est}}$, which can be regarded as another type of model estimation error,  they can be corrected via the extrapolation procedure. We note that noise could have fluctuations or drift in the experimental apparatus, which in practice could be challenging to obtain the precise noise model.
Our hybrid QEM incorporating extrapolation is therefore practically useful as this method alleviates the requirement of precise estimation of the noise model and  can be robust to the error of recovery operations.
We refer to Appendix~\ref{appendix:HybridQEM} for detailed analysis of error mitigation for the model estimation.

\section{Numerical simulation}
\vspace{0.2cm}

Now, we test our QEM schemes for analog quantum simulators and gate-based digital quantum circuits.
We first consider a 2D anisotropic Heisenberg model $H=J\sum_{\langle ij\rangle} \big[(1+\gamma)\sigma_x^{(i)}  \sigma_x^{(j)}+(1-\gamma)\sigma_y^{(i)}  \sigma_y^{(j)} + \sigma_z^{(i)}  \sigma_z^{(j)} \big] - \gamma h\sum_{ i=1}^4 \sigma_y^{(i)}$ on a $2\times 2$ square lattice, where $\langle ij\rangle $ represents nearest neighbour pairs. This model has been extensively used to investigate the quantum magnetism and criticality~\cite{lee1984time,siurakshina2000theory,hucht1995monte,torelli2018calculating,soukoulis1991spin}.
We consider analog simulation via a noisy superconducting quantum simulator with  energy relaxation $\mathcal{L}_{1}$ and dephasing  $\mathcal{L}_{2}$ noise~\cite{lamata2018digital,houck2012chip,oliver2013materials,gu2017microwave}. Here the Lindblad operator takes the form of
\begin{equation}
    \mathcal{L}_{\beta} \left[ \rho\right]=\sum_{j}\lambda_{\beta}\big(  L^{(j)}_{\beta} \rho L_{\beta}^{(j)\dagger}-\frac{1}{2}\{L_{\beta}^{(j)\dagger} L^{(j)}_{\beta}, \rho \}\big)
\end{equation} for $\beta=1,2$, $L_{1 }^{(j)}=  \sigma_{-}^{(j)}= \ket{0}\bra{1}$, and $L_{2}^{(j)} = \sigma_z^{(j)}$. Such a noise model is also relevant for other quantum simulators such as trapped ions~\cite{bohnet2016quantum,blatt2012quantum,friedenauer2008simulating,jurcevic2017direct}, NMR~\cite{li2017measuring,li2014experimental,garttner2017measuring}, ultracold atoms~\cite{hart2015observation,bernien2017probing}, optical lattices apparatus~\cite{struck2011quantum}, etc.
The noise can be characterised by measuring energy relaxation time $T_1$ and dephasing time $T_2$ without full process tomography~\cite{bylander2011noise,gu2017microwave,zhou2008relaxation,oliver2013materials} and \sun{more generally via local measurements~\cite{bairey2020learning, da2011practical}.}
We also consider physical errors for the single-qubit recovery operations as single-qubit inhomogeneous Pauli error, $\mathcal{E}_{\textrm{inh}} = (1-p_x-p_y-p_z)\mathcal{I}+p_x\mathcal{X}+p_y\mathcal{Y}+p_z\mathcal{Z}$ with $\mathcal{I}, \mathcal{X}, \mathcal{Y}, \mathcal{Z}$ being the Pauli channel and $p_{\alpha}$ being the error probability.
{In our simulation, we set $J=h=2\pi \times 4$ MHz, $\gamma=0.25$, and the noise strength  $\lambda_{1}=\lambda_{2}=0.04$ MHz~\cite{gu2017microwave,bylander2011noise,krantz2019quantum}. For model estimation error, we set $p_x=p_y=0.25\%$ and $p_z=0.5\%$, which can be achieved with current superconducting simulators~\cite{linke2017experimental,kjaergaard2019superconducting}, and consider the real noise strength to be $10\%$  greater than the estimated one, i.e., $\lambda_{\rm exp}  = 1.1 \lambda_{\rm est}$.}
We set the initial state to $\ket{+}^{\otimes 4}$ with $\ket{+}=(\ket{0}+\ket{1})/\sqrt{2}$, evolve it to time $T=16\pi/J$, and measure the expectation value of the normalised nearest-neighbour correlation function
$\sum_{\langle ij\rangle}   \sigma_x^{(i)}  \sigma_x^{(j)}/4$ with $10^6$ samples.

The numerical result without model estimation error is shown in Fig.~\ref{fig:Allmitig}~(a)(b)(c). Specifically, we compare the time evolution of the expectation value of the correlation operator
in Fig.~\ref{fig:Allmitig}~(a)(b) and the fidelity $F(\rho_{I},\rho_{\textrm{eff}})=\tr \sqrt{\rho_{\textrm{eff}}^{1 / 2} \rho_{I} \rho_{\textrm{eff}}^{1 / 2}}$ of the effective density matrix $\rho_{\textrm{eff}}$ and the ideal one $\rho_{I}$ in Fig.~\ref{fig:Allmitig}~(c). We can see that Richardson extrapolation and stochastic QEM improve the accuracy by one and two orders, respectively.
The result with model estimation error is shown in Fig.~\ref{fig:Allmitig}~(d)(e)(f).
Here, we also consider the hybrid method with both stochastic QEM and linear extrapolation, with optimised $r_0=1$ and $r_1=1.8$. We can see that stochastic QEM still outperforms Richardson extrapolation with large evolution time and the hybrid method can be further used to improve the simulation accuracy. The simulation result thus indicates that the hybrid QEM scheme can be robust to the drift of noise~\cite{muller2015interacting,meissner2018probing,neill2013fluctuations}. The performance of the QEM schemes can be made clearer without considering sampling errors. As shown in Fig.~\ref{fig:Allmitig}(g), we consider simulations of the eight-qubit anisotropic Heisenberg model on $2\times 4$ lattice under different QEM schemes with infinite samples. The result indicates that both stochastic and hybrid QEM can effectively eliminate the accumulation of errors during the evolution.
\begin{figure}[t]
\centering
\includegraphics[width=0.45\textwidth]{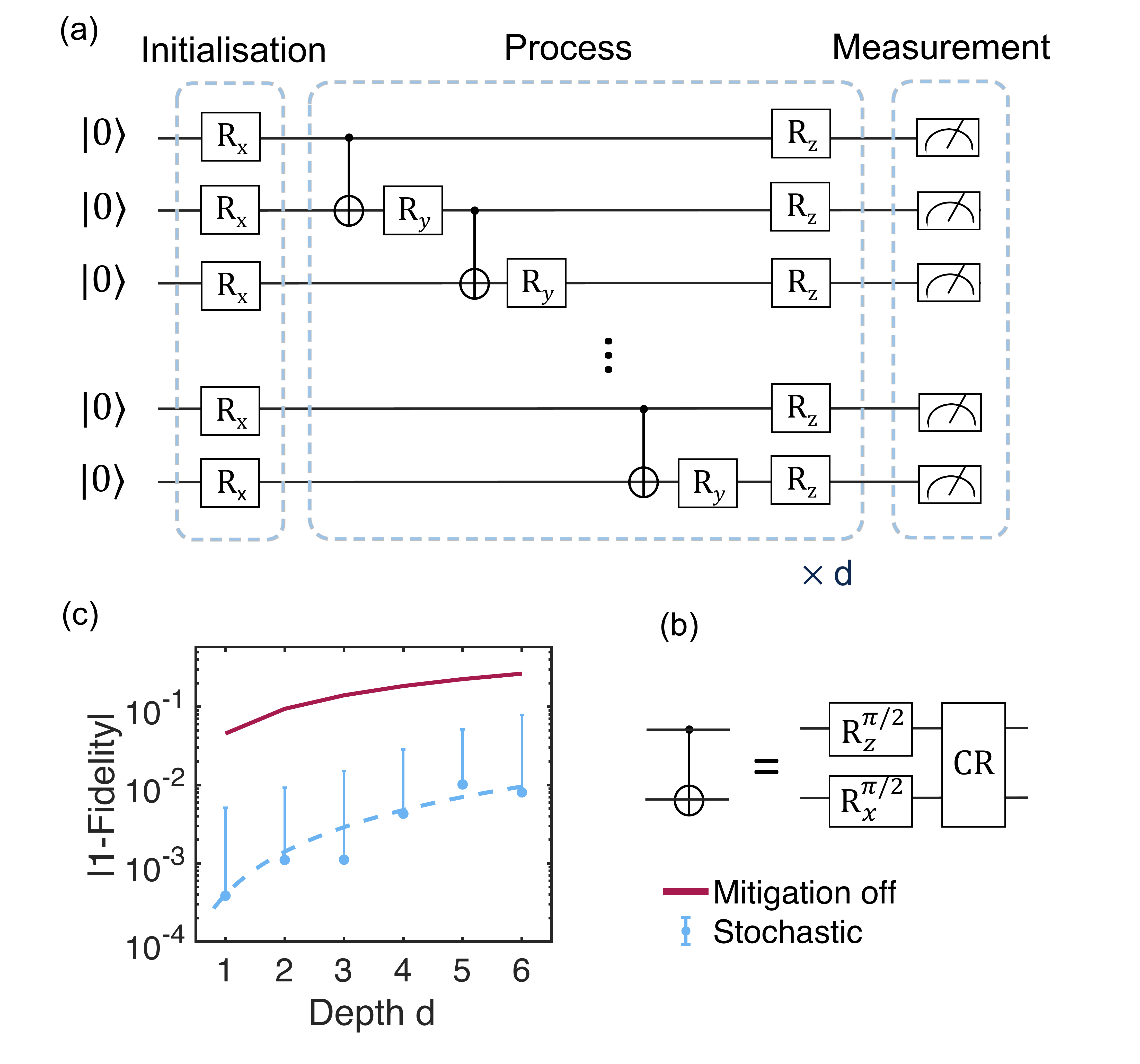}
\caption{
Stochastic QEM for eight-qubit superconducting quantum circuits with environmental and crosstalk noise.
(a) considers a $d$-depth parameterised quantum circuits with single-qubit rotations $R_{\alpha}$ ($\alpha\in \{X,Y,Z\}$ and CNOT gates. The rotation angles are randomly generated from $[0,2\pi]$.
(b) shows the realisation of the CNOT gate via the CR gate $U_{CR}=\exp(i\pi \sigma_z^{(c)} \sigma_x^{(t)} /4)$ and single-qubit gates $R_{z}^{\pi/2}$ and $R_{x}^{\pi/2}$ up to a global phase $e^{i\pi/4}$.
(c) shows the fidelity dependence of circuit depth $d$ with/without QEM.
}
\label{fig:gate_error}
\end{figure}

Next, we consider an eight-qubit, $d$-depth parameterised quantum circuit (Fig.~\ref{fig:gate_error}(c)) and show how stochastic QEM can suppress coherent errors in multi-qubit operations. Here, the controlled-NOT (CNOT) gate in the quantum circuits is generated by cross-resonance (CR) gates, which are experimentally realised by using microwaves to drive the control qubit ($c$) at the frequency of the target qubit ($t$), resulting in a driving Hamiltonian $H\approx \Omega(-\sigma_z^{(c)}\sigma_x^{(t)} +\gamma \mathbb{I}^{(c)}\sigma_x^{(t)})$~\cite{kandala2019error,krantz2019quantum,corcoles2013process,rigetti2010fully,chow2012universal}. Here, $\Omega$ is the effective qubit-qubit coupling and $\gamma$ represents the effect of crosstalk between qubits~\cite{corcoles2013process}. We consider inherent environmental noise and recovery operation error as in the above analog simulator, and additional coherent crosstalk errors $\gamma=1\%$. We set $\Omega=2\pi $ MHz, the evolution time $T=\pi/4\Omega$, and run $10^5$ samples. We mitigate the noisy two-qubit pulse sequence by inserting basis operations, and shows the fidelity dependence of circuit depth $d$ with/without QEM in Fig.~\ref{fig:gate_error}(c).  The result clearly shows that stochastic QEM improves the computing accuracy by two orders.

{In Appendix~\ref{appendix:numer}, we show numerical simulations for both Ising and frustrated quantum spin Hamiltonian, and we further demonstrate how the QEM methods can be applied to temporally correlated noise.}

\begin{figure}[t]
\centering
\includegraphics[width=0.47\textwidth]{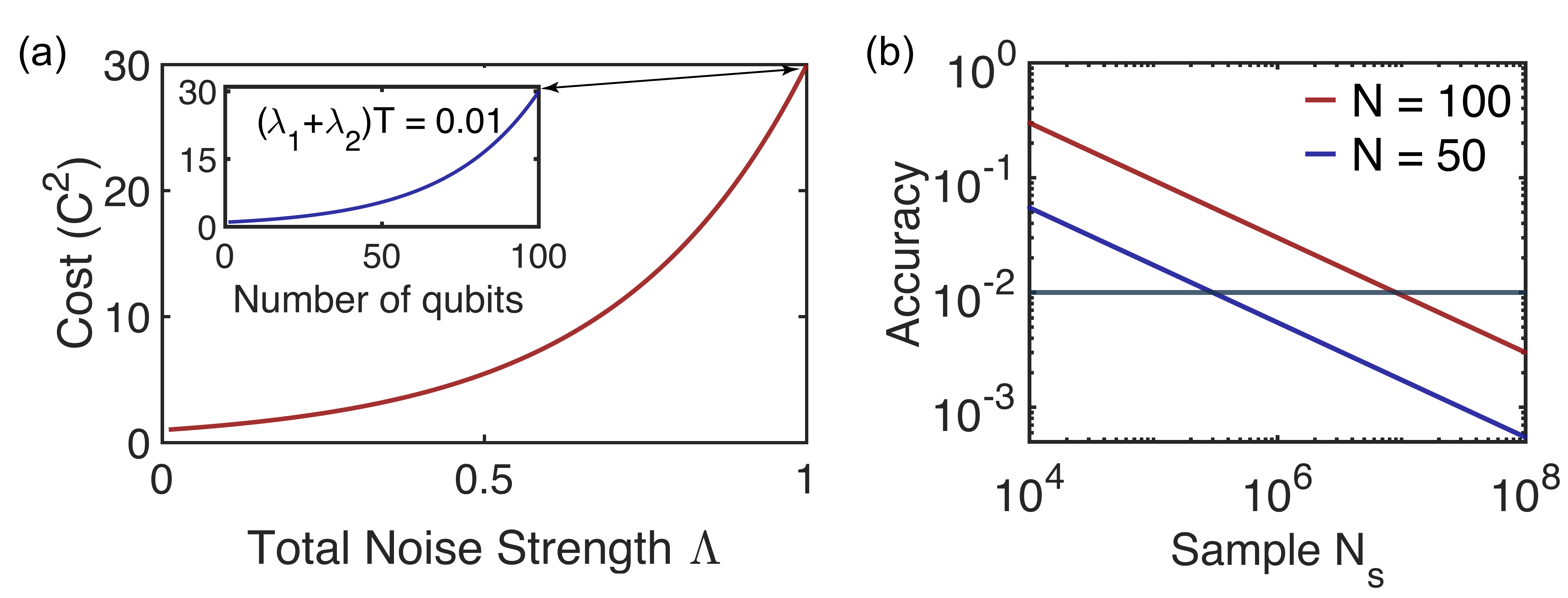}
\caption{(a) Cost versus total noise strength $\Lambda=NT(\lambda_1+\lambda_2)$. We consider a  general $N$-qubit Hamiltonian $H_{\rm sim}$ with single-qubit energy relaxation ($\lambda_1$) and dephasing noise ($\lambda_2=\lambda_1$), and evolution time $T$. The  inset shows the cost versus different number of qubits with  $(\lambda_1+\lambda_2)T=0.01$. (b) Simulation accuracy $\varepsilon\propto C/\sqrt{N_s}$ with different number of samples $N_s$ with $T=1~\mu$s, $\lambda_1+\lambda_2=0.01$ MHz, $N=100$ (red) and $N=50$ (blue). We only consider pessimistic estimation $C/\sqrt{N_s}$ and the error $\varepsilon$ can be much smaller in practice.
}
\label{fig:cost}
\end{figure}

\vspace{0.2cm}
\section{Resource cost for NISQ devices}
In this section, we estimate the resource cost for stochastic error mitigation with NISQ devices.
Given precise noise model, the stochastic error mitigation method in principle enables exact recovery of the ideal evolution. However, to achieve the same accuracy of the measurement on the ideal evolution, we need $C^2$ times more samples or experiment runs with the error-mitigated noisy evolution. The overhead  $C^2$ is likely to be prohibitively large with a significant amount of noise on a NISQ device. Nevertheless, we show that the overhead can be reasonably small (less than 100) when the total error (defined below) is less than 1.
In particular, we consider a noisy superconducting simulator with up to $N=100$ qubits, which suffers from single-qubit relaxation and dephasing noise with equal noise strength $\lambda_1=\lambda_2$.
While the noise strength is defined as the noise rate at instant time, we define the total noise strength
\begin{equation}
    \Lambda=NT(\lambda_1+\lambda_2)
\end{equation}
 as the noise strength of the whole $N$-qubit system within time $T$. The dependence of the overhead $C^2$ on the total noise strength $\Lambda$ and number of qubits is shown in Fig.~\ref{fig:cost}(a).
For a practical case with $T=1~\mu$s, $\lambda_1+\lambda_2=0.01$ MHz, $N=100$, the cost $C^2(\Lambda=1)=30$ and we further show the number of measurements needed to achieve a given simulation accuracy in Fig.~\ref{fig:cost}(b).  Note that the overhead $C^2$ is independent of the Hamiltonian $H_{\rm sim}$, so the results apply for general NISQ devices (see Appendix~\ref{section:continuouserror}).
\vspace{0.2cm}
\section{Discussion}
To summarise, we propose stochastic and hybrid quantum error mitigation schemes to mitigate noise in a continuous process.
While previous error mitigation methods for DQS regard each gate as one entity and noise as an
error channel before or after the gate, such a description becomes inadequate when the quantum gate is on
multi-qubits and the noise are inherently mixed in the realisation of the quantum gate. By regarding the implementation of each quantum gate as a continuous process, our error mitigation methods can thus be applied to mitigate errors for realisation of multi-qubit gates (which generally have large errors).
Since dominant noise in NISQ devices is from implementing multi-qubit operations or inherent noise with finite coherence time, our scheme can effectively suppress them and thus extend the computation capability of analog quantum simulators and digital gate-based quantum computers in solving practical problems~\cite{kjaergaard2019superconducting}.
We numerically test it with analog simulators for several Hamiltonian simulations under incoherent errors including energy relaxing and dephasing noise and a parameterised quantum circuit under additional coherent crosstalk noise.  We show its feasibility with general NISQ devices with up to 100 qubits. The proposed QEM schemes work for all digital and many analog quantum simulators with accurate single-qubit controls.

Furthermore, resolving the drift or temporal fluctuations of noise is challenging for conventional QEM methods. Our hybrid scheme incorporating extrapolation can mitigate model estimation error and the error of recovery operations, which alleviates the requirement of precise tomography of error model and precise control of the quantum simulators. Our method is tested to be robust to the drift of noise.
Although the discussion focused on local time-independent noise, our scheme can be potentially generalised to general non-local time-dependent noise by employing the time dependent recovery operation $\mathcal{E}_Q(t)$, and we numerically test its viability for the time-dependent noise in Appendix~\ref{appendix: timedepsimulation}. We leave the detailed discussion to a future work.

\vspace{0.1cm}
\begin{acknowledgments}
SE~and JS~acknowledge Yuichiro Matsuzaki, Hideaki Hakoshima, Tianhan Liu and Shuxiang Cao for useful discussions.  SE~acknowledges financial support from the Japan Student Services Organization (JASSO) Student Exchange Support Program (Graduate Scholarship for Degree Seeking Students).  TT~acknowledges financial support from the Masason Foundation. XY~and SCB~acknowledges EPSRC projects EP/M013243/1 and the European Quantum Technology Flagship project AQTION. VV thanks the National Research Foundation, Prime Minister’s Office, Singapore, under its Competitive Research Programme (CRP Award No. NRF- CRP14-2014-02) and administered by Centre for Quantum Technologies, National
University of Singapore.
\end{acknowledgments}

\bibliographystyle{apsrev4-1}
\bibliography{bibcontinous}

\begin{thebibliography}{100}%
\makeatletter
\providecommand \@ifxundefined [1]{%
 \@ifx{#1\undefined}
}%
\providecommand \@ifnum [1]{%
 \ifnum #1\expandafter \@firstoftwo
 \else \expandafter \@secondoftwo
 \fi
}%
\providecommand \@ifx [1]{%
 \ifx #1\expandafter \@firstoftwo
 \else \expandafter \@secondoftwo
 \fi
}%
\providecommand \natexlab [1]{#1}%
\providecommand \enquote  [1]{``#1''}%
\providecommand \bibnamefont  [1]{#1}%
\providecommand \bibfnamefont [1]{#1}%
\providecommand \citenamefont [1]{#1}%
\providecommand \href@noop [0]{\@secondoftwo}%
\providecommand \href [0]{\begingroup \@sanitize@url \@href}%
\providecommand \@href[1]{\@@startlink{#1}\@@href}%
\providecommand \@@href[1]{\endgroup#1\@@endlink}%
\providecommand \@sanitize@url [0]{\catcode `\\12\catcode `\$12\catcode
  `\&12\catcode `\#12\catcode `\^12\catcode `\_12\catcode `\%12\relax}%
\providecommand \@@startlink[1]{}%
\providecommand \@@endlink[0]{}%
\providecommand \url  [0]{\begingroup\@sanitize@url \@url }%
\providecommand \@url [1]{\endgroup\@href {#1}{\urlprefix }}%
\providecommand \urlprefix  [0]{URL }%
\providecommand \Eprint [0]{\href }%
\providecommand \doibase [0]{http://dx.doi.org/}%
\providecommand \selectlanguage [0]{\@gobble}%
\providecommand \bibinfo  [0]{\@secondoftwo}%
\providecommand \bibfield  [0]{\@secondoftwo}%
\providecommand \translation [1]{[#1]}%
\providecommand \BibitemOpen [0]{}%
\providecommand \bibitemStop [0]{}%
\providecommand \bibitemNoStop [0]{.\EOS\space}%
\providecommand \EOS [0]{\spacefactor3000\relax}%
\providecommand \BibitemShut  [1]{\csname bibitem#1\endcsname}%
\let\auto@bib@innerbib\@empty
\bibitem [{\citenamefont {Arute}\ \emph {et~al.}(2019)\citenamefont {Arute},
  \citenamefont {Arya}, \citenamefont {Babbush}, \citenamefont {Bacon},
  \citenamefont {Bardin}, \citenamefont {Barends}, \citenamefont {Biswas},
  \citenamefont {Boixo}, \citenamefont {Brandao}, \citenamefont {Buell} \emph
  {et~al.}}]{arute2019quantum}%
  \BibitemOpen
  \bibfield  {author} {\bibinfo {author} {\bibfnamefont {F.}~\bibnamefont
  {Arute}}, \bibinfo {author} {\bibfnamefont {K.}~\bibnamefont {Arya}},
  \bibinfo {author} {\bibfnamefont {R.}~\bibnamefont {Babbush}}, \bibinfo
  {author} {\bibfnamefont {D.}~\bibnamefont {Bacon}}, \bibinfo {author}
  {\bibfnamefont {J.~C.}\ \bibnamefont {Bardin}}, \bibinfo {author}
  {\bibfnamefont {R.}~\bibnamefont {Barends}}, \bibinfo {author} {\bibfnamefont
  {R.}~\bibnamefont {Biswas}}, \bibinfo {author} {\bibfnamefont
  {S.}~\bibnamefont {Boixo}}, \bibinfo {author} {\bibfnamefont {F.~G.}\
  \bibnamefont {Brandao}}, \bibinfo {author} {\bibfnamefont {D.~A.}\
  \bibnamefont {Buell}},  \emph {et~al.},\ }\href@noop {} {\bibfield  {journal}
  {\bibinfo  {journal} {Nature}\ }\textbf {\bibinfo {volume} {574}},\ \bibinfo
  {pages} {505} (\bibinfo {year} {2019})}\BibitemShut {NoStop}%
\bibitem [{\citenamefont {Preskill}(2018)}]{preskill2018quantum}%
  \BibitemOpen
  \bibfield  {author} {\bibinfo {author} {\bibfnamefont {J.}~\bibnamefont
  {Preskill}},\ }\href@noop {} {\bibfield  {journal} {\bibinfo  {journal}
  {Quantum}\ }\textbf {\bibinfo {volume} {2}},\ \bibinfo {pages} {79} (\bibinfo
  {year} {2018})}\BibitemShut {NoStop}%
\bibitem [{\citenamefont {Li}\ and\ \citenamefont
  {Benjamin}(2017)}]{li2017efficient}%
  \BibitemOpen
  \bibfield  {author} {\bibinfo {author} {\bibfnamefont {Y.}~\bibnamefont
  {Li}}\ and\ \bibinfo {author} {\bibfnamefont {S.~C.}\ \bibnamefont
  {Benjamin}},\ }\href@noop {} {\bibfield  {journal} {\bibinfo  {journal}
  {Physical Review X}\ }\textbf {\bibinfo {volume} {7}},\ \bibinfo {pages}
  {021050} (\bibinfo {year} {2017})}\BibitemShut {NoStop}%
\bibitem [{\citenamefont {Temme}\ \emph {et~al.}(2017)\citenamefont {Temme},
  \citenamefont {Bravyi},\ and\ \citenamefont {Gambetta}}]{temme2017error}%
  \BibitemOpen
  \bibfield  {author} {\bibinfo {author} {\bibfnamefont {K.}~\bibnamefont
  {Temme}}, \bibinfo {author} {\bibfnamefont {S.}~\bibnamefont {Bravyi}}, \
  and\ \bibinfo {author} {\bibfnamefont {J.~M.}\ \bibnamefont {Gambetta}},\
  }\href@noop {} {\bibfield  {journal} {\bibinfo  {journal} {Physical review
  letters}\ }\textbf {\bibinfo {volume} {119}},\ \bibinfo {pages} {180509}
  (\bibinfo {year} {2017})}\BibitemShut {NoStop}%
\bibitem [{\citenamefont {Endo}\ \emph {et~al.}(2018)\citenamefont {Endo},
  \citenamefont {Benjamin},\ and\ \citenamefont {Li}}]{endo2018practical}%
  \BibitemOpen
  \bibfield  {author} {\bibinfo {author} {\bibfnamefont {S.}~\bibnamefont
  {Endo}}, \bibinfo {author} {\bibfnamefont {S.~C.}\ \bibnamefont {Benjamin}},
  \ and\ \bibinfo {author} {\bibfnamefont {Y.}~\bibnamefont {Li}},\ }\href@noop
  {} {\bibfield  {journal} {\bibinfo  {journal} {Physical Review X}\ }\textbf
  {\bibinfo {volume} {8}},\ \bibinfo {pages} {031027} (\bibinfo {year}
  {2018})}\BibitemShut {NoStop}%
\bibitem [{\citenamefont {McClean}\ \emph {et~al.}(2017)\citenamefont
  {McClean}, \citenamefont {Kimchi-Schwartz}, \citenamefont {Carter},\ and\
  \citenamefont {de~Jong}}]{PhysRevA.95.042308}%
  \BibitemOpen
  \bibfield  {author} {\bibinfo {author} {\bibfnamefont {J.~R.}\ \bibnamefont
  {McClean}}, \bibinfo {author} {\bibfnamefont {M.~E.}\ \bibnamefont
  {Kimchi-Schwartz}}, \bibinfo {author} {\bibfnamefont {J.}~\bibnamefont
  {Carter}}, \ and\ \bibinfo {author} {\bibfnamefont {W.~A.}\ \bibnamefont
  {de~Jong}},\ }\href {\doibase 10.1103/PhysRevA.95.042308} {\bibfield
  {journal} {\bibinfo  {journal} {Phys. Rev. A}\ }\textbf {\bibinfo {volume}
  {95}},\ \bibinfo {pages} {042308} (\bibinfo {year} {2017})}\BibitemShut
  {NoStop}%
\bibitem [{\citenamefont {Huo}\ and\ \citenamefont
  {Li}(2018)}]{huo2018temporally}%
  \BibitemOpen
  \bibfield  {author} {\bibinfo {author} {\bibfnamefont {M.}~\bibnamefont
  {Huo}}\ and\ \bibinfo {author} {\bibfnamefont {Y.}~\bibnamefont {Li}},\
  }\href@noop {} {\bibfield  {journal} {\bibinfo  {journal} {arXiv preprint
  arXiv:1811.02734}\ } (\bibinfo {year} {2018})}\BibitemShut {NoStop}%
\bibitem [{\citenamefont {Bonet-Monroig}\ \emph {et~al.}(2018)\citenamefont
  {Bonet-Monroig}, \citenamefont {Sagastizabal}, \citenamefont {Singh},\ and\
  \citenamefont {O'Brien}}]{bonet2018low}%
  \BibitemOpen
  \bibfield  {author} {\bibinfo {author} {\bibfnamefont {X.}~\bibnamefont
  {Bonet-Monroig}}, \bibinfo {author} {\bibfnamefont {R.}~\bibnamefont
  {Sagastizabal}}, \bibinfo {author} {\bibfnamefont {M.}~\bibnamefont {Singh}},
  \ and\ \bibinfo {author} {\bibfnamefont {T.}~\bibnamefont {O'Brien}},\
  }\href@noop {} {\bibfield  {journal} {\bibinfo  {journal} {Physical Review
  A}\ }\textbf {\bibinfo {volume} {98}},\ \bibinfo {pages} {062339} (\bibinfo
  {year} {2018})}\BibitemShut {NoStop}%
\bibitem [{\citenamefont {Dumitrescu}\ \emph {et~al.}(2018)\citenamefont
  {Dumitrescu}, \citenamefont {McCaskey}, \citenamefont {Hagen}, \citenamefont
  {Jansen}, \citenamefont {Morris}, \citenamefont {Papenbrock}, \citenamefont
  {Pooser}, \citenamefont {Dean},\ and\ \citenamefont
  {Lougovski}}]{dumitrescu2018cloud}%
  \BibitemOpen
  \bibfield  {author} {\bibinfo {author} {\bibfnamefont {E.~F.}\ \bibnamefont
  {Dumitrescu}}, \bibinfo {author} {\bibfnamefont {A.~J.}\ \bibnamefont
  {McCaskey}}, \bibinfo {author} {\bibfnamefont {G.}~\bibnamefont {Hagen}},
  \bibinfo {author} {\bibfnamefont {G.~R.}\ \bibnamefont {Jansen}}, \bibinfo
  {author} {\bibfnamefont {T.~D.}\ \bibnamefont {Morris}}, \bibinfo {author}
  {\bibfnamefont {T.}~\bibnamefont {Papenbrock}}, \bibinfo {author}
  {\bibfnamefont {R.~C.}\ \bibnamefont {Pooser}}, \bibinfo {author}
  {\bibfnamefont {D.~J.}\ \bibnamefont {Dean}}, \ and\ \bibinfo {author}
  {\bibfnamefont {P.}~\bibnamefont {Lougovski}},\ }\href@noop {} {\bibfield
  {journal} {\bibinfo  {journal} {Physical review letters}\ }\textbf {\bibinfo
  {volume} {120}},\ \bibinfo {pages} {210501} (\bibinfo {year}
  {2018})}\BibitemShut {NoStop}%
\bibitem [{\citenamefont {Otten}\ and\ \citenamefont
  {Gray}(2019{\natexlab{a}})}]{otten2019recovering}%
  \BibitemOpen
  \bibfield  {author} {\bibinfo {author} {\bibfnamefont {M.}~\bibnamefont
  {Otten}}\ and\ \bibinfo {author} {\bibfnamefont {S.~K.}\ \bibnamefont
  {Gray}},\ }\href@noop {} {\bibfield  {journal} {\bibinfo  {journal} {Physical
  Review A}\ }\textbf {\bibinfo {volume} {99}},\ \bibinfo {pages} {012338}
  (\bibinfo {year} {2019}{\natexlab{a}})}\BibitemShut {NoStop}%
\bibitem [{\citenamefont {McArdle}\ \emph {et~al.}(2019)\citenamefont
  {McArdle}, \citenamefont {Yuan},\ and\ \citenamefont
  {Benjamin}}]{mcardle2019error}%
  \BibitemOpen
  \bibfield  {author} {\bibinfo {author} {\bibfnamefont {S.}~\bibnamefont
  {McArdle}}, \bibinfo {author} {\bibfnamefont {X.}~\bibnamefont {Yuan}}, \
  and\ \bibinfo {author} {\bibfnamefont {S.}~\bibnamefont {Benjamin}},\
  }\href@noop {} {\bibfield  {journal} {\bibinfo  {journal} {Physical review
  letters}\ }\textbf {\bibinfo {volume} {122}},\ \bibinfo {pages} {180501}
  (\bibinfo {year} {2019})}\BibitemShut {NoStop}%
\bibitem [{\citenamefont {Sagastizabal}\ \emph {et~al.}(2019)\citenamefont
  {Sagastizabal}, \citenamefont {Bonet-Monroig}, \citenamefont {Singh},
  \citenamefont {Rol}, \citenamefont {Bultink}, \citenamefont {Fu},
  \citenamefont {Price}, \citenamefont {Ostroukh}, \citenamefont
  {Muthusubramanian}, \citenamefont {Bruno} \emph
  {et~al.}}]{sagastizabal2019experimental}%
  \BibitemOpen
  \bibfield  {author} {\bibinfo {author} {\bibfnamefont {R.}~\bibnamefont
  {Sagastizabal}}, \bibinfo {author} {\bibfnamefont {X.}~\bibnamefont
  {Bonet-Monroig}}, \bibinfo {author} {\bibfnamefont {M.}~\bibnamefont
  {Singh}}, \bibinfo {author} {\bibfnamefont {M.~A.}\ \bibnamefont {Rol}},
  \bibinfo {author} {\bibfnamefont {C.}~\bibnamefont {Bultink}}, \bibinfo
  {author} {\bibfnamefont {X.}~\bibnamefont {Fu}}, \bibinfo {author}
  {\bibfnamefont {C.}~\bibnamefont {Price}}, \bibinfo {author} {\bibfnamefont
  {V.}~\bibnamefont {Ostroukh}}, \bibinfo {author} {\bibfnamefont
  {N.}~\bibnamefont {Muthusubramanian}}, \bibinfo {author} {\bibfnamefont
  {A.}~\bibnamefont {Bruno}},  \emph {et~al.},\ }\href@noop {} {\bibfield
  {journal} {\bibinfo  {journal} {Physical Review A}\ }\textbf {\bibinfo
  {volume} {100}},\ \bibinfo {pages} {010302} (\bibinfo {year}
  {2019})}\BibitemShut {NoStop}%
\bibitem [{\citenamefont {Huggins}\ \emph {et~al.}(2019)\citenamefont
  {Huggins}, \citenamefont {McClean}, \citenamefont {Rubin}, \citenamefont
  {Jiang}, \citenamefont {Wiebe}, \citenamefont {Whaley},\ and\ \citenamefont
  {Babbush}}]{huggins2019efficient}%
  \BibitemOpen
  \bibfield  {author} {\bibinfo {author} {\bibfnamefont {W.~J.}\ \bibnamefont
  {Huggins}}, \bibinfo {author} {\bibfnamefont {J.}~\bibnamefont {McClean}},
  \bibinfo {author} {\bibfnamefont {N.}~\bibnamefont {Rubin}}, \bibinfo
  {author} {\bibfnamefont {Z.}~\bibnamefont {Jiang}}, \bibinfo {author}
  {\bibfnamefont {N.}~\bibnamefont {Wiebe}}, \bibinfo {author} {\bibfnamefont
  {K.~B.}\ \bibnamefont {Whaley}}, \ and\ \bibinfo {author} {\bibfnamefont
  {R.}~\bibnamefont {Babbush}},\ }\href@noop {} {\bibfield  {journal} {\bibinfo
   {journal} {arXiv preprint arXiv:1907.13117}\ } (\bibinfo {year}
  {2019})}\BibitemShut {NoStop}%
\bibitem [{\citenamefont {Colless}\ \emph {et~al.}(2018)\citenamefont
  {Colless}, \citenamefont {Ramasesh}, \citenamefont {Dahlen}, \citenamefont
  {Blok}, \citenamefont {Kimchi-Schwartz}, \citenamefont {McClean},
  \citenamefont {Carter}, \citenamefont {de~Jong},\ and\ \citenamefont
  {Siddiqi}}]{PhysRevX.8.011021}%
  \BibitemOpen
  \bibfield  {author} {\bibinfo {author} {\bibfnamefont {J.~I.}\ \bibnamefont
  {Colless}}, \bibinfo {author} {\bibfnamefont {V.~V.}\ \bibnamefont
  {Ramasesh}}, \bibinfo {author} {\bibfnamefont {D.}~\bibnamefont {Dahlen}},
  \bibinfo {author} {\bibfnamefont {M.~S.}\ \bibnamefont {Blok}}, \bibinfo
  {author} {\bibfnamefont {M.~E.}\ \bibnamefont {Kimchi-Schwartz}}, \bibinfo
  {author} {\bibfnamefont {J.~R.}\ \bibnamefont {McClean}}, \bibinfo {author}
  {\bibfnamefont {J.}~\bibnamefont {Carter}}, \bibinfo {author} {\bibfnamefont
  {W.~A.}\ \bibnamefont {de~Jong}}, \ and\ \bibinfo {author} {\bibfnamefont
  {I.}~\bibnamefont {Siddiqi}},\ }\href {\doibase 10.1103/PhysRevX.8.011021}
  {\bibfield  {journal} {\bibinfo  {journal} {Phys. Rev. X}\ }\textbf {\bibinfo
  {volume} {8}},\ \bibinfo {pages} {011021} (\bibinfo {year}
  {2018})}\BibitemShut {NoStop}%
\bibitem [{\citenamefont {Otten}\ and\ \citenamefont
  {Gray}(2019{\natexlab{b}})}]{otten2019accounting}%
  \BibitemOpen
  \bibfield  {author} {\bibinfo {author} {\bibfnamefont {M.}~\bibnamefont
  {Otten}}\ and\ \bibinfo {author} {\bibfnamefont {S.~K.}\ \bibnamefont
  {Gray}},\ }\href@noop {} {\bibfield  {journal} {\bibinfo  {journal} {Npj
  Quantum Inf.}\ }\textbf {\bibinfo {volume} {5}},\ \bibinfo {pages} {11}
  (\bibinfo {year} {2019}{\natexlab{b}})}\BibitemShut {NoStop}%
\bibitem [{\citenamefont {McClean}\ \emph {et~al.}(2020)\citenamefont
  {McClean}, \citenamefont {Jiang}, \citenamefont {Rubin}, \citenamefont
  {Babbush},\ and\ \citenamefont {Neven}}]{mcclean2020decoding}%
  \BibitemOpen
  \bibfield  {author} {\bibinfo {author} {\bibfnamefont {J.~R.}\ \bibnamefont
  {McClean}}, \bibinfo {author} {\bibfnamefont {Z.}~\bibnamefont {Jiang}},
  \bibinfo {author} {\bibfnamefont {N.~C.}\ \bibnamefont {Rubin}}, \bibinfo
  {author} {\bibfnamefont {R.}~\bibnamefont {Babbush}}, \ and\ \bibinfo
  {author} {\bibfnamefont {H.}~\bibnamefont {Neven}},\ }\href@noop {}
  {\bibfield  {journal} {\bibinfo  {journal} {Nature Communications}\ }\textbf
  {\bibinfo {volume} {11}},\ \bibinfo {pages} {1} (\bibinfo {year}
  {2020})}\BibitemShut {NoStop}%
\bibitem [{\citenamefont {Giurgica-Tiron}\ \emph {et~al.}(2020)\citenamefont
  {Giurgica-Tiron}, \citenamefont {Hindy}, \citenamefont {LaRose},
  \citenamefont {Mari},\ and\ \citenamefont {Zeng}}]{giurgica2020digital}%
  \BibitemOpen
  \bibfield  {author} {\bibinfo {author} {\bibfnamefont {T.}~\bibnamefont
  {Giurgica-Tiron}}, \bibinfo {author} {\bibfnamefont {Y.}~\bibnamefont
  {Hindy}}, \bibinfo {author} {\bibfnamefont {R.}~\bibnamefont {LaRose}},
  \bibinfo {author} {\bibfnamefont {A.}~\bibnamefont {Mari}}, \ and\ \bibinfo
  {author} {\bibfnamefont {W.~J.}\ \bibnamefont {Zeng}},\ }\href@noop {}
  {\bibfield  {journal} {\bibinfo  {journal} {arXiv preprint arXiv:2005.10921}\
  } (\bibinfo {year} {2020})}\BibitemShut {NoStop}%
\bibitem [{\citenamefont {Keen}\ \emph {et~al.}(2020)\citenamefont {Keen},
  \citenamefont {Maier}, \citenamefont {Johnston},\ and\ \citenamefont
  {Lougovski}}]{keen2020quantum}%
  \BibitemOpen
  \bibfield  {author} {\bibinfo {author} {\bibfnamefont {T.}~\bibnamefont
  {Keen}}, \bibinfo {author} {\bibfnamefont {T.}~\bibnamefont {Maier}},
  \bibinfo {author} {\bibfnamefont {S.}~\bibnamefont {Johnston}}, \ and\
  \bibinfo {author} {\bibfnamefont {P.}~\bibnamefont {Lougovski}},\ }\href@noop
  {} {\bibfield  {journal} {\bibinfo  {journal} {Quantum Science and
  Technology}\ }\textbf {\bibinfo {volume} {5}},\ \bibinfo {pages} {035001}
  (\bibinfo {year} {2020})}\BibitemShut {NoStop}%
\bibitem [{\citenamefont {Cai}(2020)}]{cai2020multi}%
  \BibitemOpen
  \bibfield  {author} {\bibinfo {author} {\bibfnamefont {Z.}~\bibnamefont
  {Cai}},\ }\href@noop {} {\bibfield  {journal} {\bibinfo  {journal} {arXiv
  preprint arXiv:2007.01265}\ } (\bibinfo {year} {2020})}\BibitemShut {NoStop}%
\bibitem [{\citenamefont {He}\ \emph {et~al.}(2020)\citenamefont {He},
  \citenamefont {Nachman}, \citenamefont {de~Jong},\ and\ \citenamefont
  {Bauer}}]{he2020resource}%
  \BibitemOpen
  \bibfield  {author} {\bibinfo {author} {\bibfnamefont {A.}~\bibnamefont
  {He}}, \bibinfo {author} {\bibfnamefont {B.}~\bibnamefont {Nachman}},
  \bibinfo {author} {\bibfnamefont {W.~A.}\ \bibnamefont {de~Jong}}, \ and\
  \bibinfo {author} {\bibfnamefont {C.~W.}\ \bibnamefont {Bauer}},\ }\href@noop
  {} {\bibfield  {journal} {\bibinfo  {journal} {arXiv preprint
  arXiv:2003.04941}\ } (\bibinfo {year} {2020})}\BibitemShut {NoStop}%
\bibitem [{\citenamefont {Maciejewski}\ \emph {et~al.}(2020)\citenamefont
  {Maciejewski}, \citenamefont {Zimbor{\'a}s},\ and\ \citenamefont
  {Oszmaniec}}]{maciejewski2020mitigation}%
  \BibitemOpen
  \bibfield  {author} {\bibinfo {author} {\bibfnamefont {F.~B.}\ \bibnamefont
  {Maciejewski}}, \bibinfo {author} {\bibfnamefont {Z.}~\bibnamefont
  {Zimbor{\'a}s}}, \ and\ \bibinfo {author} {\bibfnamefont {M.}~\bibnamefont
  {Oszmaniec}},\ }\href@noop {} {\bibfield  {journal} {\bibinfo  {journal}
  {Quantum}\ }\textbf {\bibinfo {volume} {4}},\ \bibinfo {pages} {257}
  (\bibinfo {year} {2020})}\BibitemShut {NoStop}%
\bibitem [{\citenamefont {Chen}\ \emph {et~al.}(2019)\citenamefont {Chen},
  \citenamefont {Farahzad}, \citenamefont {Yoo},\ and\ \citenamefont
  {Wei}}]{chen2019detector}%
  \BibitemOpen
  \bibfield  {author} {\bibinfo {author} {\bibfnamefont {Y.}~\bibnamefont
  {Chen}}, \bibinfo {author} {\bibfnamefont {M.}~\bibnamefont {Farahzad}},
  \bibinfo {author} {\bibfnamefont {S.}~\bibnamefont {Yoo}}, \ and\ \bibinfo
  {author} {\bibfnamefont {T.-C.}\ \bibnamefont {Wei}},\ }\href@noop {}
  {\bibfield  {journal} {\bibinfo  {journal} {Physical Review A}\ }\textbf
  {\bibinfo {volume} {100}},\ \bibinfo {pages} {052315} (\bibinfo {year}
  {2019})}\BibitemShut {NoStop}%
\bibitem [{\citenamefont {Kwon}\ and\ \citenamefont
  {Bae}(2020)}]{kwon2020hybrid}%
  \BibitemOpen
  \bibfield  {author} {\bibinfo {author} {\bibfnamefont {H.}~\bibnamefont
  {Kwon}}\ and\ \bibinfo {author} {\bibfnamefont {J.}~\bibnamefont {Bae}},\
  }\href@noop {} {\bibfield  {journal} {\bibinfo  {journal} {arXiv preprint
  arXiv:2003.12314}\ } (\bibinfo {year} {2020})}\BibitemShut {NoStop}%
\bibitem [{\citenamefont {Strikis}\ \emph {et~al.}(2020)\citenamefont
  {Strikis}, \citenamefont {Qin}, \citenamefont {Chen}, \citenamefont
  {Benjamin},\ and\ \citenamefont {Li}}]{strikis2020learning}%
  \BibitemOpen
  \bibfield  {author} {\bibinfo {author} {\bibfnamefont {A.}~\bibnamefont
  {Strikis}}, \bibinfo {author} {\bibfnamefont {D.}~\bibnamefont {Qin}},
  \bibinfo {author} {\bibfnamefont {Y.}~\bibnamefont {Chen}}, \bibinfo {author}
  {\bibfnamefont {S.~C.}\ \bibnamefont {Benjamin}}, \ and\ \bibinfo {author}
  {\bibfnamefont {Y.}~\bibnamefont {Li}},\ }\href@noop {} {\bibfield  {journal}
  {\bibinfo  {journal} {arXiv preprint arXiv:2005.07601}\ } (\bibinfo {year}
  {2020})}\BibitemShut {NoStop}%
\bibitem [{\citenamefont {Bravyi}\ \emph {et~al.}(2020)\citenamefont {Bravyi},
  \citenamefont {Sheldon}, \citenamefont {Kandala}, \citenamefont {Mckay},\
  and\ \citenamefont {Gambetta}}]{bravyi2020mitigating}%
  \BibitemOpen
  \bibfield  {author} {\bibinfo {author} {\bibfnamefont {S.}~\bibnamefont
  {Bravyi}}, \bibinfo {author} {\bibfnamefont {S.}~\bibnamefont {Sheldon}},
  \bibinfo {author} {\bibfnamefont {A.}~\bibnamefont {Kandala}}, \bibinfo
  {author} {\bibfnamefont {D.~C.}\ \bibnamefont {Mckay}}, \ and\ \bibinfo
  {author} {\bibfnamefont {J.~M.}\ \bibnamefont {Gambetta}},\ }\href@noop {}
  {\bibfield  {journal} {\bibinfo  {journal} {arXiv preprint arXiv:2006.14044}\
  } (\bibinfo {year} {2020})}\BibitemShut {NoStop}%
\bibitem [{\citenamefont {Czarnik}\ \emph {et~al.}(2020)\citenamefont
  {Czarnik}, \citenamefont {Arrasmith}, \citenamefont {Coles},\ and\
  \citenamefont {Cincio}}]{czarnik2020error}%
  \BibitemOpen
  \bibfield  {author} {\bibinfo {author} {\bibfnamefont {P.}~\bibnamefont
  {Czarnik}}, \bibinfo {author} {\bibfnamefont {A.}~\bibnamefont {Arrasmith}},
  \bibinfo {author} {\bibfnamefont {P.~J.}\ \bibnamefont {Coles}}, \ and\
  \bibinfo {author} {\bibfnamefont {L.}~\bibnamefont {Cincio}},\ }\href@noop {}
  {\bibfield  {journal} {\bibinfo  {journal} {arXiv preprint arXiv:2005.10189}\
  } (\bibinfo {year} {2020})}\BibitemShut {NoStop}%
\bibitem [{\citenamefont {Zlokapa}\ and\ \citenamefont
  {Gheorghiu}(2020)}]{zlokapa2020deep}%
  \BibitemOpen
  \bibfield  {author} {\bibinfo {author} {\bibfnamefont {A.}~\bibnamefont
  {Zlokapa}}\ and\ \bibinfo {author} {\bibfnamefont {A.}~\bibnamefont
  {Gheorghiu}},\ }\href@noop {} {\bibfield  {journal} {\bibinfo  {journal}
  {arXiv preprint arXiv:2005.10811}\ } (\bibinfo {year} {2020})}\BibitemShut
  {NoStop}%
\bibitem [{\citenamefont {Endo}\ \emph {et~al.}(2020)\citenamefont {Endo},
  \citenamefont {Cai}, \citenamefont {Benjamin},\ and\ \citenamefont
  {Yuan}}]{endo2020hybrid}%
  \BibitemOpen
  \bibfield  {author} {\bibinfo {author} {\bibfnamefont {S.}~\bibnamefont
  {Endo}}, \bibinfo {author} {\bibfnamefont {Z.}~\bibnamefont {Cai}}, \bibinfo
  {author} {\bibfnamefont {S.~C.}\ \bibnamefont {Benjamin}}, \ and\ \bibinfo
  {author} {\bibfnamefont {X.}~\bibnamefont {Yuan}},\ }\href@noop {} {\bibfield
   {journal} {\bibinfo  {journal} {arXiv preprint arXiv:2011.01382}\ }
  (\bibinfo {year} {2020})}\BibitemShut {NoStop}%
\bibitem [{\citenamefont {Cerezo}\ \emph {et~al.}(2020)\citenamefont {Cerezo},
  \citenamefont {Arrasmith}, \citenamefont {Babbush}, \citenamefont {Benjamin},
  \citenamefont {Endo}, \citenamefont {Fujii}, \citenamefont {McClean},
  \citenamefont {Mitarai}, \citenamefont {Yuan}, \citenamefont {Cincio},\ and\
  \citenamefont {Coles}}]{cerezo2020variationalreview}%
  \BibitemOpen
  \bibfield  {author} {\bibinfo {author} {\bibfnamefont {M.}~\bibnamefont
  {Cerezo}}, \bibinfo {author} {\bibfnamefont {A.}~\bibnamefont {Arrasmith}},
  \bibinfo {author} {\bibfnamefont {R.}~\bibnamefont {Babbush}}, \bibinfo
  {author} {\bibfnamefont {S.~C.}\ \bibnamefont {Benjamin}}, \bibinfo {author}
  {\bibfnamefont {S.}~\bibnamefont {Endo}}, \bibinfo {author} {\bibfnamefont
  {K.}~\bibnamefont {Fujii}}, \bibinfo {author} {\bibfnamefont {J.~R.}\
  \bibnamefont {McClean}}, \bibinfo {author} {\bibfnamefont {K.}~\bibnamefont
  {Mitarai}}, \bibinfo {author} {\bibfnamefont {X.}~\bibnamefont {Yuan}},
  \bibinfo {author} {\bibfnamefont {L.}~\bibnamefont {Cincio}}, \ and\ \bibinfo
  {author} {\bibfnamefont {P.~J.}\ \bibnamefont {Coles}},\ }\href@noop {}
  {\bibfield  {journal} {\bibinfo  {journal} {arXiv preprint arXiv:2012.09265}\
  } (\bibinfo {year} {2020})}\BibitemShut {NoStop}%
\bibitem [{\citenamefont {Kandala}\ \emph {et~al.}(2019)\citenamefont
  {Kandala}, \citenamefont {Temme}, \citenamefont {C{\'o}rcoles}, \citenamefont
  {Mezzacapo}, \citenamefont {Chow},\ and\ \citenamefont
  {Gambetta}}]{kandala2019error}%
  \BibitemOpen
  \bibfield  {author} {\bibinfo {author} {\bibfnamefont {A.}~\bibnamefont
  {Kandala}}, \bibinfo {author} {\bibfnamefont {K.}~\bibnamefont {Temme}},
  \bibinfo {author} {\bibfnamefont {A.~D.}\ \bibnamefont {C{\'o}rcoles}},
  \bibinfo {author} {\bibfnamefont {A.}~\bibnamefont {Mezzacapo}}, \bibinfo
  {author} {\bibfnamefont {J.~M.}\ \bibnamefont {Chow}}, \ and\ \bibinfo
  {author} {\bibfnamefont {J.~M.}\ \bibnamefont {Gambetta}},\ }\href@noop {}
  {\bibfield  {journal} {\bibinfo  {journal} {Nature}\ }\textbf {\bibinfo
  {volume} {567}},\ \bibinfo {pages} {491} (\bibinfo {year}
  {2019})}\BibitemShut {NoStop}%
\bibitem [{\citenamefont {Kjaergaard}\ \emph {et~al.}(2019)\citenamefont
  {Kjaergaard}, \citenamefont {Schwartz}, \citenamefont {Braum{\"u}ller},
  \citenamefont {Krantz}, \citenamefont {Wang}, \citenamefont {Gustavsson},\
  and\ \citenamefont {Oliver}}]{kjaergaard2019superconducting}%
  \BibitemOpen
  \bibfield  {author} {\bibinfo {author} {\bibfnamefont {M.}~\bibnamefont
  {Kjaergaard}}, \bibinfo {author} {\bibfnamefont {M.~E.}\ \bibnamefont
  {Schwartz}}, \bibinfo {author} {\bibfnamefont {J.}~\bibnamefont
  {Braum{\"u}ller}}, \bibinfo {author} {\bibfnamefont {P.}~\bibnamefont
  {Krantz}}, \bibinfo {author} {\bibfnamefont {J.~I.-J.}\ \bibnamefont {Wang}},
  \bibinfo {author} {\bibfnamefont {S.}~\bibnamefont {Gustavsson}}, \ and\
  \bibinfo {author} {\bibfnamefont {W.~D.}\ \bibnamefont {Oliver}},\
  }\href@noop {} {\bibfield  {journal} {\bibinfo  {journal} {arXiv preprint
  arXiv:1905.13641}\ } (\bibinfo {year} {2019})}\BibitemShut {NoStop}%
\bibitem [{\citenamefont {Krantz}\ \emph {et~al.}(2019)\citenamefont {Krantz},
  \citenamefont {Kjaergaard}, \citenamefont {Yan}, \citenamefont {Orlando},
  \citenamefont {Gustavsson},\ and\ \citenamefont
  {Oliver}}]{krantz2019quantum}%
  \BibitemOpen
  \bibfield  {author} {\bibinfo {author} {\bibfnamefont {P.}~\bibnamefont
  {Krantz}}, \bibinfo {author} {\bibfnamefont {M.}~\bibnamefont {Kjaergaard}},
  \bibinfo {author} {\bibfnamefont {F.}~\bibnamefont {Yan}}, \bibinfo {author}
  {\bibfnamefont {T.~P.}\ \bibnamefont {Orlando}}, \bibinfo {author}
  {\bibfnamefont {S.}~\bibnamefont {Gustavsson}}, \ and\ \bibinfo {author}
  {\bibfnamefont {W.~D.}\ \bibnamefont {Oliver}},\ }\href@noop {} {\bibfield
  {journal} {\bibinfo  {journal} {Applied Physics Reviews}\ }\textbf {\bibinfo
  {volume} {6}},\ \bibinfo {pages} {021318} (\bibinfo {year}
  {2019})}\BibitemShut {NoStop}%
\bibitem [{\citenamefont {Sheldon}\ \emph {et~al.}(2016)\citenamefont
  {Sheldon}, \citenamefont {Magesan}, \citenamefont {Chow},\ and\ \citenamefont
  {Gambetta}}]{sheldon2016procedure}%
  \BibitemOpen
  \bibfield  {author} {\bibinfo {author} {\bibfnamefont {S.}~\bibnamefont
  {Sheldon}}, \bibinfo {author} {\bibfnamefont {E.}~\bibnamefont {Magesan}},
  \bibinfo {author} {\bibfnamefont {J.~M.}\ \bibnamefont {Chow}}, \ and\
  \bibinfo {author} {\bibfnamefont {J.~M.}\ \bibnamefont {Gambetta}},\
  }\href@noop {} {\bibfield  {journal} {\bibinfo  {journal} {Physical Review
  A}\ }\textbf {\bibinfo {volume} {93}},\ \bibinfo {pages} {060302} (\bibinfo
  {year} {2016})}\BibitemShut {NoStop}%
\bibitem [{\citenamefont {Plantenberg}\ \emph {et~al.}(2007)\citenamefont
  {Plantenberg}, \citenamefont {De~Groot}, \citenamefont {Harmans},\ and\
  \citenamefont {Mooij}}]{plantenberg2007demonstration}%
  \BibitemOpen
  \bibfield  {author} {\bibinfo {author} {\bibfnamefont {J.}~\bibnamefont
  {Plantenberg}}, \bibinfo {author} {\bibfnamefont {P.}~\bibnamefont
  {De~Groot}}, \bibinfo {author} {\bibfnamefont {C.}~\bibnamefont {Harmans}}, \
  and\ \bibinfo {author} {\bibfnamefont {J.}~\bibnamefont {Mooij}},\
  }\href@noop {} {\bibfield  {journal} {\bibinfo  {journal} {Nature}\ }\textbf
  {\bibinfo {volume} {447}},\ \bibinfo {pages} {836} (\bibinfo {year}
  {2007})}\BibitemShut {NoStop}%
\bibitem [{\citenamefont {Neill}\ \emph {et~al.}(2018)\citenamefont {Neill},
  \citenamefont {Roushan}, \citenamefont {Kechedzhi}, \citenamefont {Boixo},
  \citenamefont {Isakov}, \citenamefont {Smelyanskiy}, \citenamefont {Megrant},
  \citenamefont {Chiaro}, \citenamefont {Dunsworth}, \citenamefont {Arya} \emph
  {et~al.}}]{neill2018blueprint}%
  \BibitemOpen
  \bibfield  {author} {\bibinfo {author} {\bibfnamefont {C.}~\bibnamefont
  {Neill}}, \bibinfo {author} {\bibfnamefont {P.}~\bibnamefont {Roushan}},
  \bibinfo {author} {\bibfnamefont {K.}~\bibnamefont {Kechedzhi}}, \bibinfo
  {author} {\bibfnamefont {S.}~\bibnamefont {Boixo}}, \bibinfo {author}
  {\bibfnamefont {S.~V.}\ \bibnamefont {Isakov}}, \bibinfo {author}
  {\bibfnamefont {V.}~\bibnamefont {Smelyanskiy}}, \bibinfo {author}
  {\bibfnamefont {A.}~\bibnamefont {Megrant}}, \bibinfo {author} {\bibfnamefont
  {B.}~\bibnamefont {Chiaro}}, \bibinfo {author} {\bibfnamefont
  {A.}~\bibnamefont {Dunsworth}}, \bibinfo {author} {\bibfnamefont
  {K.}~\bibnamefont {Arya}},  \emph {et~al.},\ }\href@noop {} {\bibfield
  {journal} {\bibinfo  {journal} {Science}\ }\textbf {\bibinfo {volume}
  {360}},\ \bibinfo {pages} {195} (\bibinfo {year} {2018})}\BibitemShut
  {NoStop}%
\bibitem [{\citenamefont {C{\'o}rcoles}\ \emph {et~al.}(2013)\citenamefont
  {C{\'o}rcoles}, \citenamefont {Gambetta}, \citenamefont {Chow}, \citenamefont
  {Smolin}, \citenamefont {Ware}, \citenamefont {Strand}, \citenamefont
  {Plourde},\ and\ \citenamefont {Steffen}}]{corcoles2013process}%
  \BibitemOpen
  \bibfield  {author} {\bibinfo {author} {\bibfnamefont {A.~D.}\ \bibnamefont
  {C{\'o}rcoles}}, \bibinfo {author} {\bibfnamefont {J.~M.}\ \bibnamefont
  {Gambetta}}, \bibinfo {author} {\bibfnamefont {J.~M.}\ \bibnamefont {Chow}},
  \bibinfo {author} {\bibfnamefont {J.~A.}\ \bibnamefont {Smolin}}, \bibinfo
  {author} {\bibfnamefont {M.}~\bibnamefont {Ware}}, \bibinfo {author}
  {\bibfnamefont {J.}~\bibnamefont {Strand}}, \bibinfo {author} {\bibfnamefont
  {B.~L.}\ \bibnamefont {Plourde}}, \ and\ \bibinfo {author} {\bibfnamefont
  {M.}~\bibnamefont {Steffen}},\ }\href@noop {} {\bibfield  {journal} {\bibinfo
   {journal} {Physical Review A}\ }\textbf {\bibinfo {volume} {87}},\ \bibinfo
  {pages} {030301} (\bibinfo {year} {2013})}\BibitemShut {NoStop}%
\bibitem [{\citenamefont {Chow}\ \emph {et~al.}(2012)\citenamefont {Chow},
  \citenamefont {Gambetta}, \citenamefont {Corcoles}, \citenamefont {Merkel},
  \citenamefont {Smolin}, \citenamefont {Rigetti}, \citenamefont {Poletto},
  \citenamefont {Keefe}, \citenamefont {Rothwell}, \citenamefont {Rozen} \emph
  {et~al.}}]{chow2012universal}%
  \BibitemOpen
  \bibfield  {author} {\bibinfo {author} {\bibfnamefont {J.~M.}\ \bibnamefont
  {Chow}}, \bibinfo {author} {\bibfnamefont {J.~M.}\ \bibnamefont {Gambetta}},
  \bibinfo {author} {\bibfnamefont {A.~D.}\ \bibnamefont {Corcoles}}, \bibinfo
  {author} {\bibfnamefont {S.~T.}\ \bibnamefont {Merkel}}, \bibinfo {author}
  {\bibfnamefont {J.~A.}\ \bibnamefont {Smolin}}, \bibinfo {author}
  {\bibfnamefont {C.}~\bibnamefont {Rigetti}}, \bibinfo {author} {\bibfnamefont
  {S.}~\bibnamefont {Poletto}}, \bibinfo {author} {\bibfnamefont {G.~A.}\
  \bibnamefont {Keefe}}, \bibinfo {author} {\bibfnamefont {M.~B.}\ \bibnamefont
  {Rothwell}}, \bibinfo {author} {\bibfnamefont {J.~R.}\ \bibnamefont {Rozen}},
   \emph {et~al.},\ }\href@noop {} {\bibfield  {journal} {\bibinfo  {journal}
  {Physical review letters}\ }\textbf {\bibinfo {volume} {109}},\ \bibinfo
  {pages} {060501} (\bibinfo {year} {2012})}\BibitemShut {NoStop}%
\bibitem [{\citenamefont {Rigetti}\ and\ \citenamefont
  {Devoret}(2010)}]{rigetti2010fully}%
  \BibitemOpen
  \bibfield  {author} {\bibinfo {author} {\bibfnamefont {C.}~\bibnamefont
  {Rigetti}}\ and\ \bibinfo {author} {\bibfnamefont {M.}~\bibnamefont
  {Devoret}},\ }\href@noop {} {\bibfield  {journal} {\bibinfo  {journal}
  {Physical Review B}\ }\textbf {\bibinfo {volume} {81}},\ \bibinfo {pages}
  {134507} (\bibinfo {year} {2010})}\BibitemShut {NoStop}%
\bibitem [{\citenamefont {Takita}\ \emph {et~al.}(2017)\citenamefont {Takita},
  \citenamefont {Cross}, \citenamefont {C{\'o}rcoles}, \citenamefont {Chow},\
  and\ \citenamefont {Gambetta}}]{takita2017experimental}%
  \BibitemOpen
  \bibfield  {author} {\bibinfo {author} {\bibfnamefont {M.}~\bibnamefont
  {Takita}}, \bibinfo {author} {\bibfnamefont {A.~W.}\ \bibnamefont {Cross}},
  \bibinfo {author} {\bibfnamefont {A.}~\bibnamefont {C{\'o}rcoles}}, \bibinfo
  {author} {\bibfnamefont {J.~M.}\ \bibnamefont {Chow}}, \ and\ \bibinfo
  {author} {\bibfnamefont {J.~M.}\ \bibnamefont {Gambetta}},\ }\href@noop {}
  {\bibfield  {journal} {\bibinfo  {journal} {Physical review letters}\
  }\textbf {\bibinfo {volume} {119}},\ \bibinfo {pages} {180501} (\bibinfo
  {year} {2017})}\BibitemShut {NoStop}%
\bibitem [{\citenamefont {Reagor}\ \emph {et~al.}(2018)\citenamefont {Reagor},
  \citenamefont {Osborn}, \citenamefont {Tezak}, \citenamefont {Staley},
  \citenamefont {Prawiroatmodjo}, \citenamefont {Scheer}, \citenamefont
  {Alidoust}, \citenamefont {Sete}, \citenamefont {Didier}, \citenamefont
  {da~Silva} \emph {et~al.}}]{reagor2018demonstration}%
  \BibitemOpen
  \bibfield  {author} {\bibinfo {author} {\bibfnamefont {M.}~\bibnamefont
  {Reagor}}, \bibinfo {author} {\bibfnamefont {C.~B.}\ \bibnamefont {Osborn}},
  \bibinfo {author} {\bibfnamefont {N.}~\bibnamefont {Tezak}}, \bibinfo
  {author} {\bibfnamefont {A.}~\bibnamefont {Staley}}, \bibinfo {author}
  {\bibfnamefont {G.}~\bibnamefont {Prawiroatmodjo}}, \bibinfo {author}
  {\bibfnamefont {M.}~\bibnamefont {Scheer}}, \bibinfo {author} {\bibfnamefont
  {N.}~\bibnamefont {Alidoust}}, \bibinfo {author} {\bibfnamefont {E.~A.}\
  \bibnamefont {Sete}}, \bibinfo {author} {\bibfnamefont {N.}~\bibnamefont
  {Didier}}, \bibinfo {author} {\bibfnamefont {M.~P.}\ \bibnamefont
  {da~Silva}},  \emph {et~al.},\ }\href@noop {} {\bibfield  {journal} {\bibinfo
   {journal} {Science advances}\ }\textbf {\bibinfo {volume} {4}},\ \bibinfo
  {pages} {eaao3603} (\bibinfo {year} {2018})}\BibitemShut {NoStop}%
\bibitem [{\citenamefont {Eisert}\ \emph {et~al.}(2015)\citenamefont {Eisert},
  \citenamefont {Friesdorf},\ and\ \citenamefont
  {Gogolin}}]{eisert2015quantum}%
  \BibitemOpen
  \bibfield  {author} {\bibinfo {author} {\bibfnamefont {J.}~\bibnamefont
  {Eisert}}, \bibinfo {author} {\bibfnamefont {M.}~\bibnamefont {Friesdorf}}, \
  and\ \bibinfo {author} {\bibfnamefont {C.}~\bibnamefont {Gogolin}},\
  }\href@noop {} {\bibfield  {journal} {\bibinfo  {journal} {Nature Physics}\
  }\textbf {\bibinfo {volume} {11}},\ \bibinfo {pages} {124} (\bibinfo {year}
  {2015})}\BibitemShut {NoStop}%
\bibitem [{\citenamefont {Bohnet}\ \emph {et~al.}(2016)\citenamefont {Bohnet},
  \citenamefont {Sawyer}, \citenamefont {Britton}, \citenamefont {Wall},
  \citenamefont {Rey}, \citenamefont {Foss-Feig},\ and\ \citenamefont
  {Bollinger}}]{bohnet2016quantum}%
  \BibitemOpen
  \bibfield  {author} {\bibinfo {author} {\bibfnamefont {J.~G.}\ \bibnamefont
  {Bohnet}}, \bibinfo {author} {\bibfnamefont {B.~C.}\ \bibnamefont {Sawyer}},
  \bibinfo {author} {\bibfnamefont {J.~W.}\ \bibnamefont {Britton}}, \bibinfo
  {author} {\bibfnamefont {M.~L.}\ \bibnamefont {Wall}}, \bibinfo {author}
  {\bibfnamefont {A.~M.}\ \bibnamefont {Rey}}, \bibinfo {author} {\bibfnamefont
  {M.}~\bibnamefont {Foss-Feig}}, \ and\ \bibinfo {author} {\bibfnamefont
  {J.~J.}\ \bibnamefont {Bollinger}},\ }\href@noop {} {\bibfield  {journal}
  {\bibinfo  {journal} {Science}\ }\textbf {\bibinfo {volume} {352}},\ \bibinfo
  {pages} {1297} (\bibinfo {year} {2016})}\BibitemShut {NoStop}%
\bibitem [{\citenamefont {An}\ \emph {et~al.}(2015)\citenamefont {An},
  \citenamefont {Zhang}, \citenamefont {Um}, \citenamefont {Lv}, \citenamefont
  {Lu}, \citenamefont {Zhang}, \citenamefont {Yin}, \citenamefont {Quan},\ and\
  \citenamefont {Kim}}]{an2015experimental}%
  \BibitemOpen
  \bibfield  {author} {\bibinfo {author} {\bibfnamefont {S.}~\bibnamefont
  {An}}, \bibinfo {author} {\bibfnamefont {J.-N.}\ \bibnamefont {Zhang}},
  \bibinfo {author} {\bibfnamefont {M.}~\bibnamefont {Um}}, \bibinfo {author}
  {\bibfnamefont {D.}~\bibnamefont {Lv}}, \bibinfo {author} {\bibfnamefont
  {Y.}~\bibnamefont {Lu}}, \bibinfo {author} {\bibfnamefont {J.}~\bibnamefont
  {Zhang}}, \bibinfo {author} {\bibfnamefont {Z.-Q.}\ \bibnamefont {Yin}},
  \bibinfo {author} {\bibfnamefont {H.}~\bibnamefont {Quan}}, \ and\ \bibinfo
  {author} {\bibfnamefont {K.}~\bibnamefont {Kim}},\ }\href@noop {} {\bibfield
  {journal} {\bibinfo  {journal} {Nature Physics}\ }\textbf {\bibinfo {volume}
  {11}},\ \bibinfo {pages} {193} (\bibinfo {year} {2015})}\BibitemShut
  {NoStop}%
\bibitem [{\citenamefont {Schindler}\ \emph {et~al.}(2013)\citenamefont
  {Schindler}, \citenamefont {M{\"u}ller}, \citenamefont {Nigg}, \citenamefont
  {Barreiro}, \citenamefont {Martinez}, \citenamefont {Hennrich}, \citenamefont
  {Monz}, \citenamefont {Diehl}, \citenamefont {Zoller},\ and\ \citenamefont
  {Blatt}}]{schindler2013quantum}%
  \BibitemOpen
  \bibfield  {author} {\bibinfo {author} {\bibfnamefont {P.}~\bibnamefont
  {Schindler}}, \bibinfo {author} {\bibfnamefont {M.}~\bibnamefont
  {M{\"u}ller}}, \bibinfo {author} {\bibfnamefont {D.}~\bibnamefont {Nigg}},
  \bibinfo {author} {\bibfnamefont {J.~T.}\ \bibnamefont {Barreiro}}, \bibinfo
  {author} {\bibfnamefont {E.~A.}\ \bibnamefont {Martinez}}, \bibinfo {author}
  {\bibfnamefont {M.}~\bibnamefont {Hennrich}}, \bibinfo {author}
  {\bibfnamefont {T.}~\bibnamefont {Monz}}, \bibinfo {author} {\bibfnamefont
  {S.}~\bibnamefont {Diehl}}, \bibinfo {author} {\bibfnamefont
  {P.}~\bibnamefont {Zoller}}, \ and\ \bibinfo {author} {\bibfnamefont
  {R.}~\bibnamefont {Blatt}},\ }\href@noop {} {\bibfield  {journal} {\bibinfo
  {journal} {Nature Physics}\ }\textbf {\bibinfo {volume} {9}},\ \bibinfo
  {pages} {361} (\bibinfo {year} {2013})}\BibitemShut {NoStop}%
\bibitem [{\citenamefont {Landsman}\ \emph {et~al.}(2019)\citenamefont
  {Landsman}, \citenamefont {Figgatt}, \citenamefont {Schuster}, \citenamefont
  {Linke}, \citenamefont {Yoshida}, \citenamefont {Yao},\ and\ \citenamefont
  {Monroe}}]{landsman2019verified}%
  \BibitemOpen
  \bibfield  {author} {\bibinfo {author} {\bibfnamefont {K.~A.}\ \bibnamefont
  {Landsman}}, \bibinfo {author} {\bibfnamefont {C.}~\bibnamefont {Figgatt}},
  \bibinfo {author} {\bibfnamefont {T.}~\bibnamefont {Schuster}}, \bibinfo
  {author} {\bibfnamefont {N.~M.}\ \bibnamefont {Linke}}, \bibinfo {author}
  {\bibfnamefont {B.}~\bibnamefont {Yoshida}}, \bibinfo {author} {\bibfnamefont
  {N.~Y.}\ \bibnamefont {Yao}}, \ and\ \bibinfo {author} {\bibfnamefont
  {C.}~\bibnamefont {Monroe}},\ }\href@noop {} {\bibfield  {journal} {\bibinfo
  {journal} {Nature}\ }\textbf {\bibinfo {volume} {567}},\ \bibinfo {pages}
  {61} (\bibinfo {year} {2019})}\BibitemShut {NoStop}%
\bibitem [{\citenamefont {Li}\ \emph {et~al.}(2017)\citenamefont {Li},
  \citenamefont {Fan}, \citenamefont {Wang}, \citenamefont {Ye}, \citenamefont
  {Zeng}, \citenamefont {Zhai}, \citenamefont {Peng},\ and\ \citenamefont
  {Du}}]{li2017measuring}%
  \BibitemOpen
  \bibfield  {author} {\bibinfo {author} {\bibfnamefont {J.}~\bibnamefont
  {Li}}, \bibinfo {author} {\bibfnamefont {R.}~\bibnamefont {Fan}}, \bibinfo
  {author} {\bibfnamefont {H.}~\bibnamefont {Wang}}, \bibinfo {author}
  {\bibfnamefont {B.}~\bibnamefont {Ye}}, \bibinfo {author} {\bibfnamefont
  {B.}~\bibnamefont {Zeng}}, \bibinfo {author} {\bibfnamefont {H.}~\bibnamefont
  {Zhai}}, \bibinfo {author} {\bibfnamefont {X.}~\bibnamefont {Peng}}, \ and\
  \bibinfo {author} {\bibfnamefont {J.}~\bibnamefont {Du}},\ }\href@noop {}
  {\bibfield  {journal} {\bibinfo  {journal} {Physical Review X}\ }\textbf
  {\bibinfo {volume} {7}},\ \bibinfo {pages} {031011} (\bibinfo {year}
  {2017})}\BibitemShut {NoStop}%
\bibitem [{\citenamefont {G{\"a}rttner}\ \emph {et~al.}(2017)\citenamefont
  {G{\"a}rttner}, \citenamefont {Bohnet}, \citenamefont {Safavi-Naini},
  \citenamefont {Wall}, \citenamefont {Bollinger},\ and\ \citenamefont
  {Rey}}]{garttner2017measuring}%
  \BibitemOpen
  \bibfield  {author} {\bibinfo {author} {\bibfnamefont {M.}~\bibnamefont
  {G{\"a}rttner}}, \bibinfo {author} {\bibfnamefont {J.~G.}\ \bibnamefont
  {Bohnet}}, \bibinfo {author} {\bibfnamefont {A.}~\bibnamefont
  {Safavi-Naini}}, \bibinfo {author} {\bibfnamefont {M.~L.}\ \bibnamefont
  {Wall}}, \bibinfo {author} {\bibfnamefont {J.~J.}\ \bibnamefont {Bollinger}},
  \ and\ \bibinfo {author} {\bibfnamefont {A.~M.}\ \bibnamefont {Rey}},\
  }\href@noop {} {\bibfield  {journal} {\bibinfo  {journal} {Nature Physics}\
  }\textbf {\bibinfo {volume} {13}},\ \bibinfo {pages} {781} (\bibinfo {year}
  {2017})}\BibitemShut {NoStop}%
\bibitem [{\citenamefont {Houck}\ \emph {et~al.}(2012)\citenamefont {Houck},
  \citenamefont {T{\"u}reci},\ and\ \citenamefont {Koch}}]{houck2012chip}%
  \BibitemOpen
  \bibfield  {author} {\bibinfo {author} {\bibfnamefont {A.~A.}\ \bibnamefont
  {Houck}}, \bibinfo {author} {\bibfnamefont {H.~E.}\ \bibnamefont
  {T{\"u}reci}}, \ and\ \bibinfo {author} {\bibfnamefont {J.}~\bibnamefont
  {Koch}},\ }\href@noop {} {\bibfield  {journal} {\bibinfo  {journal} {Nature
  Physics}\ }\textbf {\bibinfo {volume} {8}},\ \bibinfo {pages} {292} (\bibinfo
  {year} {2012})}\BibitemShut {NoStop}%
\bibitem [{\citenamefont {Harris}\ \emph {et~al.}(2018)\citenamefont {Harris},
  \citenamefont {Sato}, \citenamefont {Berkley}, \citenamefont {Reis},
  \citenamefont {Altomare}, \citenamefont {Amin}, \citenamefont {Boothby},
  \citenamefont {Bunyk}, \citenamefont {Deng}, \citenamefont {Enderud} \emph
  {et~al.}}]{harris2018phase}%
  \BibitemOpen
  \bibfield  {author} {\bibinfo {author} {\bibfnamefont {R.}~\bibnamefont
  {Harris}}, \bibinfo {author} {\bibfnamefont {Y.}~\bibnamefont {Sato}},
  \bibinfo {author} {\bibfnamefont {A.}~\bibnamefont {Berkley}}, \bibinfo
  {author} {\bibfnamefont {M.}~\bibnamefont {Reis}}, \bibinfo {author}
  {\bibfnamefont {F.}~\bibnamefont {Altomare}}, \bibinfo {author}
  {\bibfnamefont {M.}~\bibnamefont {Amin}}, \bibinfo {author} {\bibfnamefont
  {K.}~\bibnamefont {Boothby}}, \bibinfo {author} {\bibfnamefont
  {P.}~\bibnamefont {Bunyk}}, \bibinfo {author} {\bibfnamefont
  {C.}~\bibnamefont {Deng}}, \bibinfo {author} {\bibfnamefont {C.}~\bibnamefont
  {Enderud}},  \emph {et~al.},\ }\href@noop {} {\bibfield  {journal} {\bibinfo
  {journal} {Science}\ }\textbf {\bibinfo {volume} {361}},\ \bibinfo {pages}
  {162} (\bibinfo {year} {2018})}\BibitemShut {NoStop}%
\bibitem [{\citenamefont {Li}\ \emph {et~al.}(2014)\citenamefont {Li},
  \citenamefont {Zhou}, \citenamefont {Ju}, \citenamefont {Chen}, \citenamefont
  {Zheng}, \citenamefont {Lu}, \citenamefont {Rong}, \citenamefont {Duan},
  \citenamefont {Peng},\ and\ \citenamefont {Du}}]{li2014experimental}%
  \BibitemOpen
  \bibfield  {author} {\bibinfo {author} {\bibfnamefont {Z.}~\bibnamefont
  {Li}}, \bibinfo {author} {\bibfnamefont {H.}~\bibnamefont {Zhou}}, \bibinfo
  {author} {\bibfnamefont {C.}~\bibnamefont {Ju}}, \bibinfo {author}
  {\bibfnamefont {H.}~\bibnamefont {Chen}}, \bibinfo {author} {\bibfnamefont
  {W.}~\bibnamefont {Zheng}}, \bibinfo {author} {\bibfnamefont
  {D.}~\bibnamefont {Lu}}, \bibinfo {author} {\bibfnamefont {X.}~\bibnamefont
  {Rong}}, \bibinfo {author} {\bibfnamefont {C.}~\bibnamefont {Duan}}, \bibinfo
  {author} {\bibfnamefont {X.}~\bibnamefont {Peng}}, \ and\ \bibinfo {author}
  {\bibfnamefont {J.}~\bibnamefont {Du}},\ }\href@noop {} {\bibfield  {journal}
  {\bibinfo  {journal} {Physical review letters}\ }\textbf {\bibinfo {volume}
  {112}},\ \bibinfo {pages} {220501} (\bibinfo {year} {2014})}\BibitemShut
  {NoStop}%
\bibitem [{\citenamefont {Friedenauer}\ \emph {et~al.}(2008)\citenamefont
  {Friedenauer}, \citenamefont {Schmitz}, \citenamefont {Glueckert},
  \citenamefont {Porras},\ and\ \citenamefont
  {Sch{\"a}tz}}]{friedenauer2008simulating}%
  \BibitemOpen
  \bibfield  {author} {\bibinfo {author} {\bibfnamefont {A.}~\bibnamefont
  {Friedenauer}}, \bibinfo {author} {\bibfnamefont {H.}~\bibnamefont
  {Schmitz}}, \bibinfo {author} {\bibfnamefont {J.~T.}\ \bibnamefont
  {Glueckert}}, \bibinfo {author} {\bibfnamefont {D.}~\bibnamefont {Porras}}, \
  and\ \bibinfo {author} {\bibfnamefont {T.}~\bibnamefont {Sch{\"a}tz}},\
  }\href@noop {} {\bibfield  {journal} {\bibinfo  {journal} {Nature Physics}\
  }\textbf {\bibinfo {volume} {4}},\ \bibinfo {pages} {757} (\bibinfo {year}
  {2008})}\BibitemShut {NoStop}%
\bibitem [{\citenamefont {Zhang}\ \emph {et~al.}(2017)\citenamefont {Zhang},
  \citenamefont {Pagano}, \citenamefont {Hess}, \citenamefont {Kyprianidis},
  \citenamefont {Becker}, \citenamefont {Kaplan}, \citenamefont {Gorshkov},
  \citenamefont {Gong},\ and\ \citenamefont {Monroe}}]{zhang2017observation}%
  \BibitemOpen
  \bibfield  {author} {\bibinfo {author} {\bibfnamefont {J.}~\bibnamefont
  {Zhang}}, \bibinfo {author} {\bibfnamefont {G.}~\bibnamefont {Pagano}},
  \bibinfo {author} {\bibfnamefont {P.~W.}\ \bibnamefont {Hess}}, \bibinfo
  {author} {\bibfnamefont {A.}~\bibnamefont {Kyprianidis}}, \bibinfo {author}
  {\bibfnamefont {P.}~\bibnamefont {Becker}}, \bibinfo {author} {\bibfnamefont
  {H.}~\bibnamefont {Kaplan}}, \bibinfo {author} {\bibfnamefont {A.~V.}\
  \bibnamefont {Gorshkov}}, \bibinfo {author} {\bibfnamefont {Z.-X.}\
  \bibnamefont {Gong}}, \ and\ \bibinfo {author} {\bibfnamefont
  {C.}~\bibnamefont {Monroe}},\ }\href@noop {} {\bibfield  {journal} {\bibinfo
  {journal} {Nature}\ }\textbf {\bibinfo {volume} {551}},\ \bibinfo {pages}
  {601} (\bibinfo {year} {2017})}\BibitemShut {NoStop}%
\bibitem [{\citenamefont {Hart}\ \emph {et~al.}(2015)\citenamefont {Hart},
  \citenamefont {Duarte}, \citenamefont {Yang}, \citenamefont {Liu},
  \citenamefont {Paiva}, \citenamefont {Khatami}, \citenamefont {Scalettar},
  \citenamefont {Trivedi}, \citenamefont {Huse},\ and\ \citenamefont
  {Hulet}}]{hart2015observation}%
  \BibitemOpen
  \bibfield  {author} {\bibinfo {author} {\bibfnamefont {R.~A.}\ \bibnamefont
  {Hart}}, \bibinfo {author} {\bibfnamefont {P.~M.}\ \bibnamefont {Duarte}},
  \bibinfo {author} {\bibfnamefont {T.-L.}\ \bibnamefont {Yang}}, \bibinfo
  {author} {\bibfnamefont {X.}~\bibnamefont {Liu}}, \bibinfo {author}
  {\bibfnamefont {T.}~\bibnamefont {Paiva}}, \bibinfo {author} {\bibfnamefont
  {E.}~\bibnamefont {Khatami}}, \bibinfo {author} {\bibfnamefont {R.~T.}\
  \bibnamefont {Scalettar}}, \bibinfo {author} {\bibfnamefont {N.}~\bibnamefont
  {Trivedi}}, \bibinfo {author} {\bibfnamefont {D.~A.}\ \bibnamefont {Huse}}, \
  and\ \bibinfo {author} {\bibfnamefont {R.~G.}\ \bibnamefont {Hulet}},\
  }\href@noop {} {\bibfield  {journal} {\bibinfo  {journal} {Nature}\ }\textbf
  {\bibinfo {volume} {519}},\ \bibinfo {pages} {211} (\bibinfo {year}
  {2015})}\BibitemShut {NoStop}%
\bibitem [{\citenamefont {Gong}\ and\ \citenamefont
  {Duan}(2013)}]{gong2013prethermalization}%
  \BibitemOpen
  \bibfield  {author} {\bibinfo {author} {\bibfnamefont {Z.-X.}\ \bibnamefont
  {Gong}}\ and\ \bibinfo {author} {\bibfnamefont {L.-M.}\ \bibnamefont
  {Duan}},\ }\href@noop {} {\bibfield  {journal} {\bibinfo  {journal} {New
  Journal of Physics}\ }\textbf {\bibinfo {volume} {15}},\ \bibinfo {pages}
  {113051} (\bibinfo {year} {2013})}\BibitemShut {NoStop}%
\bibitem [{\citenamefont {Struck}\ \emph {et~al.}(2011)\citenamefont {Struck},
  \citenamefont {{\"O}lschl{\"a}ger}, \citenamefont {Le~Targat}, \citenamefont
  {Soltan-Panahi}, \citenamefont {Eckardt}, \citenamefont {Lewenstein},
  \citenamefont {Windpassinger},\ and\ \citenamefont
  {Sengstock}}]{struck2011quantum}%
  \BibitemOpen
  \bibfield  {author} {\bibinfo {author} {\bibfnamefont {J.}~\bibnamefont
  {Struck}}, \bibinfo {author} {\bibfnamefont {C.}~\bibnamefont
  {{\"O}lschl{\"a}ger}}, \bibinfo {author} {\bibfnamefont {R.}~\bibnamefont
  {Le~Targat}}, \bibinfo {author} {\bibfnamefont {P.}~\bibnamefont
  {Soltan-Panahi}}, \bibinfo {author} {\bibfnamefont {A.}~\bibnamefont
  {Eckardt}}, \bibinfo {author} {\bibfnamefont {M.}~\bibnamefont {Lewenstein}},
  \bibinfo {author} {\bibfnamefont {P.}~\bibnamefont {Windpassinger}}, \ and\
  \bibinfo {author} {\bibfnamefont {K.}~\bibnamefont {Sengstock}},\ }\href@noop
  {} {\bibfield  {journal} {\bibinfo  {journal} {Science}\ }\textbf {\bibinfo
  {volume} {333}},\ \bibinfo {pages} {996} (\bibinfo {year}
  {2011})}\BibitemShut {NoStop}%
\bibitem [{\citenamefont {Wu}\ and\ \citenamefont
  {Sarma}(2016)}]{wu2016understanding}%
  \BibitemOpen
  \bibfield  {author} {\bibinfo {author} {\bibfnamefont {Y.-L.}\ \bibnamefont
  {Wu}}\ and\ \bibinfo {author} {\bibfnamefont {S.~D.}\ \bibnamefont {Sarma}},\
  }\href@noop {} {\bibfield  {journal} {\bibinfo  {journal} {Physical Review
  A}\ }\textbf {\bibinfo {volume} {93}},\ \bibinfo {pages} {022332} (\bibinfo
  {year} {2016})}\BibitemShut {NoStop}%
\bibitem [{\citenamefont {Bernien}\ \emph {et~al.}(2017)\citenamefont
  {Bernien}, \citenamefont {Schwartz}, \citenamefont {Keesling}, \citenamefont
  {Levine}, \citenamefont {Omran}, \citenamefont {Pichler}, \citenamefont
  {Choi}, \citenamefont {Zibrov}, \citenamefont {Endres}, \citenamefont
  {Greiner} \emph {et~al.}}]{bernien2017probing}%
  \BibitemOpen
  \bibfield  {author} {\bibinfo {author} {\bibfnamefont {H.}~\bibnamefont
  {Bernien}}, \bibinfo {author} {\bibfnamefont {S.}~\bibnamefont {Schwartz}},
  \bibinfo {author} {\bibfnamefont {A.}~\bibnamefont {Keesling}}, \bibinfo
  {author} {\bibfnamefont {H.}~\bibnamefont {Levine}}, \bibinfo {author}
  {\bibfnamefont {A.}~\bibnamefont {Omran}}, \bibinfo {author} {\bibfnamefont
  {H.}~\bibnamefont {Pichler}}, \bibinfo {author} {\bibfnamefont
  {S.}~\bibnamefont {Choi}}, \bibinfo {author} {\bibfnamefont {A.~S.}\
  \bibnamefont {Zibrov}}, \bibinfo {author} {\bibfnamefont {M.}~\bibnamefont
  {Endres}}, \bibinfo {author} {\bibfnamefont {M.}~\bibnamefont {Greiner}},
  \emph {et~al.},\ }\href@noop {} {\bibfield  {journal} {\bibinfo  {journal}
  {Nature}\ }\textbf {\bibinfo {volume} {551}},\ \bibinfo {pages} {579}
  (\bibinfo {year} {2017})}\BibitemShut {NoStop}%
\bibitem [{\citenamefont {Jurcevic}\ \emph {et~al.}(2017)\citenamefont
  {Jurcevic}, \citenamefont {Shen}, \citenamefont {Hauke}, \citenamefont
  {Maier}, \citenamefont {Brydges}, \citenamefont {Hempel}, \citenamefont
  {Lanyon}, \citenamefont {Heyl}, \citenamefont {Blatt},\ and\ \citenamefont
  {Roos}}]{jurcevic2017direct}%
  \BibitemOpen
  \bibfield  {author} {\bibinfo {author} {\bibfnamefont {P.}~\bibnamefont
  {Jurcevic}}, \bibinfo {author} {\bibfnamefont {H.}~\bibnamefont {Shen}},
  \bibinfo {author} {\bibfnamefont {P.}~\bibnamefont {Hauke}}, \bibinfo
  {author} {\bibfnamefont {C.}~\bibnamefont {Maier}}, \bibinfo {author}
  {\bibfnamefont {T.}~\bibnamefont {Brydges}}, \bibinfo {author} {\bibfnamefont
  {C.}~\bibnamefont {Hempel}}, \bibinfo {author} {\bibfnamefont
  {B.}~\bibnamefont {Lanyon}}, \bibinfo {author} {\bibfnamefont
  {M.}~\bibnamefont {Heyl}}, \bibinfo {author} {\bibfnamefont {R.}~\bibnamefont
  {Blatt}}, \ and\ \bibinfo {author} {\bibfnamefont {C.}~\bibnamefont {Roos}},\
  }\href@noop {} {\bibfield  {journal} {\bibinfo  {journal} {Physical review
  letters}\ }\textbf {\bibinfo {volume} {119}},\ \bibinfo {pages} {080501}
  (\bibinfo {year} {2017})}\BibitemShut {NoStop}%
\bibitem [{\citenamefont {Altman}\ \emph {et~al.}(2019)\citenamefont {Altman},
  \citenamefont {Brown}, \citenamefont {Carleo}, \citenamefont {Carr},
  \citenamefont {Demler}, \citenamefont {Chin}, \citenamefont {DeMarco},
  \citenamefont {Economou}, \citenamefont {Eriksson}, \citenamefont {Fu} \emph
  {et~al.}}]{altman2019quantum}%
  \BibitemOpen
  \bibfield  {author} {\bibinfo {author} {\bibfnamefont {E.}~\bibnamefont
  {Altman}}, \bibinfo {author} {\bibfnamefont {K.~R.}\ \bibnamefont {Brown}},
  \bibinfo {author} {\bibfnamefont {G.}~\bibnamefont {Carleo}}, \bibinfo
  {author} {\bibfnamefont {L.~D.}\ \bibnamefont {Carr}}, \bibinfo {author}
  {\bibfnamefont {E.}~\bibnamefont {Demler}}, \bibinfo {author} {\bibfnamefont
  {C.}~\bibnamefont {Chin}}, \bibinfo {author} {\bibfnamefont {B.}~\bibnamefont
  {DeMarco}}, \bibinfo {author} {\bibfnamefont {S.~E.}\ \bibnamefont
  {Economou}}, \bibinfo {author} {\bibfnamefont {M.}~\bibnamefont {Eriksson}},
  \bibinfo {author} {\bibfnamefont {K.-M.~C.}\ \bibnamefont {Fu}},  \emph
  {et~al.},\ }\href@noop {} {\bibfield  {journal} {\bibinfo  {journal} {arXiv
  preprint arXiv:1912.06938}\ } (\bibinfo {year} {2019})}\BibitemShut {NoStop}%
\bibitem [{\citenamefont {Georgescu}\ \emph {et~al.}(2014)\citenamefont
  {Georgescu}, \citenamefont {Ashhab},\ and\ \citenamefont
  {Nori}}]{georgescu2014quantum}%
  \BibitemOpen
  \bibfield  {author} {\bibinfo {author} {\bibfnamefont {I.~M.}\ \bibnamefont
  {Georgescu}}, \bibinfo {author} {\bibfnamefont {S.}~\bibnamefont {Ashhab}}, \
  and\ \bibinfo {author} {\bibfnamefont {F.}~\bibnamefont {Nori}},\ }\href@noop
  {} {\bibfield  {journal} {\bibinfo  {journal} {Reviews of Modern Physics}\
  }\textbf {\bibinfo {volume} {86}},\ \bibinfo {pages} {153} (\bibinfo {year}
  {2014})}\BibitemShut {NoStop}%
\bibitem [{\citenamefont {Poggi}\ \emph {et~al.}(2020)\citenamefont {Poggi},
  \citenamefont {Lysne}, \citenamefont {Kuper}, \citenamefont {Deutsch},\ and\
  \citenamefont {Jessen}}]{poggi2020quantifying}%
  \BibitemOpen
  \bibfield  {author} {\bibinfo {author} {\bibfnamefont {P.~M.}\ \bibnamefont
  {Poggi}}, \bibinfo {author} {\bibfnamefont {N.~K.}\ \bibnamefont {Lysne}},
  \bibinfo {author} {\bibfnamefont {K.~W.}\ \bibnamefont {Kuper}}, \bibinfo
  {author} {\bibfnamefont {I.~H.}\ \bibnamefont {Deutsch}}, \ and\ \bibinfo
  {author} {\bibfnamefont {P.~S.}\ \bibnamefont {Jessen}},\ }\href@noop {}
  {\bibfield  {journal} {\bibinfo  {journal} {arXiv preprint arXiv:2007.01901}\
  } (\bibinfo {year} {2020})}\BibitemShut {NoStop}%
\bibitem [{\citenamefont {Hauke}\ \emph {et~al.}(2012)\citenamefont {Hauke},
  \citenamefont {Cucchietti}, \citenamefont {Tagliacozzo}, \citenamefont
  {Deutsch},\ and\ \citenamefont {Lewenstein}}]{hauke2012can}%
  \BibitemOpen
  \bibfield  {author} {\bibinfo {author} {\bibfnamefont {P.}~\bibnamefont
  {Hauke}}, \bibinfo {author} {\bibfnamefont {F.~M.}\ \bibnamefont
  {Cucchietti}}, \bibinfo {author} {\bibfnamefont {L.}~\bibnamefont
  {Tagliacozzo}}, \bibinfo {author} {\bibfnamefont {I.}~\bibnamefont
  {Deutsch}}, \ and\ \bibinfo {author} {\bibfnamefont {M.}~\bibnamefont
  {Lewenstein}},\ }\href@noop {} {\bibfield  {journal} {\bibinfo  {journal}
  {Reports on Progress in Physics}\ }\textbf {\bibinfo {volume} {75}},\
  \bibinfo {pages} {082401} (\bibinfo {year} {2012})}\BibitemShut {NoStop}%
\bibitem [{\citenamefont {Bylander}\ \emph {et~al.}(2011)\citenamefont
  {Bylander}, \citenamefont {Gustavsson}, \citenamefont {Yan}, \citenamefont
  {Yoshihara}, \citenamefont {Harrabi}, \citenamefont {Fitch}, \citenamefont
  {Cory}, \citenamefont {Nakamura}, \citenamefont {Tsai},\ and\ \citenamefont
  {Oliver}}]{bylander2011noise}%
  \BibitemOpen
  \bibfield  {author} {\bibinfo {author} {\bibfnamefont {J.}~\bibnamefont
  {Bylander}}, \bibinfo {author} {\bibfnamefont {S.}~\bibnamefont
  {Gustavsson}}, \bibinfo {author} {\bibfnamefont {F.}~\bibnamefont {Yan}},
  \bibinfo {author} {\bibfnamefont {F.}~\bibnamefont {Yoshihara}}, \bibinfo
  {author} {\bibfnamefont {K.}~\bibnamefont {Harrabi}}, \bibinfo {author}
  {\bibfnamefont {G.}~\bibnamefont {Fitch}}, \bibinfo {author} {\bibfnamefont
  {D.~G.}\ \bibnamefont {Cory}}, \bibinfo {author} {\bibfnamefont
  {Y.}~\bibnamefont {Nakamura}}, \bibinfo {author} {\bibfnamefont {J.-S.}\
  \bibnamefont {Tsai}}, \ and\ \bibinfo {author} {\bibfnamefont {W.~D.}\
  \bibnamefont {Oliver}},\ }\href@noop {} {\bibfield  {journal} {\bibinfo
  {journal} {Nature Physics}\ }\textbf {\bibinfo {volume} {7}},\ \bibinfo
  {pages} {565} (\bibinfo {year} {2011})}\BibitemShut {NoStop}%
\bibitem [{\citenamefont {Bertet}\ \emph {et~al.}(2005)\citenamefont {Bertet},
  \citenamefont {Chiorescu}, \citenamefont {Burkard}, \citenamefont {Semba},
  \citenamefont {Harmans}, \citenamefont {DiVincenzo},\ and\ \citenamefont
  {Mooij}}]{bertet2005dephasing}%
  \BibitemOpen
  \bibfield  {author} {\bibinfo {author} {\bibfnamefont {P.}~\bibnamefont
  {Bertet}}, \bibinfo {author} {\bibfnamefont {I.}~\bibnamefont {Chiorescu}},
  \bibinfo {author} {\bibfnamefont {G.}~\bibnamefont {Burkard}}, \bibinfo
  {author} {\bibfnamefont {K.}~\bibnamefont {Semba}}, \bibinfo {author}
  {\bibfnamefont {C.}~\bibnamefont {Harmans}}, \bibinfo {author} {\bibfnamefont
  {D.~P.}\ \bibnamefont {DiVincenzo}}, \ and\ \bibinfo {author} {\bibfnamefont
  {J.}~\bibnamefont {Mooij}},\ }\href@noop {} {\bibfield  {journal} {\bibinfo
  {journal} {Physical review letters}\ }\textbf {\bibinfo {volume} {95}},\
  \bibinfo {pages} {257002} (\bibinfo {year} {2005})}\BibitemShut {NoStop}%
\bibitem [{\citenamefont {Martinis}\ \emph {et~al.}(2003)\citenamefont
  {Martinis}, \citenamefont {Nam}, \citenamefont {Aumentado}, \citenamefont
  {Lang},\ and\ \citenamefont {Urbina}}]{martinis2003decoherence}%
  \BibitemOpen
  \bibfield  {author} {\bibinfo {author} {\bibfnamefont {J.~M.}\ \bibnamefont
  {Martinis}}, \bibinfo {author} {\bibfnamefont {S.}~\bibnamefont {Nam}},
  \bibinfo {author} {\bibfnamefont {J.}~\bibnamefont {Aumentado}}, \bibinfo
  {author} {\bibfnamefont {K.}~\bibnamefont {Lang}}, \ and\ \bibinfo {author}
  {\bibfnamefont {C.}~\bibnamefont {Urbina}},\ }\href@noop {} {\bibfield
  {journal} {\bibinfo  {journal} {Physical Review B}\ }\textbf {\bibinfo
  {volume} {67}},\ \bibinfo {pages} {094510} (\bibinfo {year}
  {2003})}\BibitemShut {NoStop}%
\bibitem [{\citenamefont {Walther}\ \emph {et~al.}(2012)\citenamefont
  {Walther}, \citenamefont {Ziesel}, \citenamefont {Ruster}, \citenamefont
  {Dawkins}, \citenamefont {Ott}, \citenamefont {Hettrich}, \citenamefont
  {Singer}, \citenamefont {Schmidt-Kaler},\ and\ \citenamefont
  {Poschinger}}]{walther2012controlling}%
  \BibitemOpen
  \bibfield  {author} {\bibinfo {author} {\bibfnamefont {A.}~\bibnamefont
  {Walther}}, \bibinfo {author} {\bibfnamefont {F.}~\bibnamefont {Ziesel}},
  \bibinfo {author} {\bibfnamefont {T.}~\bibnamefont {Ruster}}, \bibinfo
  {author} {\bibfnamefont {S.~T.}\ \bibnamefont {Dawkins}}, \bibinfo {author}
  {\bibfnamefont {K.}~\bibnamefont {Ott}}, \bibinfo {author} {\bibfnamefont
  {M.}~\bibnamefont {Hettrich}}, \bibinfo {author} {\bibfnamefont
  {K.}~\bibnamefont {Singer}}, \bibinfo {author} {\bibfnamefont
  {F.}~\bibnamefont {Schmidt-Kaler}}, \ and\ \bibinfo {author} {\bibfnamefont
  {U.}~\bibnamefont {Poschinger}},\ }\href@noop {} {\bibfield  {journal}
  {\bibinfo  {journal} {Physical review letters}\ }\textbf {\bibinfo {volume}
  {109}},\ \bibinfo {pages} {080501} (\bibinfo {year} {2012})}\BibitemShut
  {NoStop}%
\bibitem [{\citenamefont {Pokharel}\ \emph {et~al.}(2018)\citenamefont
  {Pokharel}, \citenamefont {Anand}, \citenamefont {Fortman},\ and\
  \citenamefont {Lidar}}]{pokharel2018demonstration}%
  \BibitemOpen
  \bibfield  {author} {\bibinfo {author} {\bibfnamefont {B.}~\bibnamefont
  {Pokharel}}, \bibinfo {author} {\bibfnamefont {N.}~\bibnamefont {Anand}},
  \bibinfo {author} {\bibfnamefont {B.}~\bibnamefont {Fortman}}, \ and\
  \bibinfo {author} {\bibfnamefont {D.~A.}\ \bibnamefont {Lidar}},\ }\href@noop
  {} {\bibfield  {journal} {\bibinfo  {journal} {Physical review letters}\
  }\textbf {\bibinfo {volume} {121}},\ \bibinfo {pages} {220502} (\bibinfo
  {year} {2018})}\BibitemShut {NoStop}%
\bibitem [{\citenamefont {Harty}\ \emph {et~al.}(2014)\citenamefont {Harty},
  \citenamefont {Allcock}, \citenamefont {Ballance}, \citenamefont {Guidoni},
  \citenamefont {Janacek}, \citenamefont {Linke}, \citenamefont {Stacey},\ and\
  \citenamefont {Lucas}}]{PhysRevLett.113.220501}%
  \BibitemOpen
  \bibfield  {author} {\bibinfo {author} {\bibfnamefont {T.~P.}\ \bibnamefont
  {Harty}}, \bibinfo {author} {\bibfnamefont {D.~T.~C.}\ \bibnamefont
  {Allcock}}, \bibinfo {author} {\bibfnamefont {C.~J.}\ \bibnamefont
  {Ballance}}, \bibinfo {author} {\bibfnamefont {L.}~\bibnamefont {Guidoni}},
  \bibinfo {author} {\bibfnamefont {H.~A.}\ \bibnamefont {Janacek}}, \bibinfo
  {author} {\bibfnamefont {N.~M.}\ \bibnamefont {Linke}}, \bibinfo {author}
  {\bibfnamefont {D.~N.}\ \bibnamefont {Stacey}}, \ and\ \bibinfo {author}
  {\bibfnamefont {D.~M.}\ \bibnamefont {Lucas}},\ }\href {\doibase
  10.1103/PhysRevLett.113.220501} {\bibfield  {journal} {\bibinfo  {journal}
  {Phys. Rev. Lett.}\ }\textbf {\bibinfo {volume} {113}},\ \bibinfo {pages}
  {220501} (\bibinfo {year} {2014})}\BibitemShut {NoStop}%
\bibitem [{\citenamefont {Ballance}\ \emph {et~al.}(2016)\citenamefont
  {Ballance}, \citenamefont {Harty}, \citenamefont {Linke}, \citenamefont
  {Sepiol},\ and\ \citenamefont {Lucas}}]{PhysRevLett.117.060504}%
  \BibitemOpen
  \bibfield  {author} {\bibinfo {author} {\bibfnamefont {C.~J.}\ \bibnamefont
  {Ballance}}, \bibinfo {author} {\bibfnamefont {T.~P.}\ \bibnamefont {Harty}},
  \bibinfo {author} {\bibfnamefont {N.~M.}\ \bibnamefont {Linke}}, \bibinfo
  {author} {\bibfnamefont {M.~A.}\ \bibnamefont {Sepiol}}, \ and\ \bibinfo
  {author} {\bibfnamefont {D.~M.}\ \bibnamefont {Lucas}},\ }\href {\doibase
  10.1103/PhysRevLett.117.060504} {\bibfield  {journal} {\bibinfo  {journal}
  {Phys. Rev. Lett.}\ }\textbf {\bibinfo {volume} {117}},\ \bibinfo {pages}
  {060504} (\bibinfo {year} {2016})}\BibitemShut {NoStop}%
\bibitem [{\citenamefont {Gaebler}\ \emph {et~al.}(2016)\citenamefont
  {Gaebler}, \citenamefont {Tan}, \citenamefont {Lin}, \citenamefont {Wan},
  \citenamefont {Bowler}, \citenamefont {Keith}, \citenamefont {Glancy},
  \citenamefont {Coakley}, \citenamefont {Knill}, \citenamefont {Leibfried},\
  and\ \citenamefont {Wineland}}]{PhysRevLett.117.060505}%
  \BibitemOpen
  \bibfield  {author} {\bibinfo {author} {\bibfnamefont {J.~P.}\ \bibnamefont
  {Gaebler}}, \bibinfo {author} {\bibfnamefont {T.~R.}\ \bibnamefont {Tan}},
  \bibinfo {author} {\bibfnamefont {Y.}~\bibnamefont {Lin}}, \bibinfo {author}
  {\bibfnamefont {Y.}~\bibnamefont {Wan}}, \bibinfo {author} {\bibfnamefont
  {R.}~\bibnamefont {Bowler}}, \bibinfo {author} {\bibfnamefont {A.~C.}\
  \bibnamefont {Keith}}, \bibinfo {author} {\bibfnamefont {S.}~\bibnamefont
  {Glancy}}, \bibinfo {author} {\bibfnamefont {K.}~\bibnamefont {Coakley}},
  \bibinfo {author} {\bibfnamefont {E.}~\bibnamefont {Knill}}, \bibinfo
  {author} {\bibfnamefont {D.}~\bibnamefont {Leibfried}}, \ and\ \bibinfo
  {author} {\bibfnamefont {D.~J.}\ \bibnamefont {Wineland}},\ }\href {\doibase
  10.1103/PhysRevLett.117.060505} {\bibfield  {journal} {\bibinfo  {journal}
  {Phys. Rev. Lett.}\ }\textbf {\bibinfo {volume} {117}},\ \bibinfo {pages}
  {060505} (\bibinfo {year} {2016})}\BibitemShut {NoStop}%
\bibitem [{\citenamefont {Mezzacapo}\ \emph {et~al.}(2014)\citenamefont
  {Mezzacapo}, \citenamefont {Las~Heras}, \citenamefont {Pedernales},
  \citenamefont {DiCarlo}, \citenamefont {Solano},\ and\ \citenamefont
  {Lamata}}]{mezzacapo2014digital}%
  \BibitemOpen
  \bibfield  {author} {\bibinfo {author} {\bibfnamefont {A.}~\bibnamefont
  {Mezzacapo}}, \bibinfo {author} {\bibfnamefont {U.}~\bibnamefont
  {Las~Heras}}, \bibinfo {author} {\bibfnamefont {J.}~\bibnamefont
  {Pedernales}}, \bibinfo {author} {\bibfnamefont {L.}~\bibnamefont {DiCarlo}},
  \bibinfo {author} {\bibfnamefont {E.}~\bibnamefont {Solano}}, \ and\ \bibinfo
  {author} {\bibfnamefont {L.}~\bibnamefont {Lamata}},\ }\href@noop {}
  {\bibfield  {journal} {\bibinfo  {journal} {Scientific reports}\ }\textbf
  {\bibinfo {volume} {4}},\ \bibinfo {pages} {1} (\bibinfo {year}
  {2014})}\BibitemShut {NoStop}%
\bibitem [{\citenamefont {Garc{\'\i}a-{\'A}lvarez}\ \emph
  {et~al.}(2015)\citenamefont {Garc{\'\i}a-{\'A}lvarez}, \citenamefont
  {Casanova}, \citenamefont {Mezzacapo}, \citenamefont {Egusquiza},
  \citenamefont {Lamata}, \citenamefont {Romero},\ and\ \citenamefont
  {Solano}}]{garcia2015fermion}%
  \BibitemOpen
  \bibfield  {author} {\bibinfo {author} {\bibfnamefont {L.}~\bibnamefont
  {Garc{\'\i}a-{\'A}lvarez}}, \bibinfo {author} {\bibfnamefont
  {J.}~\bibnamefont {Casanova}}, \bibinfo {author} {\bibfnamefont
  {A.}~\bibnamefont {Mezzacapo}}, \bibinfo {author} {\bibfnamefont
  {I.}~\bibnamefont {Egusquiza}}, \bibinfo {author} {\bibfnamefont
  {L.}~\bibnamefont {Lamata}}, \bibinfo {author} {\bibfnamefont
  {G.}~\bibnamefont {Romero}}, \ and\ \bibinfo {author} {\bibfnamefont
  {E.}~\bibnamefont {Solano}},\ }\href@noop {} {\bibfield  {journal} {\bibinfo
  {journal} {Physical review letters}\ }\textbf {\bibinfo {volume} {114}},\
  \bibinfo {pages} {070502} (\bibinfo {year} {2015})}\BibitemShut {NoStop}%
\bibitem [{\citenamefont {Asaad}\ \emph {et~al.}(2016)\citenamefont {Asaad},
  \citenamefont {Dickel}, \citenamefont {Langford}, \citenamefont {Poletto},
  \citenamefont {Bruno}, \citenamefont {Rol}, \citenamefont {Deurloo},\ and\
  \citenamefont {DiCarlo}}]{asaad2016independent}%
  \BibitemOpen
  \bibfield  {author} {\bibinfo {author} {\bibfnamefont {S.}~\bibnamefont
  {Asaad}}, \bibinfo {author} {\bibfnamefont {C.}~\bibnamefont {Dickel}},
  \bibinfo {author} {\bibfnamefont {N.~K.}\ \bibnamefont {Langford}}, \bibinfo
  {author} {\bibfnamefont {S.}~\bibnamefont {Poletto}}, \bibinfo {author}
  {\bibfnamefont {A.}~\bibnamefont {Bruno}}, \bibinfo {author} {\bibfnamefont
  {M.~A.}\ \bibnamefont {Rol}}, \bibinfo {author} {\bibfnamefont
  {D.}~\bibnamefont {Deurloo}}, \ and\ \bibinfo {author} {\bibfnamefont
  {L.}~\bibnamefont {DiCarlo}},\ }\href@noop {} {\bibfield  {journal} {\bibinfo
   {journal} {npj Quantum Information}\ }\textbf {\bibinfo {volume} {2}},\
  \bibinfo {pages} {1} (\bibinfo {year} {2016})}\BibitemShut {NoStop}%
\bibitem [{\citenamefont {Weber}\ \emph {et~al.}(2017)\citenamefont {Weber},
  \citenamefont {Samach}, \citenamefont {Hover}, \citenamefont {Gustavsson},
  \citenamefont {Kim}, \citenamefont {Melville}, \citenamefont {Rosenberg},
  \citenamefont {Sears}, \citenamefont {Yan}, \citenamefont {Yoder} \emph
  {et~al.}}]{weber2017coherent}%
  \BibitemOpen
  \bibfield  {author} {\bibinfo {author} {\bibfnamefont {S.~J.}\ \bibnamefont
  {Weber}}, \bibinfo {author} {\bibfnamefont {G.~O.}\ \bibnamefont {Samach}},
  \bibinfo {author} {\bibfnamefont {D.}~\bibnamefont {Hover}}, \bibinfo
  {author} {\bibfnamefont {S.}~\bibnamefont {Gustavsson}}, \bibinfo {author}
  {\bibfnamefont {D.~K.}\ \bibnamefont {Kim}}, \bibinfo {author} {\bibfnamefont
  {A.}~\bibnamefont {Melville}}, \bibinfo {author} {\bibfnamefont
  {D.}~\bibnamefont {Rosenberg}}, \bibinfo {author} {\bibfnamefont {A.~P.}\
  \bibnamefont {Sears}}, \bibinfo {author} {\bibfnamefont {F.}~\bibnamefont
  {Yan}}, \bibinfo {author} {\bibfnamefont {J.~L.}\ \bibnamefont {Yoder}},
  \emph {et~al.},\ }\href@noop {} {\bibfield  {journal} {\bibinfo  {journal}
  {Physical Review Applied}\ }\textbf {\bibinfo {volume} {8}},\ \bibinfo
  {pages} {014004} (\bibinfo {year} {2017})}\BibitemShut {NoStop}%
\bibitem [{\citenamefont {H{\"a}ffner}\ \emph {et~al.}(2005)\citenamefont
  {H{\"a}ffner}, \citenamefont {H{\"a}nsel}, \citenamefont {Roos},
  \citenamefont {Benhelm}, \citenamefont {Chwalla}, \citenamefont {K{\"o}rber},
  \citenamefont {Rapol}, \citenamefont {Riebe}, \citenamefont {Schmidt},
  \citenamefont {Becher} \emph {et~al.}}]{haffner2005scalable}%
  \BibitemOpen
  \bibfield  {author} {\bibinfo {author} {\bibfnamefont {H.}~\bibnamefont
  {H{\"a}ffner}}, \bibinfo {author} {\bibfnamefont {W.}~\bibnamefont
  {H{\"a}nsel}}, \bibinfo {author} {\bibfnamefont {C.}~\bibnamefont {Roos}},
  \bibinfo {author} {\bibfnamefont {J.}~\bibnamefont {Benhelm}}, \bibinfo
  {author} {\bibfnamefont {M.}~\bibnamefont {Chwalla}}, \bibinfo {author}
  {\bibfnamefont {T.}~\bibnamefont {K{\"o}rber}}, \bibinfo {author}
  {\bibfnamefont {U.}~\bibnamefont {Rapol}}, \bibinfo {author} {\bibfnamefont
  {M.}~\bibnamefont {Riebe}}, \bibinfo {author} {\bibfnamefont
  {P.}~\bibnamefont {Schmidt}}, \bibinfo {author} {\bibfnamefont
  {C.}~\bibnamefont {Becher}},  \emph {et~al.},\ }\href@noop {} {\bibfield
  {journal} {\bibinfo  {journal} {Nature}\ }\textbf {\bibinfo {volume} {438}},\
  \bibinfo {pages} {643} (\bibinfo {year} {2005})}\BibitemShut {NoStop}%
\bibitem [{\citenamefont {Blatt}\ and\ \citenamefont
  {Roos}(2012)}]{blatt2012quantum}%
  \BibitemOpen
  \bibfield  {author} {\bibinfo {author} {\bibfnamefont {R.}~\bibnamefont
  {Blatt}}\ and\ \bibinfo {author} {\bibfnamefont {C.~F.}\ \bibnamefont
  {Roos}},\ }\href@noop {} {\bibfield  {journal} {\bibinfo  {journal} {Nature
  Physics}\ }\textbf {\bibinfo {volume} {8}},\ \bibinfo {pages} {277} (\bibinfo
  {year} {2012})}\BibitemShut {NoStop}%
\bibitem [{\citenamefont {Saffman}(2016)}]{saffman2016quantum}%
  \BibitemOpen
  \bibfield  {author} {\bibinfo {author} {\bibfnamefont {M.}~\bibnamefont
  {Saffman}},\ }\href@noop {} {\bibfield  {journal} {\bibinfo  {journal}
  {Journal of Physics B: Atomic, Molecular and Optical Physics}\ }\textbf
  {\bibinfo {volume} {49}},\ \bibinfo {pages} {202001} (\bibinfo {year}
  {2016})}\BibitemShut {NoStop}%
\bibitem [{\citenamefont {Greenbaum}(2015)}]{greenbaum2015introduction}%
  \BibitemOpen
  \bibfield  {author} {\bibinfo {author} {\bibfnamefont {D.}~\bibnamefont
  {Greenbaum}},\ }\href@noop {} {\bibfield  {journal} {\bibinfo  {journal}
  {arXiv preprint arXiv:1509.02921}\ } (\bibinfo {year} {2015})}\BibitemShut
  {NoStop}%
\bibitem [{\citenamefont {Bairey}\ \emph {et~al.}(2020)\citenamefont {Bairey},
  \citenamefont {Guo}, \citenamefont {Poletti}, \citenamefont {Lindner},\ and\
  \citenamefont {Arad}}]{bairey2020learning}%
  \BibitemOpen
  \bibfield  {author} {\bibinfo {author} {\bibfnamefont {E.}~\bibnamefont
  {Bairey}}, \bibinfo {author} {\bibfnamefont {C.}~\bibnamefont {Guo}},
  \bibinfo {author} {\bibfnamefont {D.}~\bibnamefont {Poletti}}, \bibinfo
  {author} {\bibfnamefont {N.~H.}\ \bibnamefont {Lindner}}, \ and\ \bibinfo
  {author} {\bibfnamefont {I.}~\bibnamefont {Arad}},\ }\href@noop {} {\bibfield
   {journal} {\bibinfo  {journal} {New Journal of Physics}\ }\textbf {\bibinfo
  {volume} {22}},\ \bibinfo {pages} {032001} (\bibinfo {year}
  {2020})}\BibitemShut {NoStop}%
\bibitem [{\citenamefont {da~Silva}\ \emph {et~al.}(2011)\citenamefont
  {da~Silva}, \citenamefont {Landon-Cardinal},\ and\ \citenamefont
  {Poulin}}]{da2011practical}%
  \BibitemOpen
  \bibfield  {author} {\bibinfo {author} {\bibfnamefont {M.~P.}\ \bibnamefont
  {da~Silva}}, \bibinfo {author} {\bibfnamefont {O.}~\bibnamefont
  {Landon-Cardinal}}, \ and\ \bibinfo {author} {\bibfnamefont {D.}~\bibnamefont
  {Poulin}},\ }\href@noop {} {\bibfield  {journal} {\bibinfo  {journal}
  {Physical Review Letters}\ }\textbf {\bibinfo {volume} {107}},\ \bibinfo
  {pages} {210404} (\bibinfo {year} {2011})}\BibitemShut {NoStop}%
\bibitem [{\citenamefont {Parra-Rodriguez}\ \emph {et~al.}(2020)\citenamefont
  {Parra-Rodriguez}, \citenamefont {Lougovski}, \citenamefont {Lamata},
  \citenamefont {Solano},\ and\ \citenamefont {Sanz}}]{parra2020digital}%
  \BibitemOpen
  \bibfield  {author} {\bibinfo {author} {\bibfnamefont {A.}~\bibnamefont
  {Parra-Rodriguez}}, \bibinfo {author} {\bibfnamefont {P.}~\bibnamefont
  {Lougovski}}, \bibinfo {author} {\bibfnamefont {L.}~\bibnamefont {Lamata}},
  \bibinfo {author} {\bibfnamefont {E.}~\bibnamefont {Solano}}, \ and\ \bibinfo
  {author} {\bibfnamefont {M.}~\bibnamefont {Sanz}},\ }\href@noop {} {\bibfield
   {journal} {\bibinfo  {journal} {Physical Review A}\ }\textbf {\bibinfo
  {volume} {101}},\ \bibinfo {pages} {022305} (\bibinfo {year}
  {2020})}\BibitemShut {NoStop}%
\bibitem [{\citenamefont {Martin}\ \emph {et~al.}(2020)\citenamefont {Martin},
  \citenamefont {Lamata}, \citenamefont {Solano},\ and\ \citenamefont
  {Sanz}}]{martin2020digital}%
  \BibitemOpen
  \bibfield  {author} {\bibinfo {author} {\bibfnamefont {A.}~\bibnamefont
  {Martin}}, \bibinfo {author} {\bibfnamefont {L.}~\bibnamefont {Lamata}},
  \bibinfo {author} {\bibfnamefont {E.}~\bibnamefont {Solano}}, \ and\ \bibinfo
  {author} {\bibfnamefont {M.}~\bibnamefont {Sanz}},\ }\href@noop {} {\bibfield
   {journal} {\bibinfo  {journal} {Physical Review Research}\ }\textbf
  {\bibinfo {volume} {2}},\ \bibinfo {pages} {013012} (\bibinfo {year}
  {2020})}\BibitemShut {NoStop}%
\bibitem [{\citenamefont {Lee}\ \emph {et~al.}(1984)\citenamefont {Lee},
  \citenamefont {Kim},\ and\ \citenamefont {Dekeyser}}]{lee1984time}%
  \BibitemOpen
  \bibfield  {author} {\bibinfo {author} {\bibfnamefont {M.~H.}\ \bibnamefont
  {Lee}}, \bibinfo {author} {\bibfnamefont {I.}~\bibnamefont {Kim}}, \ and\
  \bibinfo {author} {\bibfnamefont {R.}~\bibnamefont {Dekeyser}},\ }\href@noop
  {} {\bibfield  {journal} {\bibinfo  {journal} {Physical review letters}\
  }\textbf {\bibinfo {volume} {52}},\ \bibinfo {pages} {1579} (\bibinfo {year}
  {1984})}\BibitemShut {NoStop}%
\bibitem [{\citenamefont {Siurakshina}\ \emph {et~al.}(2000)\citenamefont
  {Siurakshina}, \citenamefont {Ihle},\ and\ \citenamefont
  {Hayn}}]{siurakshina2000theory}%
  \BibitemOpen
  \bibfield  {author} {\bibinfo {author} {\bibfnamefont {L.}~\bibnamefont
  {Siurakshina}}, \bibinfo {author} {\bibfnamefont {D.}~\bibnamefont {Ihle}}, \
  and\ \bibinfo {author} {\bibfnamefont {R.}~\bibnamefont {Hayn}},\ }\href@noop
  {} {\bibfield  {journal} {\bibinfo  {journal} {Physical Review B}\ }\textbf
  {\bibinfo {volume} {61}},\ \bibinfo {pages} {14601} (\bibinfo {year}
  {2000})}\BibitemShut {NoStop}%
\bibitem [{\citenamefont {Hucht}\ \emph {et~al.}(1995)\citenamefont {Hucht},
  \citenamefont {Moschel},\ and\ \citenamefont {Usadel}}]{hucht1995monte}%
  \BibitemOpen
  \bibfield  {author} {\bibinfo {author} {\bibfnamefont {A.}~\bibnamefont
  {Hucht}}, \bibinfo {author} {\bibfnamefont {A.}~\bibnamefont {Moschel}}, \
  and\ \bibinfo {author} {\bibfnamefont {K.}~\bibnamefont {Usadel}},\
  }\href@noop {} {\bibfield  {journal} {\bibinfo  {journal} {Journal of
  magnetism and magnetic materials}\ }\textbf {\bibinfo {volume} {148}},\
  \bibinfo {pages} {32} (\bibinfo {year} {1995})}\BibitemShut {NoStop}%
\bibitem [{\citenamefont {Torelli}\ and\ \citenamefont
  {Olsen}(2018)}]{torelli2018calculating}%
  \BibitemOpen
  \bibfield  {author} {\bibinfo {author} {\bibfnamefont {D.}~\bibnamefont
  {Torelli}}\ and\ \bibinfo {author} {\bibfnamefont {T.}~\bibnamefont
  {Olsen}},\ }\href@noop {} {\bibfield  {journal} {\bibinfo  {journal} {2D
  Materials}\ }\textbf {\bibinfo {volume} {6}},\ \bibinfo {pages} {015028}
  (\bibinfo {year} {2018})}\BibitemShut {NoStop}%
\bibitem [{\citenamefont {Soukoulis}\ \emph {et~al.}(1991)\citenamefont
  {Soukoulis}, \citenamefont {Datta},\ and\ \citenamefont
  {Lee}}]{soukoulis1991spin}%
  \BibitemOpen
  \bibfield  {author} {\bibinfo {author} {\bibfnamefont {C.}~\bibnamefont
  {Soukoulis}}, \bibinfo {author} {\bibfnamefont {S.}~\bibnamefont {Datta}}, \
  and\ \bibinfo {author} {\bibfnamefont {Y.~H.}\ \bibnamefont {Lee}},\
  }\href@noop {} {\bibfield  {journal} {\bibinfo  {journal} {Physical Review
  B}\ }\textbf {\bibinfo {volume} {44}},\ \bibinfo {pages} {446} (\bibinfo
  {year} {1991})}\BibitemShut {NoStop}%
\bibitem [{\citenamefont {Lamata}\ \emph {et~al.}(2018)\citenamefont {Lamata},
  \citenamefont {Parra-Rodriguez}, \citenamefont {Sanz},\ and\ \citenamefont
  {Solano}}]{lamata2018digital}%
  \BibitemOpen
  \bibfield  {author} {\bibinfo {author} {\bibfnamefont {L.}~\bibnamefont
  {Lamata}}, \bibinfo {author} {\bibfnamefont {A.}~\bibnamefont
  {Parra-Rodriguez}}, \bibinfo {author} {\bibfnamefont {M.}~\bibnamefont
  {Sanz}}, \ and\ \bibinfo {author} {\bibfnamefont {E.}~\bibnamefont
  {Solano}},\ }\href@noop {} {\bibfield  {journal} {\bibinfo  {journal}
  {Advances in Physics: X}\ }\textbf {\bibinfo {volume} {3}},\ \bibinfo {pages}
  {1457981} (\bibinfo {year} {2018})}\BibitemShut {NoStop}%
\bibitem [{\citenamefont {Oliver}\ and\ \citenamefont
  {Welander}(2013)}]{oliver2013materials}%
  \BibitemOpen
  \bibfield  {author} {\bibinfo {author} {\bibfnamefont {W.~D.}\ \bibnamefont
  {Oliver}}\ and\ \bibinfo {author} {\bibfnamefont {P.~B.}\ \bibnamefont
  {Welander}},\ }\href@noop {} {\bibfield  {journal} {\bibinfo  {journal} {MRS
  bulletin}\ }\textbf {\bibinfo {volume} {38}},\ \bibinfo {pages} {816}
  (\bibinfo {year} {2013})}\BibitemShut {NoStop}%
\bibitem [{\citenamefont {Gu}\ \emph {et~al.}(2017)\citenamefont {Gu},
  \citenamefont {Kockum}, \citenamefont {Miranowicz}, \citenamefont {Liu},\
  and\ \citenamefont {Nori}}]{gu2017microwave}%
  \BibitemOpen
  \bibfield  {author} {\bibinfo {author} {\bibfnamefont {X.}~\bibnamefont
  {Gu}}, \bibinfo {author} {\bibfnamefont {A.~F.}\ \bibnamefont {Kockum}},
  \bibinfo {author} {\bibfnamefont {A.}~\bibnamefont {Miranowicz}}, \bibinfo
  {author} {\bibfnamefont {Y.-x.}\ \bibnamefont {Liu}}, \ and\ \bibinfo
  {author} {\bibfnamefont {F.}~\bibnamefont {Nori}},\ }\href@noop {} {\bibfield
   {journal} {\bibinfo  {journal} {Physics Reports}\ }\textbf {\bibinfo
  {volume} {718}},\ \bibinfo {pages} {1} (\bibinfo {year} {2017})}\BibitemShut
  {NoStop}%
\bibitem [{\citenamefont {Zhou}\ \emph {et~al.}(2008)\citenamefont {Zhou},
  \citenamefont {Chu},\ and\ \citenamefont {Han}}]{zhou2008relaxation}%
  \BibitemOpen
  \bibfield  {author} {\bibinfo {author} {\bibfnamefont {Z.}~\bibnamefont
  {Zhou}}, \bibinfo {author} {\bibfnamefont {S.-I.}\ \bibnamefont {Chu}}, \
  and\ \bibinfo {author} {\bibfnamefont {S.}~\bibnamefont {Han}},\ }\href@noop
  {} {\bibfield  {journal} {\bibinfo  {journal} {Journal of Physics B: Atomic,
  Molecular and Optical Physics}\ }\textbf {\bibinfo {volume} {41}},\ \bibinfo
  {pages} {045506} (\bibinfo {year} {2008})}\BibitemShut {NoStop}%
\bibitem [{\citenamefont {Linke}\ \emph {et~al.}(2017)\citenamefont {Linke},
  \citenamefont {Maslov}, \citenamefont {Roetteler}, \citenamefont {Debnath},
  \citenamefont {Figgatt}, \citenamefont {Landsman}, \citenamefont {Wright},\
  and\ \citenamefont {Monroe}}]{linke2017experimental}%
  \BibitemOpen
  \bibfield  {author} {\bibinfo {author} {\bibfnamefont {N.~M.}\ \bibnamefont
  {Linke}}, \bibinfo {author} {\bibfnamefont {D.}~\bibnamefont {Maslov}},
  \bibinfo {author} {\bibfnamefont {M.}~\bibnamefont {Roetteler}}, \bibinfo
  {author} {\bibfnamefont {S.}~\bibnamefont {Debnath}}, \bibinfo {author}
  {\bibfnamefont {C.}~\bibnamefont {Figgatt}}, \bibinfo {author} {\bibfnamefont
  {K.~A.}\ \bibnamefont {Landsman}}, \bibinfo {author} {\bibfnamefont
  {K.}~\bibnamefont {Wright}}, \ and\ \bibinfo {author} {\bibfnamefont
  {C.}~\bibnamefont {Monroe}},\ }\href@noop {} {\bibfield  {journal} {\bibinfo
  {journal} {Proceedings of the National Academy of Sciences}\ }\textbf
  {\bibinfo {volume} {114}},\ \bibinfo {pages} {3305} (\bibinfo {year}
  {2017})}\BibitemShut {NoStop}%
\bibitem [{\citenamefont {M{\"u}ller}\ \emph {et~al.}(2015)\citenamefont
  {M{\"u}ller}, \citenamefont {Lisenfeld}, \citenamefont {Shnirman},\ and\
  \citenamefont {Poletto}}]{muller2015interacting}%
  \BibitemOpen
  \bibfield  {author} {\bibinfo {author} {\bibfnamefont {C.}~\bibnamefont
  {M{\"u}ller}}, \bibinfo {author} {\bibfnamefont {J.}~\bibnamefont
  {Lisenfeld}}, \bibinfo {author} {\bibfnamefont {A.}~\bibnamefont {Shnirman}},
  \ and\ \bibinfo {author} {\bibfnamefont {S.}~\bibnamefont {Poletto}},\
  }\href@noop {} {\bibfield  {journal} {\bibinfo  {journal} {Physical Review
  B}\ }\textbf {\bibinfo {volume} {92}},\ \bibinfo {pages} {035442} (\bibinfo
  {year} {2015})}\BibitemShut {NoStop}%
\bibitem [{\citenamefont {Mei{\ss}ner}\ \emph {et~al.}(2018)\citenamefont
  {Mei{\ss}ner}, \citenamefont {Seiler}, \citenamefont {Lisenfeld},
  \citenamefont {Ustinov},\ and\ \citenamefont {Weiss}}]{meissner2018probing}%
  \BibitemOpen
  \bibfield  {author} {\bibinfo {author} {\bibfnamefont {S.~M.}\ \bibnamefont
  {Mei{\ss}ner}}, \bibinfo {author} {\bibfnamefont {A.}~\bibnamefont {Seiler}},
  \bibinfo {author} {\bibfnamefont {J.}~\bibnamefont {Lisenfeld}}, \bibinfo
  {author} {\bibfnamefont {A.~V.}\ \bibnamefont {Ustinov}}, \ and\ \bibinfo
  {author} {\bibfnamefont {G.}~\bibnamefont {Weiss}},\ }\href@noop {}
  {\bibfield  {journal} {\bibinfo  {journal} {Physical Review B}\ }\textbf
  {\bibinfo {volume} {97}},\ \bibinfo {pages} {180505} (\bibinfo {year}
  {2018})}\BibitemShut {NoStop}%
\bibitem [{\citenamefont {Neill}\ \emph {et~al.}(2013)\citenamefont {Neill},
  \citenamefont {Megrant}, \citenamefont {Barends}, \citenamefont {Chen},
  \citenamefont {Chiaro}, \citenamefont {Kelly}, \citenamefont {Mutus},
  \citenamefont {O'Malley}, \citenamefont {Sank}, \citenamefont {Wenner} \emph
  {et~al.}}]{neill2013fluctuations}%
  \BibitemOpen
  \bibfield  {author} {\bibinfo {author} {\bibfnamefont {C.}~\bibnamefont
  {Neill}}, \bibinfo {author} {\bibfnamefont {A.}~\bibnamefont {Megrant}},
  \bibinfo {author} {\bibfnamefont {R.}~\bibnamefont {Barends}}, \bibinfo
  {author} {\bibfnamefont {Y.}~\bibnamefont {Chen}}, \bibinfo {author}
  {\bibfnamefont {B.}~\bibnamefont {Chiaro}}, \bibinfo {author} {\bibfnamefont
  {J.}~\bibnamefont {Kelly}}, \bibinfo {author} {\bibfnamefont
  {J.}~\bibnamefont {Mutus}}, \bibinfo {author} {\bibfnamefont
  {P.}~\bibnamefont {O'Malley}}, \bibinfo {author} {\bibfnamefont
  {D.}~\bibnamefont {Sank}}, \bibinfo {author} {\bibfnamefont {J.}~\bibnamefont
  {Wenner}},  \emph {et~al.},\ }\href@noop {} {\bibfield  {journal} {\bibinfo
  {journal} {Applied Physics Letters}\ }\textbf {\bibinfo {volume} {103}},\
  \bibinfo {pages} {072601} (\bibinfo {year} {2013})}\BibitemShut {NoStop}%
\bibitem [{\citenamefont {Bishop}\ \emph {et~al.}(1998)\citenamefont {Bishop},
  \citenamefont {Farnell},\ and\ \citenamefont {Parkinson}}]{bishop1998phase}%
  \BibitemOpen
  \bibfield  {author} {\bibinfo {author} {\bibfnamefont {R.~F.}\ \bibnamefont
  {Bishop}}, \bibinfo {author} {\bibfnamefont {D.~J.}\ \bibnamefont {Farnell}},
  \ and\ \bibinfo {author} {\bibfnamefont {J.~B.}\ \bibnamefont {Parkinson}},\
  }\href@noop {} {\bibfield  {journal} {\bibinfo  {journal} {Physical Review
  B}\ }\textbf {\bibinfo {volume} {58}},\ \bibinfo {pages} {6394} (\bibinfo
  {year} {1998})}\BibitemShut {NoStop}%
\bibitem [{\citenamefont {Matsuzaki}\ \emph {et~al.}(2011)\citenamefont
  {Matsuzaki}, \citenamefont {Benjamin},\ and\ \citenamefont
  {Fitzsimons}}]{matsuzaki2011magnetic}%
  \BibitemOpen
  \bibfield  {author} {\bibinfo {author} {\bibfnamefont {Y.}~\bibnamefont
  {Matsuzaki}}, \bibinfo {author} {\bibfnamefont {S.~C.}\ \bibnamefont
  {Benjamin}}, \ and\ \bibinfo {author} {\bibfnamefont {J.}~\bibnamefont
  {Fitzsimons}},\ }\href@noop {} {\bibfield  {journal} {\bibinfo  {journal}
  {Physical Review A}\ }\textbf {\bibinfo {volume} {84}},\ \bibinfo {pages}
  {012103} (\bibinfo {year} {2011})}\BibitemShut {NoStop}%
\bibitem [{\citenamefont {Matsuzaki}\ \emph {et~al.}(2010)\citenamefont
  {Matsuzaki}, \citenamefont {Saito}, \citenamefont {Kakuyanagi},\ and\
  \citenamefont {Semba}}]{matsuzaki2010quantum}%
  \BibitemOpen
  \bibfield  {author} {\bibinfo {author} {\bibfnamefont {Y.}~\bibnamefont
  {Matsuzaki}}, \bibinfo {author} {\bibfnamefont {S.}~\bibnamefont {Saito}},
  \bibinfo {author} {\bibfnamefont {K.}~\bibnamefont {Kakuyanagi}}, \ and\
  \bibinfo {author} {\bibfnamefont {K.}~\bibnamefont {Semba}},\ }\href@noop {}
  {\bibfield  {journal} {\bibinfo  {journal} {Physical Review B}\ }\textbf
  {\bibinfo {volume} {82}},\ \bibinfo {pages} {180518} (\bibinfo {year}
  {2010})}\BibitemShut {NoStop}%
\bibitem [{\citenamefont {Kakuyanagi}\ \emph {et~al.}(2007)\citenamefont
  {Kakuyanagi}, \citenamefont {Meno}, \citenamefont {Saito}, \citenamefont
  {Nakano}, \citenamefont {Semba}, \citenamefont {Takayanagi}, \citenamefont
  {Deppe},\ and\ \citenamefont {Shnirman}}]{kakuyanagi2007dephasing}%
  \BibitemOpen
  \bibfield  {author} {\bibinfo {author} {\bibfnamefont {K.}~\bibnamefont
  {Kakuyanagi}}, \bibinfo {author} {\bibfnamefont {T.}~\bibnamefont {Meno}},
  \bibinfo {author} {\bibfnamefont {S.}~\bibnamefont {Saito}}, \bibinfo
  {author} {\bibfnamefont {H.}~\bibnamefont {Nakano}}, \bibinfo {author}
  {\bibfnamefont {K.}~\bibnamefont {Semba}}, \bibinfo {author} {\bibfnamefont
  {H.}~\bibnamefont {Takayanagi}}, \bibinfo {author} {\bibfnamefont
  {F.}~\bibnamefont {Deppe}}, \ and\ \bibinfo {author} {\bibfnamefont
  {A.}~\bibnamefont {Shnirman}},\ }\href@noop {} {\bibfield  {journal}
  {\bibinfo  {journal} {Physical review letters}\ }\textbf {\bibinfo {volume}
  {98}},\ \bibinfo {pages} {047004} (\bibinfo {year} {2007})}\BibitemShut
  {NoStop}%
\bibitem [{\citenamefont {Yoshihara}\ \emph {et~al.}(2006)\citenamefont
  {Yoshihara}, \citenamefont {Harrabi}, \citenamefont {Niskanen}, \citenamefont
  {Nakamura},\ and\ \citenamefont {Tsai}}]{yoshihara2006decoherence}%
  \BibitemOpen
  \bibfield  {author} {\bibinfo {author} {\bibfnamefont {F.}~\bibnamefont
  {Yoshihara}}, \bibinfo {author} {\bibfnamefont {K.}~\bibnamefont {Harrabi}},
  \bibinfo {author} {\bibfnamefont {A.}~\bibnamefont {Niskanen}}, \bibinfo
  {author} {\bibfnamefont {Y.}~\bibnamefont {Nakamura}}, \ and\ \bibinfo
  {author} {\bibfnamefont {J.~S.}\ \bibnamefont {Tsai}},\ }\href@noop {}
  {\bibfield  {journal} {\bibinfo  {journal} {Physical review letters}\
  }\textbf {\bibinfo {volume} {97}},\ \bibinfo {pages} {167001} (\bibinfo
  {year} {2006})}\BibitemShut {NoStop}%
\end{thebibliography}%

\newpage

\widetext
\appendix

\section{Quantum error mitigation for gated-based quantum computing}
\subsection{Error Model}

Here we review the concept of quantum error mitigation (QEM) for digital quantum computing. In a digital gate-based quantum computer, the effect of noise is simplified as a quantum channel appearing either before or after each gate.  The output state is different from the ideal output, which can be described as
\begin{equation}
\begin{aligned}
\rho_{\mathrm{out}}^{\mathrm{noisy}}&=\mathcal{N}_{N_g}\circ \mathcal{U}_{N_g} \circ  \dots  \mathcal{N}_{1}\circ \mathcal{U}_{1} (\rho_{\mathrm{in}}) \\
\rho_{\mathrm{out}}^{\mathrm{ideal}}&=\mathcal{U}_{N_g} \circ  \dots  \circ \mathcal{U}_{1} (\rho_{\mathrm{in}}),
\end{aligned}
\end{equation}
where $\rho_{\mathrm{out}}^{\mathrm{noise}}$ is a noisy output and $\rho_{\mathrm{out}}^{\mathrm{ideal}}$ is a noise-free output from the quantum circuit, $\mathcal{U}_{k}$ and $\mathcal{N}_{k}$ are $k^{\rm{th}}$ quantum operation and accompanying noise to it, and $N_g$ is the number of gates.  Here, we assume the noise processes are Markovian for simplicity. Fault-tolerant error correction based on encoding of qubits can be used to compensate for the effect of noise and obtain correct computation results in principle. However, in near-term quantum computing, the number of qubits and gate operations are restricted due to imperfections of quantum devices including physical noise and limited interactions among qubits.  Therefore, fault-tolerant error correction necessitating encoding of qubits is not ideal for near-term quantum computing. Instead, QEM was introduced for mitigating errors in quantum circuits without using additional qubits. By using QEM, one cannot restore the quantum state itself, but can instead obtain an approximation of expected values of observables corresponding to the ideal density matrix, i.e.,
\begin{equation}
    \tr\left[\textrm{QEM}\left(\rho_{\mathrm{out}}^{\mathrm{noisy}}\right)O\right] \approx \tr\left[\rho_{\mathrm{out}}^{\mathrm{ideal}}O\right],
\end{equation}
for any observable $O$. Here we use  $\textrm{QEM}(\rho)$ to denote the process of error mitigation, which may not satisfy the requirements of a quantum channel.
Therefore, we generally need a classical post-processing to realise $\textrm{QEM}(\rho)$, which may introduce
a sampling overhead (cost) when measuring observables. The cost in general increases exponentially with respect to the error strength as we shortly see below. Therefore, a constant error strength is generally required in order to make QEM to work.


\subsection{Quasi-probability method}
Among  different  QEM schemes via different post-processing mechanisms, the quasi-probability error mitigation method is one of the most effective techniques. It recovers the ideal unitary processes by randomly generating noisy operations, with post processing of measurement results. Suppose the ideal quantum operation is denoted as $\mathcal{U}$, then the key idea of the quasi-probability method is to express the ideal evolution $\mathcal{U}$ as a linear combination of noisy operations $\mathcal{K}_i$ as
\begin{equation}
\begin{aligned}
\mathcal{U}&\approx\sum_i q_i \mathcal{K}_i
= C \sum_i p_i \mathrm{sgn}(q_i) \mathcal{K}_i,
\label{quasi1}
\end{aligned}
\end{equation}
where $\mathcal{U}$ and $\mathcal{K}_i$ are superoperators, and $\sum_i q_i =1$, $C=\sum_i |q_i|$, $p_i={|q_i|}/{C} $. As $q_i$ can be negative, we refer to $q_i$ as the quasi-probability, and therefore the overhead coefficient $C\geq 1$ in general. To obtain the error free expectation value of an observable ${O}$, we randomly generate noisy operation $\mathcal{K}_i$ with probability $p_i$, multiply the measured result by the parity factor $\mathrm{sgn} (q_i)$, and obtain the expectation value $\braket{O}_{\mathrm{eff}}$ as follows,
\begin{equation}
    \braket{O}_{\mathrm{eff}} = \sum_i p_i  \mathrm{sgn}(q_i) \tr[ O \mathcal{K}_i(\rho_{\textrm{in}})],
\end{equation}
Finally, the error free expectation value of $\braket{O}$ is approximated by $C\braket{O}_{\mathrm{eff}}$.
Note that the variance is amplified $C^2$ times greater, and thus $C^2$ can be interpreted as a resource cost to achieve the same accuracy as that without QEM.

As an example, we illustrate the case that the single qubit operation is affected by depolarising errors as $\mathcal{D} \mathcal{U}$. The removal of the error $\mathcal{D}$ can be formally done by applying its inverse channel $\mathcal{D}^{-1}$. Now, the depolarising channel can be expressed as
\begin{align}
\mathcal{D}(\rho)=\left(1-\frac{3}{4}p\right)\rho+\frac{p}{4}(X\rho X+Y \rho Y + Z \rho Z),
\end{align}
with the inverse channel derived as
\begin{align}
\mathcal{D}^{-1}(\rho)=C_{\mathcal{D}^{-1}}[p_1 \rho -p_2 (X\rho X+Y \rho Y + Z \rho Z)],
\end{align}
where $C_{\mathcal{D}^{-1}}=(p+2)/(2-2p)>1$, $p_1=(4-p)/(2p+4)$, and $p_2=p/(2p+4)$.

Consequently, the ideal channel  $\mathcal{U}$  can be expressed as
\begin{equation}
\begin{aligned}
\mathcal{U}&= \mathcal{D}^{-1} \mathcal{D} \mathcal{U} \\
&=C_{\mathcal{D}^{-1}} [p \mathcal{I}\mathcal{D}  \mathcal{U}-p_2 (\mathcal{X}\mathcal{D}  \mathcal{U}+\mathcal{Y}\mathcal{D}  \mathcal{U}+\mathcal{Z}\mathcal{D}  \mathcal{U})],
\label{quasi2}
\end{aligned}
\end{equation}
where $\mathcal{I}$, $\mathcal{X}$, $\mathcal{Y}$ and $\mathcal{Z}$ correspond to an identity operation, and superoperators for Pauli operators. Note that Eq.~(\ref{quasi2}) is written in the same form as Eq.~(\ref{quasi1}), and we can hence perform the quasi-probability method accordingly.

For the error mitigation method to be useful in digital quantum computing, this quasi-probability operation is applied after each noisy gate. The parity is updated depending on the generated operations, and the final outcome of the parity is applied to measurement results in the same way as a single quantum operation shown in Eq.~(\ref{quasi1}). Suppose there are $N$ gates, the total overhead $C_{N}$ can be expressed as
\begin{align}
C_{N}= \Pi_{i=1}^{N} C_i,
\end{align}
where $C_i$ is the overhead coefficient for $i^{\rm{th}}$ gate, and $N$ is the number of the gates in the quantum circuit. Suppose the error $\varepsilon_i$ for each gate is small, the cost $C_i$ is close to $1$. A first order expansion gives $C_i\approx 1+\lambda_i\varepsilon_i$ and thus the total overhead $C_{N}$ is approximated as
\begin{equation}
\begin{aligned}
C_N \approx \prod_i (1+\lambda_i\varepsilon_i).
\end{aligned}
\end{equation}
For simplicity, we assume $\lambda_i=\lambda$ and $\varepsilon_i=\varepsilon$ are independent of $i$. Then we have
\begin{equation}
\begin{aligned}
C_N \approx  (1+\lambda\varepsilon)^N = (1+\lambda\varepsilon)^{\frac{1}{\lambda\varepsilon} \lambda\varepsilon N} \approx e^{\lambda\varepsilon N} = e^{\lambda \varepsilon_N}.
\end{aligned}
\end{equation}
Here we denote $\varepsilon_N=\varepsilon N$ to be the total error rate of all the $N$ gates. Then it is not hard to see that the total cost $C_N$ increase exponentially to the total error rate $\varepsilon_N$. However, with a constant total error rate $\varepsilon_N$, we still have a constant overhead. Thus a constant total error rate is generally the assumption for error mitigation for digital quantum computing.

\section{Stochastic error mitigation}

As discussed in the above section, the QEM method assumes the noise appears either before or after each gate in a digital gate-based quantum computer, but realistic noise occurring in the experimental apparatus is more complicated. Specifically, every gate in digital circuits or every process in analog simulation is physically realised via a continuous real time evolution of a Hamiltonian and thus errors can either inherently mix with the evolution making it strongly gate or process dependent, or act on multiple number of qubits leading to highly nonlocal correlated effects (crosstalks).
Since conventional quantum error mitigation methods are restricted to gate-based digital quantum computers
and over-simplified noise models, they fail to work for realistic errors and general continuous quantum processes.
 In this section, we extend the QEM method to a more practical scenario and show how to mitigate errors for these inherent dynamics-based and nonlocal noise in practical noisy quantum devices.

\subsection{Pauli transfer matrix representation}
\label{appendix: paulitransfer}
We first introduce the Pauli transfer matrix representation of states, observables, and channels as a preliminary.
By using Pauli transfer representation, a state and an observable are mapped to a real column and row vectors respectively, as follows \begin{equation}
\begin{aligned}
\ket{\rho}\rangle &= [\dots \rho_k \dots] \\
\rho_k&=\mathrm{Tr}(P_k \rho),
\end{aligned}
\end{equation}
and
\begin{equation}
\begin{aligned}
\langle \bra{Q} &= [\dots Q_k \dots] \\
Q_k&=\frac{1}{d}\mathrm{Tr}(Q P_k ),
\end{aligned}
\end{equation}
where $P_k \in \{I, \sigma_x, \sigma_y, \sigma_z \}^{\otimes n}$, $n$ is the number of qubits, and $d=2^n$. Furthermore, for a process, i.e., $\mathcal{E}(\rho)=\sum_k K_k \rho K_k^\dag$, the Pauli transfer matrix representation can be expressed as
\begin{align}
E_{k,j}=\frac{1}{d}\mathrm{Tr}\big(P_k \mathcal{E}(P_j)  \big).
\end{align}
By using the Pauli transfer representation, we have
\begin{align}
\mathrm{Tr}\big(Q \mathcal{E}(\rho) \big) = \langle \bra{Q} E \ket{\rho}\rangle.
\end{align}

\subsection{Continuous error mitigation scheme}\label{section:continuouserror}

We first illustrate the detailed procedure of continuous error mitigation. We can rewrite the evolution of noisy and ideal quantum states by using infinitesimal $\delta t$ as
\begin{equation}
\begin{aligned}
\rho_N(t+\delta t)&= \rho_N(t)+\delta t \big \{ -i [H(t), \rho_N (t)] + \lambda\mathcal{L}\big[\rho_N(t) \big] \big \}
\\
\rho_I(t+\delta t)&= \rho_I(t)+\delta t \big \{ -i [H(t), \rho_I (t)]  \big \},
\label{infinitestimal1}
\end{aligned}
\end{equation}
where $H(t)$ denotes the ideal Hamiltonian with $\mathcal{L}$ corresponding to the noisy evolution. {In the presence of Markovian stochastic noise involved with environment,
\begin{align}
\mathcal{L}[\rho ]= \sum_k (2L_k \rho L_k^\dag -L_k^\dag L_k \rho - \rho L_k^\dag L_k ),
\end{align}
while the dynamics induced with the undesired Hamiltonian $H_C(t)$ which causes coherent errors can be described as
\begin{align}
\mathcal{L}[\rho]=-i [H_C(t), \rho].
\end{align}
The latter case occurs due to the imperfection of the analog quantum simulators and implementation of quantum logic gates from physical Hamiltonians~\cite{krantz2019quantum,hauke2012can}.
For systems with finite-range interactions, Bairey \textit{et al } and Silva \textit{et al } proposed methods that uses only local measurements to reconstruct local Markovian dynamical process~\cite{bairey2020learning,da2011practical}.
We will show how to eliminate these errors by using continuous error mitigation method.}

By using the Pauli transfer matrix representation, Eq. (\ref{infinitestimal1}) is mapped to $
\ket{\rho_{\alpha}(t+\delta t)}\rangle = (I + E_{\alpha}(t) \delta t)	\ket{\rho_{\alpha}(t)}\rangle $
where $\ket{\rho_{\alpha}(t)}\rangle$ ($\alpha=N,I$) is the vectorised density matrix of $\rho_\alpha(t)$ and $E_{\alpha}(t)$ corresponds to the second term of Eqs. (\ref{infinitestimal1}).
Equivalently, the superoperartor representation  of the evolution gives  ${\rho_{\alpha}(t+\delta t)} = \mc E_{\alpha}(t)	{\rho_{\alpha}(t)}$. In the following, we will use these two equivalent representations interchangeably.
Note that the evolution induced by $\mathcal{E}_{\alpha}$ in the main text becomes $I + E_\alpha \delta t$ in the Pauli transfer representation.  We introduce the recovery operation $I+E_Q \delta t $ to obtain the ideal dynamics, which can be expressed as
\begin{align}
(I+E_Q \delta t )(I+E_N \delta t)= I + E_I \delta t	+ \mc O(\delta t^2)
\label{continuousquasi1}
\end{align}
such that $E_Q=E_I-E_N$. Note that $(I+E_Q \delta t )$ corresponds to $\mathcal{E}_Q$ in the main text. Due to the linearity of the representation, we can see that $E_Q$ corresponds to the Pauli transfer matrix representation of $-\lambda\mathcal{L}\big[\rho_N(t) \big] $. In this framework,  $\mathcal{E}_I(t) \approx \mathcal{E}_N(t)$ holds within a sufficiently small time step $\delta t$.


The experimental errors including the interactions with the open environment, undesired couplings and imperfections in the quantum simulators are generally local and we therefore assume $\mathcal{E}_Q$ can be decomposed into local operators as $
\mathcal{E}_Q=\sum_{S =1}^{N_S}\mathcal{E}_Q^{(S)}$, where $\mathcal{E}_Q^{(S)}$ operates on polynomial subsystems of the $N$-qubit quantum system. We now decompose the operation $\mathcal{E}_Q$ into the set of basis operations as
\begin{equation}
\begin{aligned}
\mathcal{E}_Q^{(S)} =\sum_{j\geq 0} q_j^{(S)} \mathcal{B}_j ^{(S)},
\end{aligned}
\label{subsystem}
\end{equation}
where $q_j^{(S)}$ is the quasi-probability and $\mathcal{B}_j ^{(S)}$ is the basis operation for compensating the errors.
Note that $\mathcal{B}_j ^{(S)}$ only acts on the same small subsystem as $\mathcal{E}_Q^{(S)}$.
By performing basis operations for $\mathcal{E}_Q^{(S)}$ with corresponding quasi-probability distributions in Eq. (\ref{subsystem}), we can implement the overall quasi-probability operations corresponding to $\mathcal{E}_Q$ as shown below.
Therefore, we can extend the quasi-probability operations into a large-scale system.
We remark that this quasi-probability approach works for any errors and we can mitigate  correlated stochastic noise and unwanted interactions between (a small number of) multiple qubits.  {In addition, this argument can be naturally applied to multi-level systems when we can prepare basis operations for them.}

{In particular, the quasi-probability operation  at time $t$ takes the form of }
\begin{equation}\label{eqn:quasidecomp}
	\begin{aligned}
\mc E_Q  &= (1+q_0 \delta t) \mc I +\sum_{i \geq 1} q_{i}\delta t\mc B_i,\\
&=c\left(p_0 \mc I + \sum_{i \geq 1} \alpha_i p_{i}\mc B_i\right)
\end{aligned}
\end{equation}
where $\mc B_0$ is set to be an identity operation and we also omit the superscript $(S)$ for simplicity.
The probability to generate the identity operation $\mathcal{I}$ and $\mathcal{B}_i ~(i \geq 1)$ is $p_0 = 1-\sum_{i\ge 1} p_i$ and $p_i=|q_i|\delta t  /c ~(i \geq 1)$,
where $c=\sum_{i\ge 0} p_i=1+(q_0+\sum_{i\ge 1} |q_i|)\delta t$. In addition, the parity $\alpha_0$ for $\mathcal{B}_0=\mathcal{I}$ is always unity, and the parity $\alpha_i$ corresponding to $\mathcal{B}_i ~(i \geq 1)$ equals to $\mathrm{sign}(q_i)$.

The overhead coefficient $c$ corresponding to $\mc E_Q^{(S)}$ is given by $c=1+C_1^{(S)}\delta t$, with $C_1^{(S)}:=(q_0+\sum_{i \geq 1}|q_i|)$. As we have discussed above, this coefficient introduces a sampling overhead.
The overhead coefficient from $t=0$ to $t=T$ within infinitely small discretisation $\delta t$ is   \begin{equation}
\begin{aligned}
C(T)=\lim_{\delta t \rightarrow 0} \prod_{S} \prod_{k=0}^{T/\delta t}  (1+C_1^{(S)}\delta t)  =  \prod_{S} \mathrm{exp} \left( C_1^{(S)} T \right).
\label{eqn:overhead}
\end{aligned}
\end{equation}
Note that $|q_i|\propto \lambda$, therefore we have $C_1\propto\lambda$, and the overall overhead is
\begin{equation}
	C(T) = \exp(O(\lambda T)).
\end{equation}
{Here we choose a proper normalisation $\lambda$ so that the contribution of $\mc L$ is bounded by a constant $l$: $\| \mc L_{\exp} \|_1 \leq l$. Here, we define the super-operator norm by $\|\Phi \|_{1}=\mathrm{sup}_{{A}} \{\|\Phi({A}) \|_1 / \| {A}\|_1 : {A} \neq 0 \}$ with $\|A\|_1=\tr|A|$. Therefore, given a finite number of samples in the experiments the condition that the scheme works efficiently with a constant resource cost is $\lambda T  = O(1)$. By interpreting $\lambda T$ as the total noise strength, the requirement is thus consistent with the case of DQS. }

It is also possible to consider time-dependent recovery operation for suppressing time-dependent noise. In this case, the quasi-probability becomes time-dependent and can be obtained by Eq. (\ref{subsystem}). Therefore, the overall overhead for time-dependent noise is

\begin{equation}
\begin{aligned}
C(T)=\lim_{\delta t \rightarrow 0} \prod_{S} \prod_{k=0}^{T/\delta t}  (1+C_1^{(S)}(k \delta t)\delta t)  =  \prod_{S} \mathrm{exp} \left(\int_0^T C_1^{(S)}(t)dt \right).
\end{aligned}
\end{equation}

\subsection{Comparison with conventional error mitigation}
\label{appendix: comparison}
 {
Errors, occurring in the continuous time evolution, can inherently mix and propagate with the evolution leading to highly nonlocal correlated effects. For instance,  dominant errors in superconducting  qubits  are inherent system dephasing or relaxation,  and coherent errors (or crosstalk) when applying entangling gates.  Analog quantum simulators may not even implement discretised quantum gates. Therefore, conventional quantum error mitigation methods fail to work for realistic errors and general continuous quantum processes.
Our work addresses the problems by first considering a more general scenario of a continuous process with realistic noise models. {More concretely, we consider the time-independent Lindblad master equation}
\begin{equation}
\begin{aligned}
\frac{d \rho}{dt}&= (\mathcal{H} + \mathcal{L})(\rho)
\end{aligned}
\end{equation}
with dynamics of Hamiltonian (including coherent errors) and incoherent Markovian process
\begin{equation}
\begin{aligned}
    \mathcal{H}(\rho)&=-i [H+\delta H, \rho] \\
    \mathcal{L}(\rho)&= \sum_k (2L_k \rho L_k^\dag -L_k^\dag L_k \rho -\rho L_k^\dag L_k ),
\end{aligned}
\end{equation}
which describes either gate synthesis in digital quantum computing or the continuous evolution of a analog quantum simulator. Here $\delta H$ and $\mathcal{L}$ describe coherent errors (such as crosstalk or imperfections of Hamiltonian) and inherent coupling with the environment (such as dephasing and damping), respectively.
We note that even though the coherent error $\delta H$ and the Lindblad operators $L_k$ act locally on the quantum system, the effect of errors propagates to the entire system after the evolution. Therefore, such global effects of noise cannot be effectively mitigated using the conventional quasi-probability method, which assumes simple gate-independent error model described by single- or two-qubit error channels before or after each gate.}

 {
Our work proposes two key techniques to overcome this problem.
\begin{enumerate}
    \item First, we discretise the continuous time into small time steps so that we sequentially apply error mitigation for the noisy evolution at each time step. We emphasise that discretised evolution is yet not equivalent to discretised digital computing with local single and two qubit gates. This is because the continuous evolution even with a small time step could be a joint evolution (effectively a joint quantum gate) on all the qubits. Therefore, we directly mitigate errors of all the evolved qubits in each small time evolution, whereas conventional error mitigation methods operate on each local gate. This also explains why we can mitigate crosstalk of multiple qubits, whereas conventional methods only mitigate the effective noise channel for each gate. In practice, one can choose a sufficiently small time step so that the error mitigation works in a 'continuous' way.
    \item However, continuous error mitigation with small discretised time requires to constantly pause the original evolution to apply recovery operations and the time discretisation also introduces additional errors. To resolve these problems, we further introduce stochastic error mitigation, which equivalently simulates the continuous error mitigation procedure with infinitely small time step. The stochastic error mitigation method thus simultaneously solves all the issues and provides our final solution for error mitigation of a continuous process. We note that stochastic error mitigation only requires to apply a small number of single-qubit recovery operations at certain times. We can thus pre-engineer the recovery operations into the original evolution Hamiltonian without interrupting the simulation.
    \end{enumerate}
}
\sun{
To summarise, the first contribution of our work is to solve a major open problem of mitigating realistic (inherently gate- or process-mixed and nonlocal) noise for both digital and analog quantum simulators, which has strong applications in achieving quantum advantage with near-term noisy quantum devices. The techniques we introduce for the stochastic error mitigation method are highly non-trivial and do represent significant advances in our understanding of mitigating multi-qubit errors for processes beyond discretised gates and over-simplified noise models.
}

\subsection{Complete basis operation set}
\label{appendix: basisoperation}
In Ref~\cite{endo2018practical}, it is shown that every single qubit operation can be emulated by using $16$ basis operations. This is because every single qubit operation (including projective measurements) can be expressed with square matrices with $4 \times 4 = 16$ elements by using the Pauli transfer representation~\cite{greenbaum2015introduction}. Therefore, $16$ linearly independent operations are sufficient to emulate arbitrary single qubit operations. In Table~\ref{tab:bases}, we show one efficient set of basis operations for a single qubit in Ref~\cite{endo2018practical}.

\begin{table}[h!t!]
\begin{center}
\begin{tabular}{|c|l||c|l||c|l||c|l|}
\hline
1 & ~~$[I]$ (no operation) &
2 & ~~$[\sigma^{\rm x}] $ &
3 & ~~$[\sigma^{\rm y}] $ &
4 & ~~$[\sigma^{\rm z}] $ \\
\hline
5 & ~~$[R_{\rm x}] = [\frac{1}{\sqrt{2}}(I +i \sigma^{\rm x})]$ &
6 & ~~$[R_{\rm y}] = [\frac{1}{\sqrt{2}}(I +i \sigma^{\rm y})] $ &
7 & ~~$[R_{\rm z}] = [\frac{1}{\sqrt{2}}(I+i \sigma^{\rm z})] $ &
8 & ~~$[R_{\rm yz}] = [\frac{1}{\sqrt{2}}(\sigma^{\rm y} + \sigma^{\rm z})]$ \\
\hline
9 & ~~$[R_{\rm zx}] = [\frac{1}{\sqrt{2}}(\sigma^{\rm z} + \sigma^{\rm x})] $ &
10 & ~~$[R_{\rm xy}] = [\frac{1}{\sqrt{2}}(\sigma^{\rm x} + \sigma^{\rm y})]$ &
11 & ~~$[\pi_{\rm x}] = [\frac{1}{2}(I + \sigma^{\rm x})] $ &
12 & ~~$[\pi_{\rm y}] = [\frac{1}{2}(I + \sigma^{\rm y})]$ \\
\hline
13 & ~~$[\pi_{\rm z}] = [\frac{1}{2}(I + \sigma^{\rm z})] $ &
14 & ~~$[\pi_{\rm yz}] = [\frac{1}{2}(\sigma^{\rm y} +i \sigma^{\rm z})]$ &
15 & ~~$[\pi_{\rm zx}] = [\frac{1}{2}(\sigma^{\rm z} +i \sigma^{\rm x})]$ &
16 & ~~$[\pi_{\rm xy}] = [\frac{1}{2}(\sigma^{\rm x} +i \sigma^{\rm y})] $ \\
\hline
\end{tabular}
\end{center}
\caption{
Sixteen basis operations. These operations are composed of single qubit rotations and measurements. $[I]$ denotes an identity operation (no operation), $[\sigma^{i}]~( i=x,y,x)$ corresponds to operations applying Pauli matrices. $[\pi]$ corresponds to projective measurements.
}
\label{tab:bases}
\end{table}

For multiple qubit systems, tensor products of single qubit operations, e.g., $\mathcal{B}_i \otimes \mathcal{B}_j$ also forms a complete basis set for composite systems. Therefore, if we can implement the complete basis operations for a single qubit, we can also emulate arbitrary operations for multiple qubits systems. Moreover, we can also apply the error mitigation to multi-level systems if we can prepare the corresponding basis operations.


By using only observables within spatial domain, we can recover the Lindbladian acting on this domain and reconstruct the local Markovian dynamics~\cite{bairey2020learning}.
 Here, we provide the recovery operations for several typical Markovian processes during the quantum simulation and coherent errors in implementing CNOT gates.

The recovery operations can be analytically expressed as  $\mathcal{E}_Q=\mathcal{I}-\lambda\mathcal{L}\delta t$, where $\mathcal{L}$ represents the noise superoperator and $\lambda$ is the noise strength.
For depolarising, dephasing and amplitude damping,  the recovery operations $\mathcal{E}_Q$ can be respectively decomposed as
  \begin{equation}
  \begin{aligned}
&  \mathcal{E}_Q^{\rm depolarise}=(1+\frac{3}{4}\lambda\delta t) \mathcal{I}-\frac{\lambda}{4}(\mathcal{X} + \mathcal{Y} +\mathcal{Z}) \delta t\\
&\mathcal{E}_Q^{\rm dephase}= (1+\lambda\delta t) \mathcal{I}-\lambda\mathcal{Z} \delta t \\
&  \mathcal{E}_Q^{\rm amp}=(1+\frac{1}{4}\lambda\delta t) \mathcal{I}+\lambda(-\frac{1}{2}\mathcal{X} -\frac{1}{2}\mathcal{Y} -\frac{1}{4}\mathcal{Z} + [R_{xy} ]+ [\pi _{xy}]) \delta t.
  \end{aligned}
\end{equation}
In the parameterised quantum circuits, the CNOT gates or more general entangling gates are prepared by cross-resonance drive, with the drive Hamiltonian
\begin{equation}
H = \Omega(\sigma_z^{(c)}\sigma_x^{(t)}+ \gamma \mathbb{I}^{(c)}\sigma_x^{(t)} +  H_{\Delta})
\end{equation}
  where $ \Omega$ is the effective  qubit-qubit  coupling, $\gamma$ represents  the effect of  crosstalk between qubits and $H_{\Delta}$ corresponds to additional errors whose strengths can be revealed by Hamiltonian tomography. On the IBM's quantum devices, for example, $H_{\Delta}$ includes $\mu \sigma_z^{(c)}\mathbb I^{(t)} $ with $\mu  $  corresponding to the drive-induced Stark-shift.
 In the cross-resonance drive, one dominant error is from crosstalk, and the corresponding recovery operation  is
\begin{equation}
\mathcal{E}_Q=(1+\lambda \delta t )I +\lambda \delta t \mathcal{X} -2\lambda \delta t [R_x]
\end{equation}
with $\lambda = \gamma \Omega$.
The additional error, for example, the drive-induced Stark-shift can be mitigated by the recovery operation
$\mathcal{E}_Q=(1+\lambda \delta t )I +\lambda \delta t \mathcal{Z} -2\lambda \delta t [R_z] $.

\section{HYBRID ERROR MITIGATION}
\label{appendix:HybridQEM}

\sun{
In this section, we show how to apply the extrapolation method to mitigate model estimation error and the errors associated with imperfect recovery operations. Combined with stochastic error mitigation, we thus propose a hybrid error mitigation method for errors in practical NISQ devices. }
\subsection{Boosting model estimation error}
\label{appendix: extra}

 We first show how to boost model estimation error, which will be used for its mitigation.
Assume that the evolution of the quantum system is described by the open-system master equation
\begin{align}
\frac{d}{dt} \rho_{\lambda} = -i[H(t), \rho_{\lambda}]+\lambda \mathcal{L}_{\mathrm{exp}}\left[\rho_{\lambda}\right].
\end{align}
The evolution of the system under a scaled Hamiltonian drive $\frac{1}{r} H\big(\frac{t}{r}\big)$ takes the form of
\begin{align}
\frac{d}{dt} \rho_{\lambda}' = -i\bigg[\frac{1}{r}H\bigg(\frac{t}{r}\bigg), \rho_{\lambda}'\bigg]+\lambda \mathcal{L}_{\mathrm{exp}}\left[\rho_{\lambda}'\right].
\end{align}
Assuming the noise superoperator $\mathcal{L}$ is invariant under rescaling, we have
\begin{equation}
\begin{aligned}
\frac{d}{d t} \rho_{\lambda}^{\prime}(r t) &=\left.\frac{d t^{\prime}}{d t} \frac{\partial}{\partial t^{\prime}} \rho_{\lambda}^{\prime}\left(t^{\prime}\right)\right|_{t^{\prime}=r t} \\
&=\left.r \left\{-i\left[\frac{1}{r} H\left(\frac{t^{\prime}}{r}\right), \rho_{\lambda}^{\prime}\left(t^{\prime}\right)\right]+\lambda \mathcal{L}\left[\rho_{\lambda}^{\prime}\left(t^{\prime}\right)\right]\right\}\right|_{t^{\prime}=r t} \\
&=-i\left[H(t), \rho_{\lambda}^{\prime}(r t)\right]+r \lambda \mathcal{L}\left[\rho_{\lambda}^{\prime}(r t)\right]. \\
\end{aligned}
\label{eqn:rho_cl1}
\end{equation}

On the other hand, the density matrix $\rho_{r\lambda}( t)$ with enhanced noise strength $r\lambda$ is given by
\begin{equation}
\frac{d}{d t} \rho_{r \lambda}(t)=-i\left[H(t), \rho_{r \lambda}(t)\right]+r \lambda \mathcal{L}\left[\rho_{r \lambda}(t)\right].
\label{eqn:rho_cl2}
\end{equation}
Comparing Eqs. (\ref{eqn:rho_cl1}) and (\ref{eqn:rho_cl2}), one finds that $\rho_{\lambda}^{\prime}(r t)$ and $\rho_{r \lambda}(t)$ follow the same differential equation, and thus with the initial conditions $\rho_{\lambda}^{\prime}(0)=\rho_{r \lambda}(0)$ we prove $\rho_{\lambda}^{\prime}(r t)=\rho_{r \lambda}(t)$. This indicates by evolving the re-scaled Hamiltonian for time $rt$, we can effectively boost physical errors of quantum systems.

Now, we discuss how to boost the model estimation error. By applying stochastic error mitigation, we obtain
\begin{align}
\frac{d }{dt} \rho_{\lambda}^{(Q)}(t)  =-i[H(t), \rho_{\lambda}^{(Q)}(t)] + \lambda \Delta \mc L \big[ \rho_{\lambda}^{(Q)}(t) \big],
\label{appendix:noisysupple}
\end{align}
where $\rho_{\lambda}^{(Q)}(t)$ is the error mitigated effective density matrix after stochastic error mitigation. Assuming $\Delta \mc L= \mathcal{L}_{\rm exp}-\mathcal{L}_{\rm est}$ is invariant under re-scaling of the Hamiltonian, we can similarly obtain
\begin{align}
\frac{d }{dt} \rho_{r \lambda}^{(Q)}(t)  =-i[H(t), \rho_{r \lambda}^{(Q)}(t)] + r \lambda \Delta \mc L \big[ \rho_{r \lambda}^{(Q)}(t) \big].
\label{appendix:rescalesupple}
\end{align}
This can be experimentally achieved by applying stochastic error mitigation for a re-scaled time $rt$ under the re-scaled Hamiltonian.

\sun{
It is worth noting that even if the noise model is time dependent, our method can still work as long as the evolution can be described by a Lindblad equation and its dependence on time is known.
}
For example, when we consider a time dependent noisy process with stochastic error mitigation described by
\begin{equation}
\frac{d \rho_\lambda^{(Q)} (t)}{dt} = -i [H(t), \rho_\lambda ^{(Q)} (t)]+\lambda t \Delta \mc L_0 [\rho_\lambda ^{(Q)} (t)],
\end{equation}
where $ \Delta \mc L_0$ is time independent. Then, the re-scaled dynamical equation becomes
\begin{equation}
\frac{d \rho^{\prime (Q)}_{\lambda} (rt)}{dt} = -i [H(t), \rho^{\prime (Q)}_{ \lambda}(rt)]+r^2 \lambda t \Delta \mc L_0[\rho^{\prime (Q)}_{ \lambda}(rt)].
\label{eqn:appendixc2}
\end{equation}
In this case, we can interpret that the noise rate is boosted by a factor of $r^2$. We will later show how such a time dependent noise process can be mitigated in Sec.~\ref{appendix: timedepsimulation}.

\subsection{Richardson's extrapolation for physical errors and model estimation errors}\label{appendix: model}
In this section, we briefly review the extrapolation method proposed in Ref.~\cite{temme2017error,li2017efficient}. We assume the open system evolution is described by
\begin{align}
\frac{d \rho_N(t)}{dt}=-i[H_{\rm sim}(t), \rho_N(t)]+\lambda \mathcal{L}_{\mathrm{exp}}\big[\rho _N(t)\big].
\label{eqn:noisyeqnsupple}
\end{align}
In Ref.~\cite{temme2017error}, it is shown that the expectation value of an observable $O$ can be expressed as
\begin{align}
\braket{O(\lambda)}=\braket{O(0)}+\sum_{k=1}^n \alpha_k \lambda^k + B_{n+1}(\lambda, \mathcal{L},T),
\end{align}
where $\alpha_k \approx O(N^k T^k)$ and $B_{n+1}(\lambda, \mathcal{L},T)$ is upper bounded by
\begin{align}
B_{n+1}(\lambda, \mathcal{L},T) &\leq \|O \| a_{n+1} \frac{\lambda^{n+1} T^{n+1}}{(n+1)!},
\end{align}
where $\|O \|=\max_{\psi}\braket{\psi|O|\psi}$ is the spectra norm of $O$.
Here, in the case that $\mathcal{L}$ is a Lindblad type operator, one can have the bound for $a_{n+1}$ as
\begin{align}
a_{n+1} \leq \|\mathcal{L}_{\mathrm{exp}} \|_{1}^{n+1}.
\end{align}
Now, we have
\begin{align}
B_{n+1}(\lambda, \mathcal{L},T) &\leq  \| O \|
\frac{ (\lambda T \|\mathcal{L}_{\mathrm{exp}}\|_{1})^{n+1} }{(n+1)!}.
\end{align}

In order to employ the extrapolation method, we need to obtain the expectation value of observable $\braket{O (r_j \lambda)}$ ($j=0,1,..,n$, $r_0=1$) at time $t=T$ corresponding to the equation
\begin{align}
\frac{d}{d t} \rho_{\lambda}(t)=-i\left[H(t), \rho_{\lambda}(t)\right]+ r_j \lambda \mathcal{L}\left(\rho_{\lambda}(t)\right),
\label{eqn:scale}
\end{align}
which can be obtained by using the re-scaling of the Hamiltonian as described in section~\ref{appendix: extra}.
Then we can obtain the approximation of the noise free expectation value of the observable $O$ as
\begin{align}
\braket{O(0)}_n^*=\sum_{j=0}^n \beta_j \braket{O}_{r_j \lambda}',
\end{align}
where $\braket{O(0)}^*$ is the estimated noise free expectation value up to an error of order $O(\lambda^{n+1})$, and $\braket{O}_{r \lambda}'$ are the measurement outcome corresponding to the state $\rho_{r \lambda}(T)$.
 Here the coefficients $\beta_j=\prod_{l \neq j}r_l (r_l-r_j)^{-1}$ are defined by the solution of the following equations
\begin{align}
\sum_{j=0}^n \beta_j=1, \,
\sum_{j=0}^n \beta_j r_j^k= 0, k=1,...,n.
\end{align}

In Ref.~\cite{temme2017error}, it has been shown that the difference between the estimator and the error free expectation value is bounded by
\begin{equation}\label{bound1}
\begin{aligned}
 |\braket{O(0)}_n^*-\braket{O}_I |
 \leq \gamma_n \bigg(\frac{r_{\max}^{n+1} \Delta_{\mathrm{max}}}{\sqrt{N_{\mathrm{sample}}}}+\| O \|
\frac{ (r_{\max}\lambda T \|\mathcal{L}_{\mathrm{exp}} \|_{1})^{n+1} }{(n+1)!} \bigg),
\end{aligned}
\end{equation}
where $\gamma_n=\sum_{j=0}^n |\beta_j|$, $r_{\max}=\max_j r_j$, and $\Delta_{\mathrm{max}}/\sqrt{N_{\mathrm{sample}}}$ is the largest experimental errors due to shot noises with $N_{\mathrm{sample}}$ being the number of samples. From Eq. (\ref{bound1}), we can see that extrapolation methods requires
\begin{align}
r_{\max}\lambda T \|\mathcal{L}_{\mathrm{exp}}\|_{1} = \mc O(1).
\label{condition1}
\end{align}

Now, under the stochastic error mitigation for a continuous process, Eq. (\ref{eqn:noisyeqnsupple}) is modified to
\begin{align}
\frac{d }{dt} \rho_{\lambda}^{(Q)}(t)  =-i[H(t), \rho_{\lambda}^{(Q)}(t)] + \lambda \Delta \mc L\big[ \rho_{\lambda}^{(Q)}(t) \big],
\label{noisysupple}
\end{align}
where $\Delta \mc L= \mathcal{L}_{\mathrm{exp}}- \mathcal{L}_{\mathrm{est}}$.  Similar to the mitigation of physical errors via Richardson's extrapolation, we can obtain the approximation of the noise free expectation value of the observable $O$ as
\begin{align}
\braket{O(0)}_n^*=\sum_{j=0}^n \beta_j \braket{O}_{r_j \lambda},
\end{align}
where $\braket{O(0)}^*$ is the estimated noise free expectation value up to an error of order $O(\lambda^{n+1})$, and $\braket{O}_{r \lambda}$ is the measurement outcome after stochastic error mitigation, corresponding to $\rho_{r \lambda}^{(Q)}(T)$.

Hence, under stochastic error mitigation, the inequality of (\ref{bound1}) is modified  to
\begin{equation}
\begin{aligned}
 |\braket{O(0)}_n^*- \braket{O}_I |
 &\leq \gamma_n \bigg(\frac{ C(r_{\mathrm{max}}T) r_{\max}^{n+1} \Delta_{\mathrm{max}}}{\sqrt{N_{\mathrm{sample}}}}+\| O \|
\frac{ (r_{\mathrm{max}}\lambda T \|\Delta \mathcal{L}\|_{1})^{n+1} }{(n+1)!} \bigg),
\label{bound_general}
\end{aligned}
\end{equation}
with Eq. (\ref{condition1}) changed into
\begin{align}
r_{\mathrm{max}}\lambda T  \|\Delta \mathcal{L} \|_{1} = \mc O(1).
\label{condition2}
\end{align}
Here, we used the fact that the variance of the error-mitigated expectation value of the observable is amplified with the overhead coefficient $C$.

\sun{
From Eq.(\ref{bound_general}), the deviation between the ideal measurement outcome and the error-mitigated one is bounded independently with the Hamiltonian,
i.e., the to-be-simulated problem. The bound only relies on the noise model, the evolution time, the number of samples, and the parameters used in extrapolation.}

\section{Numerical simulation}
\label{appendix:numer}

\sun{
As we show in the section~\ref{appendix: model},
the variation of the performance of our error mitigation methods in terms of different Hamiltonians and noise models is theoretically well bounded, which indicates that the theory does apply for general Hamiltonian simulation with NISQ devices.
In this section, we report additional numerical simulation for the transverse field Ising model and frustrated spin-half model as $J_1-J_2$ model to verify the viability of our theory.
}

\begin{figure*}[htbp]
\centering
\includegraphics[width=1\textwidth]{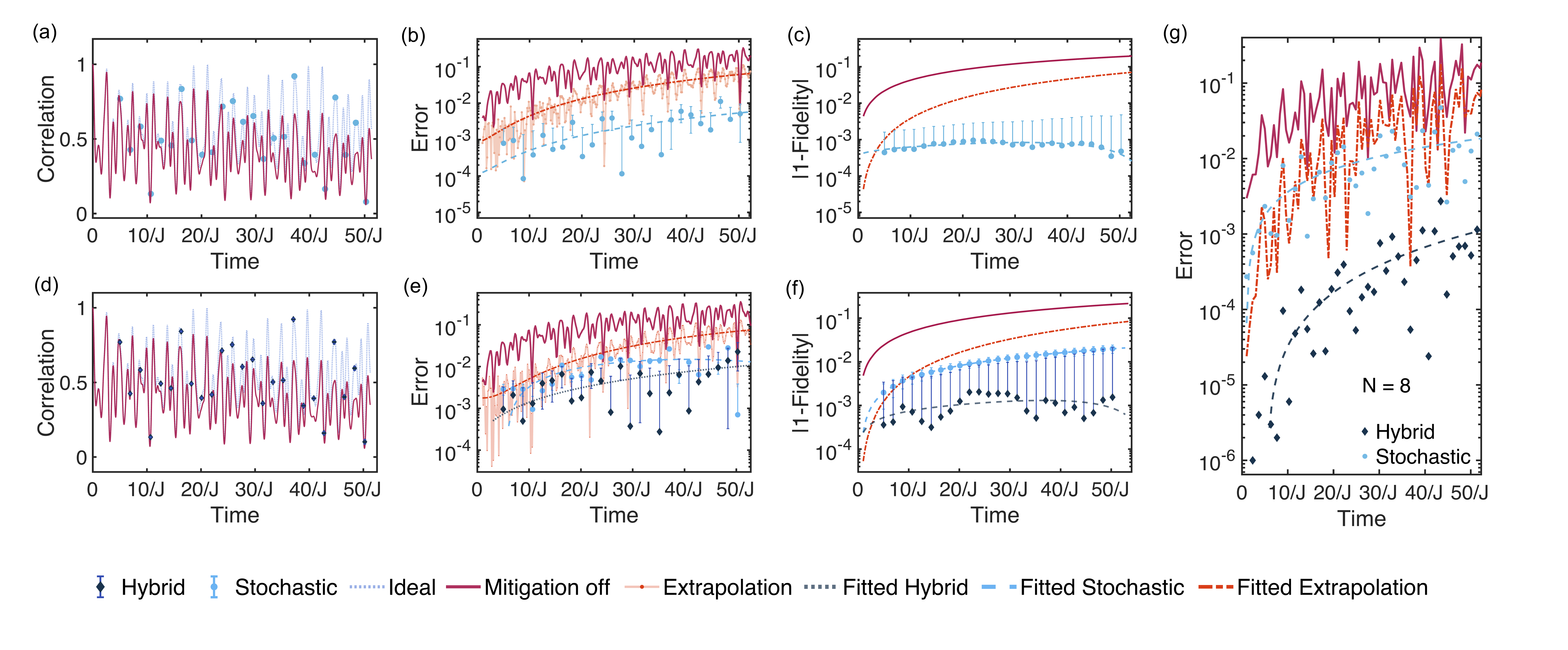}
\caption{
Numerical test of the performance of error mitigation schemes for transverse field Ising model without model estimation error ((a)(b)(c)) and with $10\%$ model estimation error $\lambda_{\textrm{exp}}  = 1.1 \lambda_{\textrm{est}}$ ((d)(e)(f)(g)).
We consider time evolution of the Ising Hamiltonian with  energy relaxation and dephasing. (a)$-$(f) considers four-qubit Hamiltonian with finite $(10^6)$ number of samples.
(g) considers eight-qubit Ising spin chain Hamiltonian with infinite number of samples.
(a) and (d) compares the time evolved normalised next-nearest-neighbour correlation function.
(b) and (e) shows the error between the exact value and the practical value.
(c) and (f) shows the fidelity of the effective density matrix $\rho^{\alpha}_{\textrm{eff}}$ and the ideal one $\rho_{I}$ under different error mitigation scheme $\alpha$.
(g) shows that the performance of stochastic and hybrid QEM.  Hybrid error mitigation scheme suppresses the error up to about four orders of magnitude even with $10\%$ model estimation error.
}
\label{fig:Ising}
\end{figure*}

\sun{
We first consider a four-qubit 1D transverse field Ising model
\begin{equation}
H= J\sum_{i=1}^4  \sigma_z^{(i)}  \sigma_z^{(i+1)} + h\sum_{j=1}^{4}   {\sigma}_{x}^{(i)}.
\end{equation}
We consider the quantum critical point at  $J=h=2\pi \times 4$ MHz where correlations exhibits power law decay instead of exponential decay.
The noise strength  $\lambda_{1}=\lambda_{2}=0.04$ MHz and errors of single-qubit operation $p_x=p_y=0.25\%$ and $p_z=0.5\%$ ,which are the same as in the main text for comparison.
We set the initial state to $(\ket{+})^{\otimes 4}$ with $\ket{+}=(\ket{0}+\ket{1})/\sqrt{2}$, evolve it to time $T=16\pi/J$, and measure the expectation value of the normalised next-nearest-neighbour correlation function
$\sum_{\langle\langle ij\rangle\rangle}   \sigma_x^{(i)}  \sigma_x^{(j)} $ . The total number of samples of the measurement is fixed to be $10^6$. To demonstrate the performance of stochastic and hybrid error mitigation much clearer, we consider eight-qubit Hamiltonian with infinite number of samples, as shown in Fig.~\ref{fig:Ising}(g).
}

\begin{figure*}[htb]
\centering
\includegraphics[width=1\textwidth]{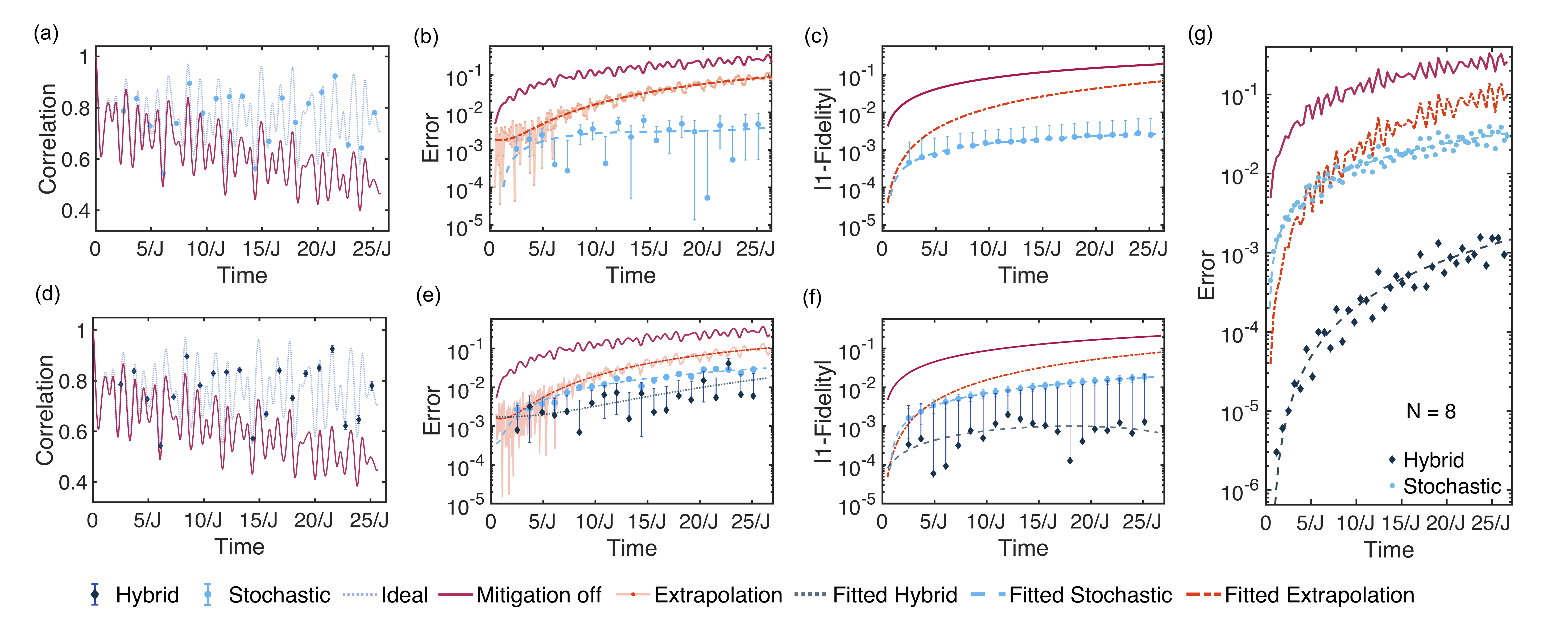}
\caption{
Numerical test of the performance of error mitigation schemes for frustrated quantum spin model without model estimation error ((a)(b)(c)) and with $10\%$ model estimation error $\lambda_{\textrm{exp}}  = 1.1 \lambda_{\textrm{est}}$ ((d)(e)(f)(g)).
The parameter settings are the same in the main text.
}
\label{fig:J1-J2}
\end{figure*}

\sun{
Next, we consider a four-qubit Hamiltonian simulation for  $J_1-J_2$ frustrated model with field
\begin{equation}
    H= J_{1}\sum_{\braket{ij}}  \sigma_z^{(i)}  \sigma_z^{(j)} + J_{2}\sum_{\braket{\braket{ij}}}  \sigma_z^{(i)}  \sigma_z^{(j)} - h\sum_{j=1}^{4}   \hat{\sigma}_{x}^{(i)}
\end{equation}
 where $\braket{ij}$  and $\braket{\braket{ij}}$ represent nearest-neighbour (NN) and next-nearest-neighbour (NNN) interactions, respectively.
This model has been widely investigated to describe the magnetism and phase transitions, and no exact solutions have been found with  general values of the coupling constants with  general values of the coupling constants $J_1$ and $J_2$.
At the point of $J_2 /J_1 =0.5$, the ground state of zero-field $J_1-J_2$ model is spin dimers and the antiferromagnetic to frustrated phase transition is believed to be near to  $J_2 /J_1 \sim 0.5$~\cite{bishop1998phase}. Therefore, a scale-up simulation of these models with error mitigated analog quantum simulators could be applied for discovering new physics.
We consider the quantum critical point at $J_2/J_1 = 0.5$ and set $J_1=h=2\pi \times 2$ MHz in the simulation.
We set the initial state to $(\ket{+})^{\otimes 4}$, evolve it to time $T=8\pi/J$, and measure the expectation value of the normalised NN correlation function
$\sum_{\langle ij\rangle}   \sigma_x^{(i)}  \sigma_x^{(j)} $ . The total number of samples of the measurement and error rates for are set the same as in the main text.
}

\sun{
From the simulation results shown in Fig.~\ref{fig:Ising} and~\ref{fig:J1-J2}, we clearly see that our stochastic and  hybrid algorithms can effectively suppress the errors during the evolution, which is consistent with the numerical simulation in the main text.
}

\section{Time dependent noise model}\label{appendix: timedepsimulation}
Here we show an example that our method also works for time-dependent noise when the time dependence is well characterised.
We consider the Ising model under low frequency noise~\cite{matsuzaki2011magnetic,matsuzaki2010quantum,kakuyanagi2007dephasing,yoshihara2006decoherence}, which is  described by
\begin{equation}
H=  \sum_{\langle ij\rangle}^n J_{ij}\hat{\sigma}_z ^{(i)} \hat{\sigma}_z ^{(j)}
 + \sum_{j=1}^{n} \left( h_j \hat{\sigma}_{z}^{(j)} +\lambda^{\prime} f_{j}(t) \hat{\sigma}_{z}^{(j)} \right),
\end{equation}
where $n$ is the number of qubits,  $\langle ij\rangle$ denotes the interactions between neighbour $i$ and $j$, $h_j$ is the coupling of the external magnetic field, and $f_j(t)$ describes the (noisy) interaction to the environment.
We assume that $f_j(t)$ is a classical Gaussian noise which satisfies $\overline{f_j(t)}=0$, $\overline{f_j(t) f_{j'}(t)}= \delta_{j j'}$, and higher order correlations are zero, where $\overline{f}$ denotes the ensemble average for a random variable $f$. After taking the ensemble average, the evolution of the density matrix averaged over trajectories can be described as
\begin{align}
\frac{\overline{d \rho(t)}}{d t} \simeq -i[H_0,\rho] + 2\lambda^{\prime 2} t \sum_{j=1}^n \left( \hat{\sigma}_{z}^{(j)}  \rho  \hat{\sigma}_{z}^{(j)}- \rho\right),
\end{align}
where $H_{0} =\sum_{\langle ij\rangle}^n J_{ij}\hat{\sigma}_z ^{(i)} \hat{\sigma}_z ^{(j)}
+ \sum_{j=1}^{n}  h_j \hat{\sigma}_{z}^{(j)} $. We refer a detailed derivation to Sec.~\ref{appendix: noisemodel}.

\begin{figure}[htb]
\centering
\includegraphics[width=0.98\textwidth]{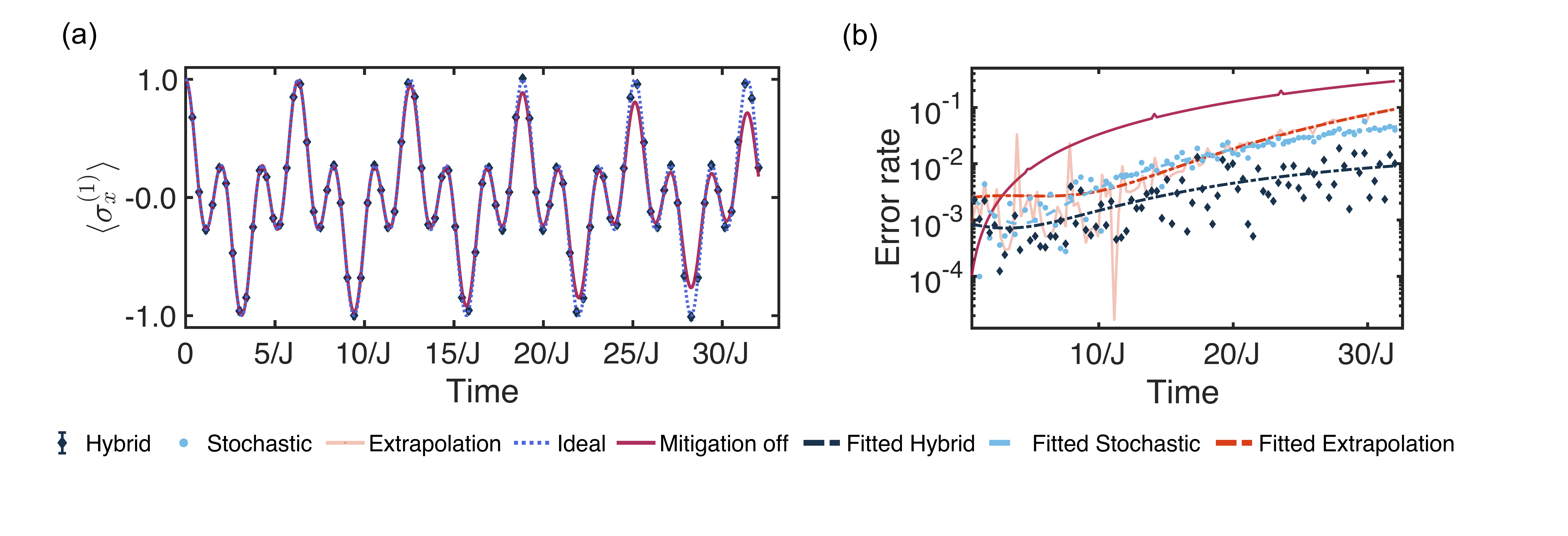}
\caption{Numerical test of the performance of error mitigation schemes for a two site quantum system. The system is affected by the time-dependent environmental noise, where the Hamiltonian reads $H= J \hat{\sigma}_z ^{(1)} \hat{\sigma}_z ^{(2)}
 - \sum_{j=1}^{2} \left( h_j \hat{\sigma}_{z}^{(j)} +\lambda^{\prime} f_{j}(t) \hat{\sigma}_{z}^{(j)} \right)$. We consider a low-frequency noise derived in section~\ref{appendix: noisemodel}, the dynamical equation is expressed as $\frac{\overline{d \rho(t)}}{d t} \simeq  -i[H_0,\rho] + 2\lambda^{\prime 2} t \sum_{j=1}^2 \left( \hat{\sigma}_{z}^{(j)}  \rho  \hat{\sigma}_{z}^{(j)}- \rho\right).$ The time correlated Lindblad noise operator is $L_{\textrm{dep}}=\sqrt{t}L$, where $L$ has the same form as in Fig. 2.  We set the coupling $J=2\pi\times 3$ MHz, field strength $h=0.5J$, noise strength $2\lambda^{\prime 2}=0.1$ MHz, time-dependent model estimation error $\lambda_{\rm exp}=1.2\lambda_{\rm est}$ and the sampling numbers $10^5$. As proven in Eq. (\ref{eqn:appendixc2}), noise rate is boosted by a factor $r_j^2$, and we used the scaling factor $r_1=1$, $r_2=1.5$ for Richardson extrapolation and the hybrid error mitigation.
 }
\label{fig:TimeDep}
\end{figure}

This result indicates that the averaged trajectory is equivalent to the time-dependent noisy evolution. By applying the hybrid error mitigation method, a combination of stochastic error mitigation and linear extrapolation, we show in Fig.~\ref{fig:TimeDep} that the time-dependent noise can be mitigated without detailed knowledge of the noise strength and noise type.

\subsection{Derivation of the time dependent noise model}
\label{appendix: noisemodel}
We consider a generic Ising Hamiltonian with time-dependent environmental noise as
\begin{equation}
H=  \sum_{\langle ij\rangle}^n J_{ij}\hat{\sigma}_z ^{(i)} \hat{\sigma}_z ^{(j)}
 + \sum_{j=1}^{n} \left( h_j \hat{\sigma}_{z}^{(j)} +\lambda^{\prime} f_{j}(t) \hat{\sigma}_{z}^{(j)} \right),
\end{equation}
where $\langle ij\rangle$ denotes the interactions between neighbour $i$ and $j$, and $ f_j(t)$ describes the interaction to the environment.

In the interaction picture, we divide the Schrodinger picture Hamiltonian into two parts:
\begin{equation}
\begin{aligned}
H_{0} &=\sum_{\langle ij\rangle}^n J_{ij}\hat{\sigma}_z ^{(i)} \hat{\sigma}_z ^{(j)} + \sum_{j=1}^{n}  h_j  \hat{\sigma}_{z}^{(j)} \\
H_{I} &=\sum_{j=1}^{n} \lambda^{\prime} f_{j}(t) \hat{\sigma}_{z}^{(j)}.
\end{aligned}
\end{equation} The interaction picture is defined through $\rho_{I}=e^{i H_{0} t} \rho e^{-i H_{0} t}$, and
the evolution equation now reads
\begin{equation}
\frac{d \rho_I}{d t}=-i[H_I, \rho_I].
\end{equation}
By taking a series expansion, we have
\begin{equation}
\begin{aligned}
\rho_I(t) &=\rho_I(0)-i \int_{0}^{t}\left[H\left(t^{\prime}\right), \rho_I\right] d t^{\prime}-\int_{0}^{t} \int_{0}^{t^{\prime}} d t^{\prime} d t^{\prime \prime}\left[H\left(t^{\prime}\right),\left[H\left(t^{\prime \prime}\right), \rho_I\left(t^{\prime \prime}\right)\right]\right] \\ & \simeq \rho_I(0)-i \int_{0}^{t}\left[H\left(t^{\prime}\right), \rho_I\right] d t^{\prime}-\int_{0}^{t} \int_{0}^{t^{\prime}} d t^{\prime} d t^{\prime \prime}\left[H\left(t^{\prime}\right),\left[H\left(t^{\prime \prime}\right), \rho_I(0)\right]\right]
\end{aligned}
\end{equation}By taking an ensemble average of the random variable, we have

\begin{equation}
\overline{\rho_{I}(t)} \simeq \rho(0)-\lambda^{\prime 2} \sum_{j, j^{\prime}=1}^{n} \int_{0}^{t} \int_{0}^{t^{\prime}} d t^{\prime} d t^{\prime \prime} \overline{f_{j}\left(t^{\prime}\right) f_{j}^{\prime}\left(t^{\prime \prime}\right)}\left[\hat{\sigma}_{z}^{(j)},\left[\hat{\sigma}_{z}^{\left(j^{\prime}\right)}, \rho(0)\right]\right]
\label{eqn:timedepnoise}
\end{equation}
where we have used
\begin{equation}
\overline{f_{j}(t)}=0, \overline{f_{j}\left(t^{\prime}\right) f_{j}\left(t^{\prime \prime}\right)}=\overline{f_{j}\left(t^{\prime}-t^{\prime \prime}\right) f(0)},  \overline{f_{j}(t) f_{j^{\prime}}(t)} \propto \delta_{j, j^{\prime}}\left(j, j^{\prime}=1,2, \cdots, n\right).
\end{equation}
For the white noise, $\overline{f_j\left(t^{\prime}\right) f_{j^{\prime}}\left(t^{\prime \prime}\right)}=\tau_{c} \delta_{t^{\prime},t^{\prime \prime}}\delta_{j, j^{\prime}}$, Eq.~(\ref{eqn:timedepnoise}) takes the form of
\begin{equation}
\begin{aligned}
\overline{\rho_{I}(t)} &\simeq \rho(0)-\sum_{j=1}^{n} \lambda^{\prime 2}t\tau_c\left(\rho(0)-\hat{\sigma}_{z}^{(j)} \rho(0) \hat{\sigma}_{z}^{(j)}\right)\\
&
=\left(1-\lambda^{\prime 2}  t\tau_c\right) \rho(0)+\sum_{j=1}^{n}  \lambda^{\prime 2} t\tau_c\hat{\sigma}_{z}^{(j)} \rho(0) \hat{\sigma}_{z}^{(j)} .\\
\end{aligned}
\end{equation}
By taking a small $t$, we obtain
\begin{equation}
\begin{aligned}
\frac{d \overline{\rho_I(t)}}{d t}  &\simeq-\lambda^{\prime 2} \frac{\tau_{c}}{2} \sum_{j=1}^{n} \left[ \hat{\sigma}_{z}^{(j)},\left[\hat{\sigma}_{z}^{(j)}, \overline{\rho_I(t)}\right]\right] \\
\end{aligned}
\end{equation}
In the Schr\"odinger picture, we have
\begin{equation}
\frac{d\rho}{dt}=-i[H_0,\rho]+e^{-iH_0t}\frac{d\rho_I}{dt}e^{iH_0t}
\end{equation}Because $H_0$ commutes with $\sigma_z^{(j)}$ , we have the expression for the dynamical equation
\begin{equation}
\begin{aligned}
\frac{\overline{d \rho(t)}}{d t} &\simeq -i[H_0,\rho] - \lambda^{\prime 2}\frac{\tau_{c}}{2}\sum_{j=1}^n \left[   \hat{\sigma}_{z}^{(j)}, \left[   \hat{\sigma}_{z}^{(j)}, \overline{\rho(t)}\right]\right] \\
& = -i[H_0,\rho] + \lambda^{\prime 2} \tau_{c} \sum_{j=1}^n \left(  \hat{\sigma}_{z}^{(j)}  \rho \hat{\sigma}_{z}^{(j)}- \rho\right)
\end{aligned}
\end{equation}
In the low frequency regime where  $\overline{f_j\left(t^{\prime}\right) f_{j^{\prime}}\left(t^{\prime \prime}\right)}=\delta_{j, j^{\prime}}$, similarly we have
\begin{equation}
\begin{aligned}
\overline{\rho_{I}(t)} &\simeq \rho(0)-\sum_{j=1}^{n} \lambda^{\prime 2} t^{2}\left(\rho(0)-\hat{\sigma}_{z}^{(j)} \rho(0) \hat{\sigma}_{z}^{(j)}\right),\\
&=\left(1-\lambda^{\prime2}\right) t^{2} \rho(0)+\sum_{j=1}^{n}  \lambda^{\prime 2} t^{2} \hat{\sigma}_{z}^{(j)} \rho(0) \hat{\sigma}_{z}^{(j)} \\
\end{aligned}
\end{equation}
Following similar transformation, we have the expression of the evolution under low frequency noise
\begin{equation}
\begin{aligned}
\frac{\overline{d \rho(t)}}{d t} &\simeq -i[H_0,\rho] - \lambda^{\prime 2} t \sum_{j=1}^{n} \left[\hat{\sigma}_{z}^{(j)},\left[\hat{\sigma}_{z}^{(j)}, \overline{\rho(t)}\right]\right] \\
&=-i[H_0,\rho] + 2\lambda^{\prime 2} t \sum_{j=1}^n \left( \hat{\sigma}_{z}^{(j)}  \rho  \hat{\sigma}_{z}^{(j)}- \rho\right).\\
\end{aligned}
\end{equation}

\end{document}